\documentclass[a4paper,11pt]{article}
\usepackage{amsfonts,amsmath,amsthm,amssymb}
\usepackage{epsfig}
\usepackage{color}

\newcommand{\difang}{\lambda}

\begin{document}

\title{The integer quantum Hall plateau transition \\
is a current algebra after all}

\author{Martin R.\ Zirnbauer, \\
Institut f\"ur Theoretische Physik, Universit\"at zu K\"oln, \\
    Z\"ulpicher Stra{\ss}e 77a, 50937 K\"oln, Germany}

\date{January 6, 2019}
\maketitle

\begin{abstract}
The scaling behavior near the transition between plateaus of the Integer Quantum Hall Effect (IQHE) has traditionally been interpreted on the basis of a two-parameter renormalization group (RG) flow conjectured from Pruisken's non-linear sigma model (NL$\sigma$M). Yet, the conformal field theory (CFT) describing the critical point remained elusive, and only fragments of a quantitative analytical understanding existed up to now. In the present paper we carry out a detailed analysis of the current-current correlation function for the conductivity tensor, initially in the Chalker-Coddington network model for the IQHE plateau transition and then in its exact reformulation as a supersymmetric vertex model. We develop a heuristic argument for the continuum limit of the non-local conductivity response function at criticality and thus identify a non-Abelian current algebra at level $n = 4$. Based on precise lattice expressions for the CFT primary fields we predict the multifractal scaling exponents of critical wavefunctions to be $\Delta_q = q(1-q)/4$. The Lagrangian of the RG-fixed point theory for $r$ retarded and $r$ advanced replicas is proposed to be the $\mathrm{GL}(r \vert r)_{n=4}$ Wess-Zumino-Witten model deformed by a truly marginal perturbation. The latter emerges from the NL$\sigma$M by a natural scenario of spontaneous symmetry breaking.
\end{abstract}

\section{Introduction}

Disordered two-dimensional electron gases at low temperatures and in a strong magnetic field exhibit the Integer Quantum Hall Effect (IQHE): they show plateaus in the Hall conductance as a function of the magnetic field strength, with plateau values that are integers in units of the conductance quantum $e^2 / h$. The present paper revisits the long-standing problem of the transition between adjacent plateaus. From real experiments and numerical simulations one knows that there exists a critical point where the single-electron localization length diverges (up to a cut-off set by the finite system size or temperature). Thus one is dealing with a critical phenomenon of the type of an Anderson localization-delocalization transition. What exactly happens at the critical point has eluded analytical understanding so far, in spite of considerable efforts that have been expended over the years. Recent developments have rekindled the interest in the plateau transition, as one expects it to be a paradigm for similar transitions that occur between different ground states of topological insulators and superconductors. An exciting scenario is that plateau-transition critical states might be realized on the $2d$ surface of disordered topological materials in three dimensions.

\subsection{Non-linear sigma model}\label{sect:1.1}

The first attempt to gain an analytical understanding of the IQHE plateau transition was made by Pruisken and collaborators \cite{Pruisken83}. Building on the theory of weak localization in a weak magnetic field, they wrote down a non-linear sigma model that has the dissipative conductivity $\sigma_{xx}$ and the Hall conductivity $\sigma_{xy}$ for its two coupling constants. Pruisken's key observation was that the breaking of parity symmetry by the magnetic field makes it possible for a so-called topological $\theta$-term (with topological angle $\theta$, and $\sigma_{xy} = \theta / 2\pi$ in units of $e^2 / h$) to appear in the field-theory Lagrangian.

Pruisken's Lagrangian is usually written in terms of a matrix field
\begin{equation}\label{eq:NLsM-Q}
    Q = u\, \Sigma_3\, u^{-1} , \quad \Sigma_3 = \mathrm{diag} (+\mathbf{1} \,, - \mathbf{1}) ,
\end{equation}
where the diagonal matrix $\Sigma_3$ accounts for the distinction between the two sectors that originate from the retarded ($+$) and advanced ($-$) single-electron Green's functions. Alternatively, one can express the Lagrangian $\mathcal{L}_{{\rm NL} \sigma {\rm M}}$ of Pruisken's theory in terms of the (gauge-dependent) matrix-valued one-forms $j = u^{-1} \partial u$ and $\bar{j} = u^{-1} \bar\partial u$, where $\partial = dz\, \partial_z$ and $\bar \partial = d\bar{z}\, \partial_{\bar z} \,$. The precise nature of $u$ depends on whether one handles the disorder by invoking fermionic replicas, or bosonic replicas, or the Wegner-Efetov supersymmetry method {\cite{Efetov, Wegner-Book}}. In the three cases one has $u \in U$ with $U = \mathrm{U}(2r)$, or $U = \mathrm{U} (r,r)$, or $U = \mathrm{U} (r,r|2r)$, respectively, with $r$ the number of replicas. We prefer the last method, in which case $\Sigma_3 = \mathrm{diag} (+ \mathbf{1}_{r\vert r} \,, -\mathbf{1}_{r \vert r})$ and
\begin{align}\label{eq:NLsM}
    - \mathrm{i} \mathcal{L}_{{\rm NL} \sigma {\rm M}} &= (\sigma_{xx} + \mathrm{i}\, \sigma_{xy})\, \mathrm{STr}\, j^{+ -} \wedge \bar{j}^{- +} \cr &+ (\sigma_{xx} - \mathrm{i}\, \sigma_{xy})\, \mathrm{STr}\, j^{- +} \wedge \bar{j}^{+ -} \,,
\end{align}
where $\mathrm{STr}$ denotes the supertrace, and the index pairs $+-$ and $-+$ indicate off-diagonal blocks defined by the decomposition
\begin{equation*}
    j = \begin{pmatrix} j^{++} &j^{+-}\cr j^{-+} &j^{--} \end{pmatrix}
\end{equation*}
with respect to the eigenspaces of the retarded-advanced signature $\Sigma_3\,$. It should be stressed that $\mathcal{L}_{{\rm NL} \sigma {\rm M}}$ is really meant to be a Lagrangian with target space not the group $U$ but rather a Riemannian symmetric (super-)\-space $U / K$ where $K \subset U$ is the centralizer of $\Sigma_3\,$. While the principal bundle $U \to U/K$ is non-trivial (obstructing the existence of a global section $u$) the lift assumed in Eq.\ (\ref{eq:NLsM}) from $U/K$ to $U$ still makes sense, since the off-diagonal blocks of $j$ and $\bar{j}$ transform as covariant field strengths under local gauge transformations $u \mapsto u k$ for $k\in K$. Note that the low-energy degrees of freedom in the $U/K$-valued field $Q = u \Sigma_3 u^{-1}$ are Goldstone modes whose \emph{raison d'etre} is to restore the symmetry under $U$ which is spontaneously broken to $K$ by the appearance of a non-zero density of states {\cite{Wegner,SchaeferWegner,McKane-Stone}}.

Assuming Pruisken's Lagrangian in conjunction with natural expectations of what should happen in limiting cases, a renormalization group (RG) flow for the running couplings $\sigma_{xx}$ and $\sigma_{xy}$ was conjectured {\cite{Pruisken83, Khmelnitskii83}}. Its central feature is a fixed point at $\sigma_{xy} = 1/2$ (mod 1) and an unknown value of $\sigma_{xx}\,$. The fixed point is expected to be isolated and universal; it is supposed to be stable in the $\sigma_{xx}$ direction and unstable in the $\sigma_{xy}$ direction.

While this so-called Pruisken-Khmelnitskii scaling picture of the IQHE plateau transition inspired a lot of further activity and provided an appealing framework in which to think about the transition, it has to be said that Pruisken's Lagrangian (for electrons in the lowest Landau level, or in a strong magnetic field, where the bare coupling $\sigma_{xx}$ is small) was never derived in a mathematically controlled manner, much less has it led to a quantitative understanding of what is the exact nature of the critical point. Indeed, the obvious and urgent question of what is the conformal field theory describing the scaling limit of the critical point, remained open for 35 years.

\subsection{The puzzle}

Pruisken's Lagrangian (\ref{eq:NLsM}) features an invariance under constant transformations $Q({\bf r}) \mapsto u\, Q({\bf r})\, u^{-1}$ for $u \in U = \mathrm{U} (r,r|2r)$. This global symmetry is dictated by the supersymmetry method applied to the disordered electron problem at hand. It is also at the heart of an apparent paradox, as follows.

A hallmark of conformal field theory in two dimensions is the separation of the energy-momentum tensor into holomorphic and anti-holomorphic parts, giving rise to two Virasoro algebras, one for the left-moving modes and another one for the right-movers; we speak of ``holomorphic factorization'' for short. In the presence of Lie group symmetries, one expects the organization by conformal blocks for the holomorphic and anti-holomorphic fields to be refined by an underlying current algebra. For the case of  theories with non-Abelian conserved currents, holomorphic factorization is well understood to be realized by Wess-Zumino-Witten models \cite{Witten, Witten-holofact}.

Now since Pruisken's non-linear sigma model enjoys invariance under the global symmetry group $U = \mathrm{U}(r,r|2r)$, it has a Noether current:
\begin{equation}
    \partial^\nu J_\nu = 0 \,, \quad J_\nu = u \begin{pmatrix} 0 &(u^{-1} \partial_\nu u)^{+-} \cr (u^{-1} \partial_\nu u)^{-+} &0 \end{pmatrix} u^{-1} .
\end{equation}
An equivalent formulation of that conservation law is
\begin{displaymath}
    \bar\partial J + \partial \bar{J} = 0 \,, \quad J = - \mathrm{i}\, u
    \begin{pmatrix} 0 &j^{+-} \cr j^{-+} &0 \end{pmatrix} u^{-1} , \quad \bar{J} = \mathrm{i}\, u \begin{pmatrix} 0 &\bar{j}^{+-} \cr \bar{j}^{-+} &0 \end{pmatrix} u^{-1} .
\end{displaymath}
Holomorphic factorization at the level of current algebra would amount to the stronger statement of \emph{two} conservation laws:
\begin{displaymath}
    \bar\partial J = 0 = \partial \bar{J} ,
\end{displaymath}
with holomorphic and anti-holomorphic currents $J$ and $\bar{J}$, respectively. There exists no known scenario by which such a factorization or doubling of conservation laws might happen for a non-linear sigma model like that of Pruisken. This is why we must abandon it (as a description of the RG-fixed point for the phase transition) and replace it by something else.

That something else, unfortunately, is not easy to come by. In fact, Chamon, Mudry and Wen \cite{CMW, MCW} demonstrated some time ago that a current algebra with Lie supergroup symmetry $U$ (for disordered Dirac fermions modeling critical points similar to that of the IQHE plateau transition) suffers from the existence of infinitely many $\mathrm{Ad} (U)$-invariant perturbations which grow under renormalization. [In the formulation by a Wess-Zumino-Witten (WZW) model, these are given by traces of powers of the WZW field.] Thus, unless there exists some robust mechanism of protection against these perturbations, the WZW models for our case with symmetry $\mathrm{U}(r,r|2r)$ will be unstable under renormalization and, therefore, cannot describe the universal fixed point in question. Later on, exact results by Read and Saleur \cite{ReadSaleur} for field-theoretic models closely related to the IQHE plateau transition, ruled out any description by a CFT with a chiral algebra of holomorphic currents matching the Lie supergroup $U$ of global symmetries.

Perhaps the most direct indication of the difficulty is the following.
The theory (\ref{eq:NLsM}) needs regularization for the zero modes generated by the non-compact symmetry $\mathrm{U}(r,r)$. The usual scheme is to add a small imaginary part to the energy argument of the retarded and advanced Green's functions $G^\pm$ or to consider a physical system with a flux-absorbing boundary. For any such choice of $K$-invariant regulator, the mono-type current-current correlators $\langle J^{++}({\bf r}) J^{++}({\bf r}^\prime) \rangle$ and $\langle J^{--}({\bf r}) J^{--}({\bf r}^\prime) \rangle$ (computed in any acceptable model for the IQHE plateau transition) turn out to be trivial, i.e.\ short-ranged, owing to the supersymmetries in the unprobed sector. On the other hand, the poly-type correlators $\langle J^{+-}({\bf r}) J^{-+}({\bf r}^\prime) \rangle$ and $\langle J^{++}({\bf r}) J^{--}({\bf r}^\prime) \rangle$ do exhibit critical behavior at the transition. These irrefutable facts \cite{EversMirlin} are at odds with any standard scenario of $U$-current algebra and confirm that the symmetry doubling $U \to U_L \times U_R$ argued in \cite{Affleck} cannot take place here.

One might now ponder the idea to look for a gauged WZW model, appealing to a procedure known as the Goddard-Kent-Olive (GKO) coset construction. Alas, gauging by a group $H$ breaks the given $U$-symmetry \emph{explicitly}, unless $H$ and $U$ centralize each other. This leaves but the center $\mathrm{U}(1)$ of $U$ as a possible group for gauging in our case. Yet, taking the GKO coset by the central subgroup $\mathrm{U} (1)$ fails to match the known properties of the critical point at hand.
Indeed, gauging $\mathrm{SU}(r,r|2r)$ by $\mathrm{U}(1)$ amounts to projecting to the target space
\begin{equation*}
    \mathrm{SU}(r,r|2r) / \mathrm{U}(1) = \mathrm{PSU}(r,r|2r) ,
\end{equation*}
and the Wess-Zumino-Witten model on $\mathrm{PSU}(r,r|2r)$ would predict some trivial correlation functions of the type $\langle J^{++} J^{++} \rangle$ and $\langle J^{--} J^{--} \rangle$ to be non-trivial -- a severe clash with known facts. Incidentally, this discrepancy rules out the proposal of \cite{Tsvelik-PSL} (following earlier ideas by the author). To summarize, it has been a long-standing puzzle how to reconcile the CFT axiom of holomorphic factorization of conserved currents with the established phenomenology for the critical point of the IQHE plateau transition.

\subsection{Proposed resolution of the puzzle}

In the present paper, we start from the Chalker-Coddington network model \cite{Chalker} for the IQHE plateau transition and its exact reformulation \cite{BWZ2} as a supersymmetric (SUSY) vertex model due to N.\ Read. Based on formulas taken from the Kubo theory of linear response, we argue that the non-local response function, a.k.a.\ the conductivity tensor, exhibits a form of holomorphic factorization at the critical point. By expressing the critical response function as a current-current correlator in the SUSY vertex model, we are led to propose lattice candidates for the continuum fields of a non-Abelian current algebra. In order to match the known phenomenology, the level of that current algebra has to be $n = 4\,$. Most importantly, our current algebra is $\widehat{\mathfrak{gl}} (r|r)_4\,$, which has only half the rank of the complexification $\mathfrak {gl} (2r|2r)$ of $\mathrm{Lie}\, \mathrm{U}(r,r|2r)$. The underlying heuristic is that for a suitable choice of basis, $\mathfrak{gl}(2r|2r)$ splits into four blocks of equal size $(r|r)$ such that the two diagonal blocks give rise to two chiral algebras $\widehat{ \mathfrak{gl}} (r|r)$ for the left- and right-moving modes, while the off-diagonal blocks bosonize to a WZW field $M$ and its inverse $M^{-1}$. (The expert reader may recognize some similarity with the free-boson representation of the $\mathrm{SU} (2)_1$ WZW model for the 6-vertex model at its $\mathrm{SU} (2)$-invariant point.) The off-diagonal position of $M$ in the global symmetry algebra $\mathfrak{gl}(2r|2r)$ obliterates the Chamon-Mudry-Wen objection of RG-instability, as it prevents the relevant perturbations $\mathrm{STr}\, M^q$ from appearing in the Lagrangian. We should add the remark that the position of our blocks $\mathfrak{gl}(r|r)$ inside $\mathfrak{gl}(2r|2r)$ is not unique but depends on a choice of maximal commutative subalgebra in the tangent space $T_{eK}(U/K)$.

The operator product expansions (OPE) for the currents $J$, $\bar{J}$ and the field $M$ are those of an integrable deformation of the $\mathrm{GL} (r|r)_{n=4}$ WZW model. The deformation has the important effect of setting the conformal weights of $M$ to zero -- a feature required by the physics of Anderson transitions in symmetry class $A$ \cite{EversMirlin}. Let us stress that the deformation has a direct physical meaning: it reflects the existence of critical current-current correlation functions of two different types, namely $\langle J^{+-} J^{-+} \rangle$ as well as $\langle J^{++} J^{--} \rangle$. It also predicts a universal amplitude ratio for these, which is expected to be verifiable by numerical simulation of the network model.

By the OPE between the currents and the WZW field, powers of the boson-boson block $M_{00}$ of $M$ are Kac-Moody primary fields (this is literally true for $r = 1$ and still true in adapted form for $r > 1$). It then follows from the Sugawara form of the deformed energy-momentum tensor that the scaling dimension of $M_{00}^q$ is $q(1-q)/n$. Since $M_{00}^q$ represents the $q^{\rm th}$ moment of a critical wavefunction for the network model \cite{BWZ2}, we are led to predict (with $n = 4$) that the multifractal scaling exponents of critical wavefunctions are $\Delta_q = q(1-q)/4\,$, in good agreement with recent computer results.

Let us finish this introduction and overview by offering more perspective on our current algebra scenario. The logic begins with the reminder that we are \emph{compelled} to regularize the field theory by a perturbation that explicitly breaks the non-compact symmetry $U$; we do so with a $K$-invariant regulator such as $\mu \int d^2 r \, \mathrm{STr} \, \Sigma_3 \, Q$ in the non-linear sigma model. If the physical system were in a metallic phase of delocalized states (realized in space dimensions $d \geq 3$) the broken symmetry $U$ would remain spontaneously broken in the limit of a vanishing regulator $\mu \to 0\,$. On the contrary, in an insulating phase of localized states (as realized, e.g., by the Chalker-Coddington network model off the critical point) the $U$-symmetry is known to be restored for $\mu \to 0\,$. The symmetry restoration happens over a scale set by the localization length (or inverse mass scale), making all current-current correlators short-ranged in the infrared limit. Now, at our phase-transition point some of these correlators, $\langle J^{+-} J^{-+} \rangle$ and $\langle J^{++} J^{--} \rangle$, are critical, while some others, $\langle J^{++} J^{++} \rangle$ and $\langle J^{--} J^{--} \rangle$, are trivial by first principles and cannot ever become critical. The upshot is that the current-current correlation functions at criticality determine on the Lie superalgebra $\mathfrak{gl}(2r|2r)$ of symmetries a bilinear form which is \emph{degenerate and fails to be $U$-invariant}. In short, the broken $U$-symmetry is not restored but remains spontaneously broken at the critical point. Hence any attempt at a $U$-invariant Lagrangian formulation of the RG fixed-point theory leads to novel and exotic CFT mathematics (as communicated by the author in various talks over the past years).

Adopting the formalism of current algebras (or affine Lie superalgebras, to be accurate), the only way to make do with a conventional form thereof is to choose some decomposition $\mathfrak{gl} (r|r)_L \oplus \mathfrak{gl}(r|r)_R \subset \mathfrak{gl}(2r|2r)$ (explicitly breaking the unbroken $K$-invariance) such that the current-current bilinear form becomes non-degenerate on restriction to both $\mathfrak{gl}(r|r)_L$ and $\mathfrak{gl}(r|r)_R\,$. The latter option is what we develop in the present paper. More specifically, we outline a plausible scenario of spontaneous symmetry breaking by the RG flow approaching a $U$-orbit of \emph{nilpotent} elements in $\mathrm{Lie}\, \mathrm{U}(r,r|2r)$. A very powerful feature of our scenario is that integration over the stiff Goldstone modes due to the broken symmetry yields \emph{precisely} the deformation of $\mathrm{GL}(r|r)_4$ which is needed in order to match the known phenomenology.

\subsection{Summary of contents}

The contents are summarized as follows. Section 2 begins with a short introduction to the Chalker-Coddington network model for the IQHE plateau transition. We point out that the model has a $\mathbb{Z}_4$ spectral symmetry, with the consequence that eigenvalues and eigenfunctions of the one-step time-evolution operator come as quadruples modeled after the fourth roots of unity (Sect.\ \ref{sect:Z4}). By introducing a suitable spinor basis, we get a clear view of the Dirac fermion that emerges at long wavelengths in the absence of disorder (Sect.\ \ref{sect:spinor}). In order to handle the strong random phase disorder of the network model, we review Read's method (Sect.\ \ref{sect:Read}) leading to the exact reformulation as a supersymmetric (SUSY) vertex model (Sect.\ \ref{sect:SUSY-VM}). In Section 3 we investigate a toy model of $n$ Dirac species coupled to a random $\mathfrak{su}(n)$ gauge field. The motivation here is to review some necessary background on current algebra (Sects.\ \ref{sect:Howe-pair}--\ref{sect:3.3}) and explain what we mean by a $\mathrm{GL}(r|r)$ Wess-Zumino-Witten model (Sect.\ \ref{sect:GL-WZW}).

The new insights and main results are presented in Section 4. There, we start from the observation (Sect.\ \ref{sect:tell}) that the $\widehat{\mathfrak{gl}}(p|q)_n$ current algebra for $p = q$ admits a truly marginal deformation to make the conformal weights of the fundamental field vanish. We also review the current algebra conundrum (Sect.\ \ref{sect:puzzle}) and the general argument for symmetry doubling in a CFT with Noether-conserved currents (Sect.\ \ref{sect:Affleck}). In the sequel, we return to the analysis of the critical point of the IQHE plateau transition. Our method is to go back and forth between the network model and the SUSY vertex model, using identities and results known on one side to complement those known on the other side. We begin by adapting the symmetry doubling argument to the $\mathrm{U}(1)$ conserved current of the network model (Sect.\ \ref{sect:SD-adapted}). We then continue with a study of the current-current correlation function for the conductivity tensor (Sect.\ \ref{sect:Kubo}). Arguing heuristically, we pass to the continuum limit of the non-local conductivity response function at criticality (Sect.\ \ref{sect:critical}). The outcome is interpreted in the framework of the SUSY vertex model (Sect.\ \ref{sect:resp-VM}), leading to concrete lattice expressions for the holomorphic currents in the continuum limit (Sect.\ \ref{sect:4.6}). We go on to indicate how a Wess-Zumino-Witten field emerges from the lattice theory (Sect.\ \ref{sect:bosonize}), and how the WZW model is deformed by a truly marginal perturbation (Sect.\ \ref{sect:deform}) to match the known phenomenology. We then compute the spectrum of multifractal scaling exponents -- doing so twice, first from the operator formalism (Sect.\ \ref{sect:multif}) and then from the functional integral of the deformed $\mathrm{GL}(1|1)$ WZW model (Sect.\ \ref{sect:GFF}). In Sect.\ \ref{sect:SBS} we explain, first in a simplified setting (Sect.\ \ref{sect:4.13.1}) and then for the full model (Sect.\ \ref{sect:4.13.2}), how RG flow toward a nilpotent $U$-orbit leads to spontaneous symmetry breaking. In particular, we demonstrate (in Sects.\ \ref{sect:4.13.3}, \ref{sect:4.13.4}) that by integrating out the stiff Goldstone modes due the broken symmetry one obtains exactly the truly marginal deformation of the WZW model. We finish with the CFT prediction for the mean conductance (Sect.\ \ref{sect:mean-G}).

\section{Analysis of the network model}\label{sect:2}

\begin{figure}
    \begin{center}
        \epsfig{file=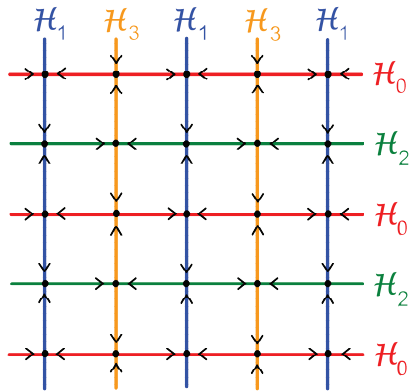,height=6.5cm}
    \end{center}
    \caption{
    A portion of the square lattice of the Chalker-Coddington network model. The direction of motion is determined by the arrows on the links. The Hilbert space of the model is a sum of subspaces $\oplus_{i=0}^3 \mathcal{H}_i$ generated by the link vectors for even rows ($\mathcal{H}_0$), even columns ($\mathcal{H}_1$), odd rows ($\mathcal{H}_2$), and odd columns ($\mathcal{H}_3$).}
    \label{fig:CC}
\end{figure}

The network model, as conceived by Chalker and Coddington in \cite{Chalker}, simulates the IQHE single-electron dynamics on a two-dimensional array of directed links connecting the sites of a square lattice; see Fig.\ \ref{fig:CC}.
The Hilbert space is $\mathcal{H} = \bigoplus_\ell \mathbb{C}_\ell$ with one copy $\mathbb{C}_\ell \cong \mathbb{C}$ for each link $\ell$. Proceeding by discrete steps in time, the quantum dynamics ($\psi_{t+1} = U \psi_t$) is generated by a unitary evolution operator $U = U_{\rm r} U_{\rm s}$ composed of a random unitary, $U_{\rm r}\,$, and a deterministic unitary, $U_{\rm s}\,$. These are defined as follows. Firstly, fixing a basis $\{ \vert \ell \rangle \}$ for $\mathcal{H}$ of unit vectors $\vert \ell \rangle \in \mathbb{C}_\ell \,$, one takes $U_{\rm r}$ to be diagonal: $U_{\rm r} \vert \ell \rangle = \vert \ell \rangle\, \mathrm{e}^{ \mathrm{i} \phi_\ell}$ with random phases $\phi_\ell$ that are uniformly distributed and statistically independent of each other. Secondly, to define $U_{\rm s}$ let $\ell$ be any link and denote by $\ell_\pm$ the two links that follow from $\ell$ by making a left turn ($+$) or right turn ($-$). Then, at the critical point to be studied here,
\begin{equation}\label{eq:Us}
    U_{\rm s} \vert \ell \rangle = \vert \ell_+ \rangle a_+ + \vert \ell_- \rangle a_-\, , \quad a_\pm = \mathrm{e}^{\pm \mathrm{i} \pi / 4} / \sqrt{2} \,.
\end{equation}
(The model becomes non-critical for $\vert a_+ \vert \not= \vert a_- \vert$.) Known as Kac-Ward amplitudes \cite{KacWard}, these conventions define an operator $U_{\rm s}$ which is invariant under rotations of the square lattice by $\pi/2$ and integer multiples thereof.

Most of the discussion below concerns the specific setting of a torus or rectangular network with periodic boundary conditions in both directions.

\subsection{$\mathbb{Z}_4$ spectral symmetry}\label{sect:Z4}

The spectrum of $U$ for the network model, on a torus with an even number of links in both directions of the square lattice, is already determined by the spectrum of its fourth power, $U^4$. To demonstrate this fact, let the Hilbert space be decomposed as $\mathcal{H} = \bigoplus_{l=0}^3 \mathcal{H}_l$ where the subspaces $\mathcal{H}_0\,,\ldots, \mathcal{H}_3$ are spanned by the basis vectors $\vert \ell \rangle$ for links $\ell$ on even rows, even columns, odd rows and odd columns of the square lattice, respectively; see Fig.\ \ref{fig:CC}. Then, taking $l$ modulo $4$, one observes that $U$ maps $\mathcal{H}_l$ to $\mathcal{H}_{l+1}\,$, and its fourth power decomposes into four maps $U^4 :\; \mathcal{H}_l \to \mathcal{H}_l$ ($l = 0, \ldots, 3$). It directly follows that if $\psi \in \mathcal{H}_0$ is an eigenvector of $U^4$ with eigenvalue $\mathrm{e}^{ \mathrm{i} \varepsilon}$, then
\begin{equation}\label{eq:extend}
    \Psi_m = \sum_{l=0}^3 \big( \mathrm{e}^{-\mathrm{i} (\varepsilon + 2\pi m) / 4} U \big)^l \, \psi \qquad (m = 0, \ldots, 3)
\end{equation}
is an eigenvector of $U$ with eigenvalue $\mathrm{e}^{\mathrm{i} (\varepsilon + 2\pi m) / 4}$. Thus, eigenvalues and eigenvectors are grouped into quadruplets modeled after the fourth roots $\mathrm{e}^{ \mathrm{i} \pi m / 2}$ of unity. Note that $m$ in the formula above may be viewed as a discrete angular momentum (or spin), while $2\pi l/4$ is a discrete angle.

\subsection{Spinor basis}\label{sect:spinor}

A key step in identifying the critical theory is to pass from the discrete setting to the continuum. To motivate the construction of good variables in which to take the continuum limit, we temporarily turn off the disorder, setting $U_{\rm r} \equiv 1$. The resulting operator $U \equiv U_{\rm s}$ is translation-invariant and can be diagonalized by hand using standard Bloch theory as follows.

\begin{figure}
    \begin{center}
        \epsfig{file=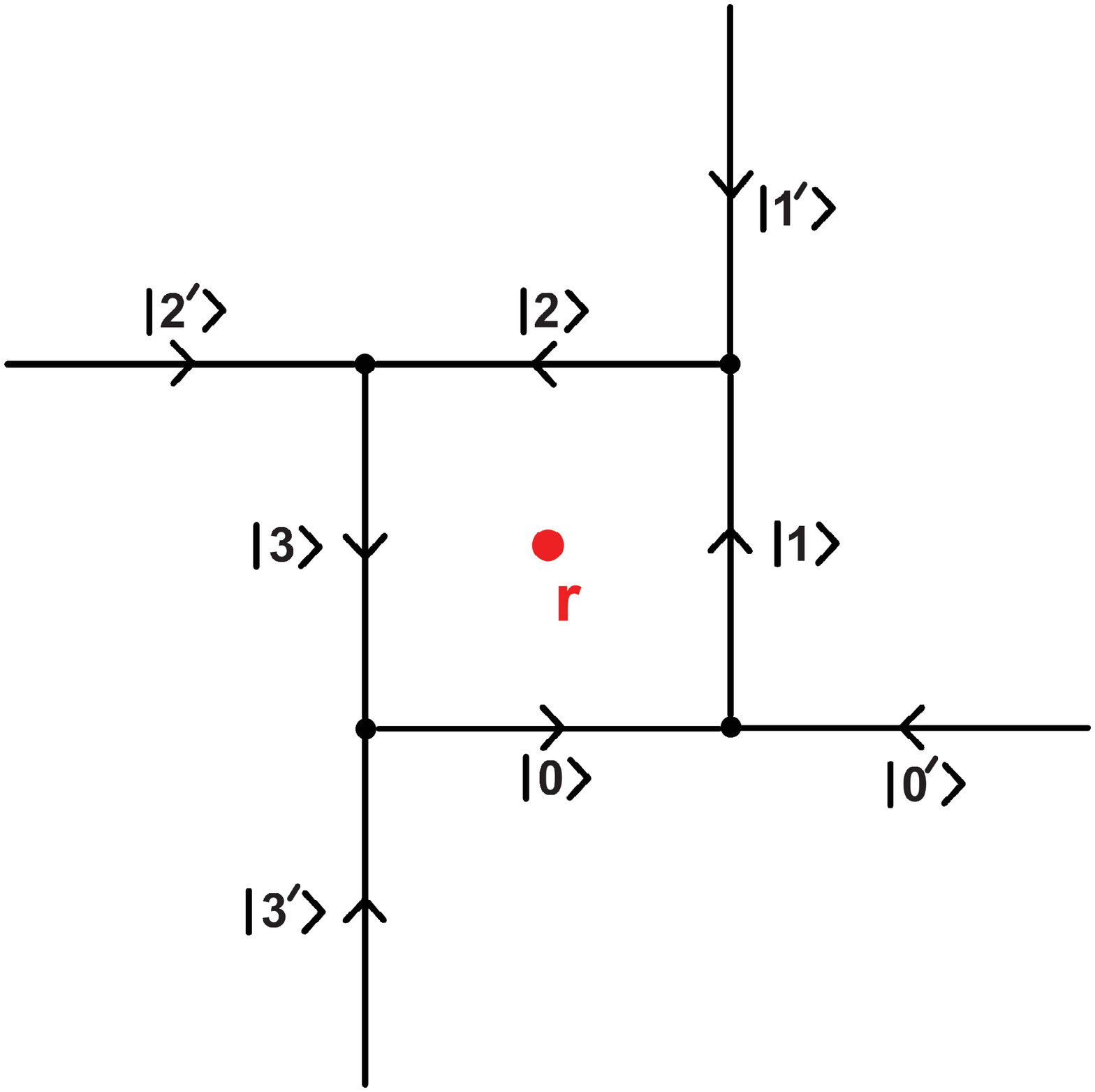,height=6cm}
    \end{center}
    \caption{
    Network model unit cell adopted in Sect.\ \ref{sect:spinor}. The unit cell consists of 8 links, four internal ones (without prime) circulating around and four external ones (with prime) directed at the central plaquette. The position ${\bf r}$ of a unit cell is given by the midpoint of its central plaquette.}
    \label{fig:unit-cell}
\end{figure}

The unit cell we choose consists of eight links, four of which are ``internal'' (circulating around a central plaquette); the other four are ``external'' (impinging on the central plaquette); see Fig.\ \ref{fig:unit-cell}.
Recalling the decomposition $\mathcal{H} = \bigoplus_{l=0}^3 \mathcal{H}_l \,$, let the 8-dimensional subspace for a given unit cell be spanned by $\vert l^\mathrm{int} \rangle \equiv \vert l \rangle \in \mathcal{H}_l$ and $\vert l^\mathrm{ext} \rangle \equiv \vert l^\prime \rangle \in \mathcal{H}_l$ ($l = 0, \ldots, 3$). We then introduce an orthonormal spinor basis $e_l^\pm$ as
\begin{equation}\label{eq:OSB}
    e_l^+ = \frac{\vert l \rangle + \vert l^\prime \rangle}{\sqrt{2}} \,, \quad e_l^- = \frac{\vert l \rangle - \vert l^\prime \rangle}{\sqrt{2}} \,  \mathrm{i}^l .
\end{equation}
Unit cells are labeled by the position ${\bf r}$ of their plaquette center. The eight basis vectors for the unit cell ${\bf r}$ are denoted by $e_l^\pm({\bf r})$, and an orthonormal plane-wave basis for the entire network is provided by
\begin{equation}\label{eq:spin-base}
    \tilde{e}_l^\sigma({\bf k}) = \frac{1}{\sqrt{\rm vol}} \sum_{\bf r} \mathrm{e}^{-\mathrm{i} {\bf k}\cdot{\bf r}}\, e_l^\sigma({\bf r}) \qquad (\sigma = \pm) .
\end{equation}
In this basis of good momentum ${\bf k}\,$, the matrix of the deterministic factor $U_\mathrm{s}$ of the network model operator $U$ block-diagonalizes to four $2 \times 2$ blocks, one for each transition $U_{\rm s} :\; \mathcal{H}_l \to \mathcal{H}_{l+1}\,$. To make the momentum dependence of the blocks explicit, we write ${\bf k}\! \cdot \!{\bf r} = k_x x + k_y y$ with ${\bf r} = (x,y) \in \mathbb{Z}^2$ and ${\bf k} = (k_x,k_y) \in [0,2\pi]^2$ where $k_x$ (resp.\ $k_y$) is the wave number along the axis of $\mathcal{H}_0$ and $\mathcal{H}_2$ (resp.\ $\mathcal{H}_1$ and $\mathcal{H}_3$). For the first block of transition ($\mathcal{H}_0 \to \mathcal{H}_1$) we then obtain
\begin{align}
    &\left\langle \tilde{e}_1^\sigma ({\bf k}) \vert U_\mathrm{s} \vert \tilde{e}_0^{\tau} ({\bf k}) \right\rangle = u_{1 0}^{\sigma\tau}(k) , \cr
    &\begin{pmatrix} u_{1 0}^{++} &u_{1 0}^{+ -}\cr u_{1 0}^{- +} &u_{1 0}^{- -} \end{pmatrix} = \frac{1}{2} \begin{pmatrix} 1 + \mathrm{e}^{-\mathrm{i} k_y} &\mathrm{i} (1-\mathrm{e}^{-\mathrm{i} k_y}) \cr - \mathrm{i} (1-\mathrm{e}^{-\mathrm{i} k_y}) &1 + \mathrm{e}^{-\mathrm{i} k_y} \end{pmatrix} \equiv u_{1 0}(k).
\end{align}
The three other blocks, namely $u_{2 1}\,$, $u_{3 2}\,$, and $u_{0 3}\,$, follow from the invariance of $U_{\rm s}$ under rotations by integer multiples of $\pi/2$. They are
\begin{equation}\label{eq:transit}
    u_{l,\,l-1}(k) = \frac{1}{2} \begin{pmatrix} 1 + \mathrm{e}^{\mathrm{i} \kappa} &\mathrm{i}^l (1-\mathrm{e}^{\mathrm{i} \kappa}) \cr \mathrm{i}^{-l} (1-\mathrm{e}^{\mathrm{i} \kappa}) &1 + \mathrm{e}^{\mathrm{i} \kappa} \end{pmatrix} ,
\end{equation}
where $\kappa = - k_y$, $k_x$, $k_y$, $-k_x$ for $l = 1, 2, 3, 0$ respectively. For the cyclic products of all four blocks (recall $l \sim l+4$) one easily verifies the result
\begin{equation}\label{eq:cyclic}
    u_{l,\,l-1}(k) \cdots u_{l+1,\,l}(k) = \begin{pmatrix} 1 &\mathrm{i}\, k_x - k_y \cr \mathrm{i}\, k_x + k_y &1 \end{pmatrix} + O(k^2) .
\end{equation}
Note that up to linear order in $k$ (i.e.\ neglecting terms of order $k^2$ and higher) the cyclic product does not depend on $l$.

To summarize, in the spinor basis (\ref{eq:spin-base}) the unitary evolution operator $U_{\rm s} :\; \mathcal{H}_l \to \mathcal{H}_{l+1}$ at momentum $k = {\bf 0}$ simply acts as an $l$-shift operator
\begin{equation}
    U_{\rm s} \, \tilde{e}_l^\sigma ({\bf 0}) = \tilde{e}_{l+1}^\sigma ({\bf 0}) .
\end{equation}
The correction terms linear in the momentum act like a Dirac operator; cf.\ \cite{HoChalker}. More precisely, the long wavelength limit of $\frac{1}{2} (U_{\rm s}^4 - 1)$ is a massless Dirac operator
\begin{equation}
    D = \begin{pmatrix} 0 &\partial_{\bar{z}} \cr \partial_z &0 \end{pmatrix} = - D^\dagger .
\end{equation}
The $\mathbb{Z}_4$ spectral symmetry principle (\ref{eq:extend}) extends this structure to four Dirac cones, all at $k = 0$, with quasi-energies at the fourth roots of unity.

\subsection{Read's method}\label{sect:Read}

We now re-instate the disordered factor $U_{\rm r}$ to return to the full operator $U = U_{\rm r} U_{\rm s}\,$. The objects of our theoretical analysis are disorder-averaged observables constructed from retarded Green's functions $\langle \ell \vert (1 - U)^{-1} \vert \ell^\prime \rangle$ where $\parallel\! U\! \parallel \, < 1$, and their advanced analogs, where $\parallel\! U ^{-1} \! \parallel \, < 1$. These can be computed by Read's variant \cite{BWZ2} of the Wegner-Efetov supersymmetry method. To sketch the idea of the method, let $u = \mathrm{e}^{\mathrm{i} \varphi}$ be a sub-unitary number (${\rm Im}\, \varphi > 0\,$; retarded case) and $b, b^\dagger$ be a canonical pair of boson annihilation and creation operators acting on the standard Fock space for bosons. Then the quantum statistical trace
\begin{equation}
    \mathrm{Tr}\; \mathrm{e}^{\mathrm{i} \varphi b^\dagger b} = \sum_{n=0}^\infty u^n \quad (|u| < 1)
\end{equation}
computes $(1-u)^{-1}$. Similarly, for $u = \mathrm{e}^{\mathrm{i} \varphi}$ a super-unitary number (${\rm Im}\, \varphi < 0\,$; advanced) one has a convergent trace $\mathrm{Tr}\; \mathrm{e}^{- \mathrm{i} \varphi b b^\dagger} = \sum_{n = 1 }^\infty u^{-n} = (u-1)^{-1}$. By adding fermions for the retarded/advanced ($+/-$) sectors one arrives at
\begin{align}\label{eq:bRead}
    &\mathrm{STr}\, \exp \big( \xi_0\, b_+^\dagger b_+^{\vphantom{\dagger}} + \xi_1\, f_+^\dagger f_+^{\vphantom{\dagger}} - \xi_2\, b_-^{\vphantom{\dagger}} b_-^\dagger + \xi_3\, f_-^{\vphantom{\dagger}} f_-^\dagger \big) \cr &= \frac{(1- \mathrm{e}^{\xi_1})(1 - \mathrm{e}^{\xi_3})}{(1 - \mathrm{e}^{\xi_0}) (1 - \mathrm{e}^{\xi_2})} \qquad ({\rm Re}\, \xi_0 < 0 < {\rm Re}\, \xi_2) .
\end{align}
The symbol $\mathrm{STr}$ stands for the supertrace (giving a minus sign to the contribution from states with odd fermion number) over the tensor product of four Fock spaces, two of bosonic ($b_\pm$) and fermionic  ($f_\pm$) type each.

The identity (\ref{eq:bRead}) has a direct generalization from numbers $u = \mathrm{e}^\xi$ to operators $U = \mathrm{e}^X$. Let $U$ be given as $U = \exp \sum \vert \ell \rangle \langle \ell \vert X \vert \ell^\prime \rangle \langle \ell^\prime \vert \,$. Then we define its second-quantized representation $\rho(U)$ as
\begin{equation}\label{eq:XRead}
    \rho(U) = \exp \sum c^\ast_\alpha(\ell) \langle \ell \vert X \vert \ell^\prime \rangle c^\alpha(\ell^\prime) ,
\end{equation}
where, suppressing the link label $\ell$, we introduce the notation
\begin{align}
    &c_0^\ast = b_+^\dagger \,, \quad c^0 = b_+ \,, \quad c_1^\ast = f_+^\dagger \,, \quad c^1 = f_+ \,, \cr
    - &c_2^\ast = b_- \,, \quad c^2 = b_-^\dagger \,, \quad c_3^\ast = f_-\,, \quad c^3 = f_-^\dagger \,. \label{eq:16mz}
\end{align}
Assuming that $X$ is anti-Hermitian, the quantum statistical trace of $\rho(U)$ over the Fock space, $\mathcal{F}$, is totally oscillatory and can be made absolutely convergent by adding in the exponent an infinitesimal chemical potential $\mu N$ where $\mu < 0$ and
\begin{equation}
    N = \sum\nolimits_\ell \big( n_+(\ell) + n_-(\ell) \big) , \quad
    n_\sigma(\ell) = b_\sigma^\dagger(\ell) b_\sigma^{\vphantom{\dagger}}(\ell) + f_\sigma^\dagger(\ell) f_\sigma^{\vphantom{\dagger}}(\ell) ,
\end{equation}
counts the total number of particles. The partition function then exists and is trivial, $\mathrm{STr}_\mathcal{F}\, \rho(U) = 1$, due to the cancelation between bosons and fermions; cf.\ Eq.\ (\ref{eq:bRead}). A noteworthy property of the second-quantization map $U \mapsto \rho(U)$ is its multiplicativity: $\rho(U_{\rm r} U_{\rm s}) = \rho(U_{\rm r}) \rho(U_{\rm s})$.

The Green's function observables of interest are obtained by inserting suitable operators under the trace. To spell out the details, we define the $\mu$-regularized expectation value of an operator $A$ as
\begin{equation}\label{eq:F-trace}
    \langle A\, \rangle_\mathcal{F} = \mathrm{STr}_\mathcal{F} \, \mathrm{e}^{\mu N} \rho(U) A \,.
\end{equation}
Since Wick's rule applies in the free-particle setting at hand, all such expectation values are determined by $\langle 1 \rangle_\mathcal{F} = 1$ and two basic Wick contractions:
\begin{align}
    &\big\langle b_+^\dagger (\ell_1) b_+ (\ell_2) \big\rangle_\mathcal{F} = \big\langle \ell_2 \vert T (1-T)^{-1} \vert \ell_1 \rangle = - \langle f_+^\dagger (\ell_1) f_+ (\ell_2) \rangle_\mathcal{F} \,, \cr
    &\big\langle b_- (\ell_1) b_-^\dagger (\ell_2) \big\rangle_\mathcal{F} = \langle \ell_2 \vert (1-T^\dagger)^{-1} \vert \ell_1 \rangle = + \langle f_- (\ell_1) f_-^\dagger(\ell_2) \rangle_\mathcal{F} \,,
    \label{eq:basic-Wick}
\end{align}
where we introduced the $\mu$-regularized time-evolution operator $T \equiv \mathrm{e}^{\mu} U$ and $T^\dagger = U^{-1} \mathrm{e}^\mu$.

\subsection{SUSY vertex model}\label{sect:SUSY-VM}

An attractive feature of Read's method is that it makes the step of random-phase averaging very easy and leads to a transparent outcome: the disorder-averaged Fock space model is translation-invariant and has the structure of a supersymmetric (SUSY) vertex model. A quick summary is as follows.

By utilizing the multiplicative property $\rho(U_{\rm r} U_{\rm s}) = \rho(U_{\rm r}) \rho(U_{\rm s})$, we see that computing the disorder average $\mathbb{E}( \rho(U))$ amounts to computing $\mathbb{E}( \rho(U_{\rm r}))$. Since $U_{\rm r} = \sum \vert \ell \rangle\, \mathrm{e}^{\mathrm{i} \phi_\ell} \langle \ell \vert$ is diagonal in the link basis and the random phases $\phi_\ell$ for different links are uncorrelated, the disorder average can be carried out for each link separately. For a single link $\ell$ we have
\begin{equation}
    \mathbb{E} \Big( \rho \big( \vert \ell \rangle\, \mathrm{e}^{\mathrm{i} \phi_\ell} \langle \ell \vert \big) \Big) =
    \mathbb{E} \big( \mathrm{e}^{\mathrm{i}\phi_\ell c_\alpha^\ast(\ell) c^{\,\alpha}(\ell)} \big) = \int_0^{2\pi} \frac{d\phi_\ell}{2\pi} \, \mathrm{e}^{\mathrm{i}\phi_\ell ( n_+(\ell) - n_-(\ell) )} .
\end{equation}
This integral is unity for $n_+(\ell) = n_-(\ell)$ and zero otherwise. Thus the step of taking the disorder average simply projects on the sector of Fock space where for every link $\ell$ there are as many retarded as advanced particles. We denote by $\mathcal{V}_\ell$ the subspace of the Fock space at $\ell$ which is selected by the constraint $n_+(\ell) = n_-(\ell)$. The total Hilbert space of the model after disorder averaging then is the tensor product $\mathcal{V} = \bigotimes_\ell \mathcal{V}_\ell \,$.

Disorder-averaged observables of the network model are now computed by restricting the trace (\ref{eq:F-trace}) to the subspace $\mathcal{V}$:
\begin{equation}\label{eq:V-trace}
    \langle A\, \rangle_\mathcal{V} \equiv \mathbb{E} \big( \langle A\, \rangle_\mathcal{F} \big) =
    \mathrm{STr}_\mathcal{V} \, \mathrm{e}^{\mu N} \rho(U_{\rm s}) A \,.
\end{equation}
This re-formulation of our problem is called the SUSY vertex model. The name communicates the fact \cite{MRZ-network} that $\rho(U_{\rm s})$ projected to $\mathcal{V}$ factors as a product over vertices, where each factor is made from the Fock operators of the four links connecting to the given vertex.

The SUSY vertex model will furnish much of the basis for our heuristic reasoning in Section \ref{sect:4}. Note that while most past approaches addressed the transfer matrix (singling out one of two axes of the network model as ``imaginary time'') and its anisotropic limit (to set up a SUSY spin chain of anti-ferromagnetic type), we will argue with the representation (\ref{eq:V-trace}) directly.

Let us finish this brief exposition by mentioning that $\mathcal{V}_\ell$ is an irreducible highest-weight module for the Lie superalgebra $\mathfrak{gl}(2|2)$ generated by the set of operators $c_\alpha^\ast c^{\,\beta}$ ($\alpha \,, \beta = 0, 1, 2, 3$) at the link $\ell$. A significant feature of $\mathcal{V}_\ell$ is that the vanishing of the first $\mathfrak{gl}(2|2)$ Casimir invariant ($c_\alpha^\ast c^{\,\alpha} = 0$) is accompanied by the vanishing of \underline{all} Casimir invariants \cite{HPZ}:
\begin{align}
    &(-1)^{\beta} c_\alpha^\ast c^{\,\beta} c_\beta^\ast c^\alpha \big\vert_{\mathcal{V}_\ell} = 0 \,, \cr &(-1)^{\beta + \gamma} c_\alpha^\ast c^{\,\beta} c_\beta^\ast c^\gamma c_\gamma^\ast c^\alpha \big\vert_{\mathcal{V}_\ell} = 0 \,, \quad \text{etc.}
\end{align}
(Here and in the following the summation convention is implied.) The big question then is to what extent these identities survive under renormalization as constraints on the conformal field theory for the critical point.

\section{Toy model: random $\mathfrak{su}(n)$ gauge field}\label{sect:3}

Our strategy will be to understand the scaling limit of the IQHE plateau transition via the SUSY vertex model (\ref{eq:V-trace}) at criticality.
Now, invoking the principle of universality at generic critical points,
it was proposed some time ago \cite{LFSG} that one possible approach to the IQHE plateau transition would be to augment the massless Dirac theory with marginal perturbations as given by a random gauge potential, a random scalar potential, and a random Dirac mass. While this proposal seems reasonable, it is not in concord with the situation for the network model: the random-phase disorder in $U_{\rm r}$ turns out to be strongly relevant \cite{footnote1} at the massless free Dirac limit of $U_{\rm s}\,$.

Nevertheless, as a preparation for our work ahead, we now invest some time to revisit the continuum approximation by massless Dirac-type fields coupled to a random gauge potential. Recall that $\alpha \in \{ 0, \ldots, 3\}$ is an index labeling retarded bosons ($\alpha = 0$), retarded fermions ($\alpha = 1$), advanced bosons ($\alpha = 2$) and advanced fermions ($\alpha = 3$). For generality, we assume that $n \geq 1$ Dirac species are present in the low-energy effective theory; we label them by $l = 0, 1, \ldots, n-1$. Since long wavelength implies low frequency for massless Dirac fields, the original nature of our $c_{l \alpha}^\ast$ and $c^{\alpha l}$ as quantum operators on Fock space is expected to give way to low-energy effective behavior as bosonic and fermionic integration variables (or \emph{classical} fields) in a functional integral that computes the statistical trace. In this vein, we now consider the toy model of a functional integral $\int \mathrm{e}^{-S}$, $S = \int \mathcal{L}\,$, with (graded-)commutative fields $c^{\alpha l} , c_{l \alpha}^\ast$ and continuum-theory Lagrangian
\begin{align}\label{eq:DiracL}
    \pi \mathcal{L} &= c^{\ast L}_{l \alpha} \, \partial_{\bar{z}} c_L^{ \alpha l} + c^{\ast R}_{l \alpha} \, \partial_{z} c_R^{\alpha l} \cr &+ \mathrm{i}\, (A_{l,\,m} c^{\ast L}_{l \alpha} c_L^{\alpha m} + \overline{A}_{l,\,m} c^{\ast R}_{m \alpha} c_R^{\alpha l} ).
\end{align}
By convention, holomorphic fields (with equation of motion $\partial_{ \bar{z}} c_L = 0$) are left-moving, while anti-holomorphic fields ($\partial_z c_R = 0$) are right-moving. The coefficients $A_{l,\,m}$ are complex random fields distributed as Gaussian white noise; they effectively represent some microscopic disorder (which differs from that of the network model). The bar means the complex conjugate.

In view of the definitions (\ref{eq:16mz}) we stipulate that $c_{l 0}^{ \ast L} = \overline{c_R^{0 l}}\,$, $c_{l 0}^{\ast R} = \overline{c_L^{0 l} }$ and $c_{l 2}^{\ast L} = - \overline{c_R^{2 l}}\,$, $c_{l 2}^{\ast R} = - \overline{c_L^{2 l}}$. The numerical (or bosonic) part of $S = \int \mathcal{L}$ for the Dirac/ghost Lagrangian (\ref{eq:DiracL}) then takes values in the imaginary numbers, rendering the functional integral totally oscillatory. Such was the case for the statistical sum $\mathrm{STr}\, \rho(U)$ in the initial setting of the supersymmetry method, and it is natural to require the same property to hold for the toy model as well. As before, the functional integral is made convergent by including an infinitesimal chemical potential term $\mu N$ ($\mu < 0$). The functional integral expression for the total particle number is
\begin{equation}
    N = \int d^2r \, (\Sigma_3)_\alpha \, (c^{\ast L}_{l \alpha} c_R^{ \alpha l} + c^{\ast R}_{l \alpha} c_L^{\alpha l} ) ({\bf r})\,,
\end{equation}
where $(\Sigma_3)_0 = (\Sigma_3)_1 = +1$ and $(\Sigma_3)_2 = (\Sigma_3)_3 = -1$. The presence of this regularization not only ensures the existence of the functional integral, but it also keeps the partition function trivial:
\begin{equation}
    Z = \int \mathrm{e}^{-S} = 1, \qquad S = \int\mathcal{L}\, d^2r ,
\end{equation}
by the symmetry between bosons and fermions.

The field theory (\ref{eq:DiracL}) is a free theory in that it is Gaussian in the Dirac-type fields $c\,, c^\ast$. However, by the rules of the game all observables of the toy model (just like in the network model) are averages over the disorder, i.e., over the random gauge field $A$. Thus $A$ is meant to be integrated out. Since it is a Gaussian-distributed field, this can be done explicitly and leads to interaction terms that are quartic in the Dirac/ghost fields $c\,, c^\ast$.

For simplicity of the outcome, we assume that $A$ is traceless: $\sum A_{l,\,l} = 0$. In order words, $A$ and $\bar{A}$ (or rather their Cartesian components $A^x$ and $A^y$) take values in $\mathfrak{su}(n)$. In the strong-coupling limit of broadly distributed $A$ the resulting field theory will turn out (Sects.\ \ref{sect:3.3}, \ref{sect:GL-WZW})
to be the $\mathrm{GL}(r|r)_n$ Wess-Zumino-Witten model for a number $r$ of replicas. The latter (with a reduced value of $r$) plays a key role in Section \ref{sect:4}, where we continue our analysis of the SUSY vertex model formulation of the network model.

\subsection{A pair of level-rank dual current algebras}
\label{sect:Howe-pair}

To convey the best possible view of our constructions and their meanings, we now make a generalization: instead of the minimal number of replicas assumed so far (one retarded and advanced boson and fermion each), we will consider the general case of any number of replicas. For the toy model with random $\mathfrak{su}(n)$ gauge field, the distinction between the retarded and advanced sectors actually does not matter; so, we are free to simply speak of $r$ bosonic and $s$ fermionic replicas. Later, when we return to the full problem of the IQHE plateau transition, the distinction will become relevant.

In $2d$ conformal field theory one has factorization into a holomorphic and an anti-holomorphic sector. In particular, the energy-momentum (or stress-energy) tensor is a sum of two pieces, one for each sector. To construct it, we may  focus on the holomorphic part, $T(z)$, of the tensor. With this focus understood, we temporarily simplify the notation: $c \equiv c_L\,$, $c^\ast \equiv c^{\ast L}$.

The main basis of the analysis is the operator product expansion (OPE) for the $r$ bosonic and $s$ fermionic replicas of our Dirac-type fields:
\begin{equation}\label{eq:basicOPE}
    c^{\alpha l}(z) c_{m \beta}^\ast (w) \sim \frac{\delta_m^l \, \delta_\beta^\alpha}{z-w} \,, \quad c_{l \alpha}^\ast (z) c^{\beta m}(w) \sim (-1)^{|\alpha|+1} \frac{\delta_\alpha^\beta \,\delta_l^m}{z-w} \,,
\end{equation}
which follows directly from the free part of the Dirac/ghost Lagrangian (\ref{eq:DiracL}). The symbol $\sim$ means that we are writing only the singular part of the OPE. The symbol $\vert \alpha \vert$ stands for the (super-)parity: $\vert \alpha \vert = 0$ for even fields (bosonic replicas) and $\vert \alpha \vert = 1$ for odd fields (fermionic replicas).

In the following, the normal-ordered product of two local fields $J$ and $K$ will be denoted by $(JK)$. As usual in this context \cite{diFrancesco}, the normal-ordered product is defined by subtracting the singular parts of the operator product and then taking the limit of coinciding points. For example,
\begin{equation}
    \big( c_{l \alpha}^\ast \, c^{\beta m} \big) (w) = \lim_{z \to w}
    \Big( c_{l \alpha}^\ast (z) c^{\beta m} (w) - (-1)^{|\alpha|+1}
    \frac{\delta_\alpha^\beta \, \delta_l^m}{z-w} \Big) .
\end{equation}
The normal-ordered products $(c_{l \alpha}^\ast \, c^{\beta m})$ are called currents. It is known \cite{Bernard} that the totality of all the currents $(c_{l \alpha}^\ast \, c^{\beta m})$, $c_{l \alpha}^\ast \, c_{m \beta}^\ast$ and $c^{\alpha l} c^{\beta m}$ generates an orthosymplectic current algebra at level one.

Two current sub-algebras will play a role. Still assuming the summation convention for notational convenience, these are generated by
\begin{equation}
    K_{l}^{\;m} = (c_{l \alpha}^\ast \, c^{\alpha m}) , \quad
    J^{\alpha}_{\;\; \beta} = (c^{\alpha l} c_{l \beta}^\ast) .
\end{equation}
In order to specify their operator product expansions in a concise way, we associate with $A \in \mathfrak{gl}(n)$ the current $K^A = A_{\;m}^l K_l^{\;m}$, where the coefficients $A_{\;m}^l$ are the matrix elements of $A = A_{\;m}^l e_l \otimes e^m$. Then with $[A,B] = AB - BA$ the commutator in $\mathfrak{gl}(n)$ we find
\begin{equation}\label{eq:OPE-gl(n)}
    K^A(z) K^B(w) \sim \frac{(s-r)\mathrm{Tr} AB}{(z-w)^2} +
    \frac{K^{[A,\,B]}(w)}{z-w} \,.
\end{equation}
This is referred to as the current algebra $\widehat{\mathfrak{gl}} (n)_{s-r}$ of $\mathfrak{gl}(n)$ at level $s-r$.

Turning to the currents $J^{\alpha}_{\;\; \beta}\,$, let us adjust our conventions a little bit for clean formulas and a smooth presentation.
To facilitate the later passage to a $\mathrm{GL}(r|s)$ Wess-Zumino-Witten model by non-Abelian bosonization, we think of the $J^{\alpha}_{\; \beta}$ as the matrix elements of a supermatrix field $J = e_\alpha J^{\alpha}_{\; \; \beta} \otimes e^{\beta}$ over the Lie superalgebra $\mathfrak{gl} (r|s)$. Similarly, let $X = e_\alpha X^{\!\alpha}_{\; \beta} \otimes e^\beta$ be a supermatrix with matrix elements $X^{\alpha}_{\; \beta}$ drawn from a parameter Grassmann algebra for $\mathfrak{gl}(r|s)$. Moreover, let $J^X \equiv \mathrm{STr}(JX) = \mathrm{STr}(XJ) = (-1)^{|\alpha|} X^{\! \alpha}_{\; \beta} J^{\beta}_{\;\; \alpha}\,$. Then an easy computation using (\ref{eq:basicOPE}) gives
\begin{equation}
    J^X(z) J^Y(w) \sim - \frac{n\, \mathrm{STr} XY}{(z-w)^2} +
    \frac{J^{[X,\,Y]}(w)}{z-w} , \label{eq:OPE-gl(r|s)}
\end{equation}
where $[X,Y] = XY - YX$ is still the commutator (albeit of supermatrices). The OPE (\ref{eq:OPE-gl(r|s)}) is that of a level-$n$ current superalgebra denoted by $\widehat{\mathfrak{gl}}(r|s)_n\,$.

Our two current sub-algebras $\widehat{\mathfrak{gl}} (n)_{s-r}$ and
$\widehat{\mathfrak{gl}}(r|s)_n$ commute with one another, i.e., mixed operator products do not give rise to any singular terms. In the classical setting of Lie algebras (as opposed to current algebras) one says that $\mathfrak{gl}(n)$ and $\mathfrak{gl}(r|s)$ form a so-called Howe pair inside the orthosymplectic Lie superalgebra $\mathfrak{osp}(2nr|2ns)$. A distinctive feature of this Howe pair is that $\mathfrak{gl}(n)$ and $\mathfrak{gl}(r|s)$ have a non-trivial intersection:
\begin{equation}
    \mathfrak{gl}(1) = \mathfrak{gl}(n) \cap \mathfrak{gl}(r|s)
\end{equation}
by their common center, $\mathfrak{gl}(1)$. In the present setting, this means that the two current algebras share one current, which we denote by $E$:
\begin{equation}\label{eq:traceless}
    K_l^{\;l} = (-1)^{|\alpha|} J^{\alpha}_{\;\; \alpha} \equiv E .
\end{equation}

In conformal field theory one does not usually speak of a Howe pair. Rather, one says that the current algebras $\widehat{\mathfrak{gl}} (r|s)_n$ and $\widehat{\mathfrak{gl}} (n)_{s-r}$ are level-rank dual to one another. Indeed, the level $n$ of $\widehat {\mathfrak{gl}}(r|s)_n$ is the rank of $\widehat{\mathfrak{gl}}(n)_{s-r}$ and, conversely, the level $s-r$ of the latter is the rank (i.e.\ the super-dimension of a Cartan subalgebra) of the former.

Level-rank duality leads to a useful relation between the normal-ordered quadratic Casimir elements of the two current algebras. Let $C_{KK} = (K_l^{\;m} K_m^{\; l})$ and $C_{JJ} = (-1)^{|\alpha|+1} (J^\alpha_{\;\; \beta} J^\beta_{\;\; \alpha})$. (To appreciate the overall sign of the latter case, note that the theory must become unitary upon restriction to the fermion-fermion sector). Then from the basic operator product expansions (\ref{eq:basicOPE}) one derives the identity
\begin{equation}
    C_{KK} + C_{JJ} = 2(n+s-r)\, T_{\rm free} \,,
\end{equation}
where
\begin{equation}\label{eq:T-free}
    T_{\rm free} = - {\textstyle{\frac{1}{2}}} (c_{l \alpha}^\ast \, \partial_z c^{\alpha l}) + {\textstyle{\frac{1}{2}}} (\partial_z c_{l \alpha}^\ast \cdot c^{\alpha l})
\end{equation}
is the energy-momentum tensor of the free theory. In other words, the quartic interaction terms that appear in the normal-ordered products $C_{KK}$ and $C_{JJ}$ are exact negatives of each other and thus cancel in the sum, leaving only a multiple of $T_{\rm free}\,$.

\subsection{Virasoro decomposition of $T_{\rm free}$}
\label{sect:Virasoro}

Let us now embark on a brief digression to prepare the coset construction of the next section. Writing $C_{KK} = 2(n+s-r) T_{KK}$ and $C_{JJ} = 2(n+s-r) T_{JJ}\,$, one has
\begin{equation}\label{eq:notVirasoro}
    T_{\rm free} = T_{KK} + T_{JJ} \,.
\end{equation}
This decomposition is not Virasoro, which is to say that the individual summands on the right-hand side do not obey the operator product expansion
\begin{equation}\label{eq:OPE-TT}
    T(z) T(w) \sim \frac{c/2}{(z-w)^4} + \frac{2 T(w)}{(z-w)^2} +
    \frac{T^\prime (w)}{z-w}
\end{equation}
for a Virasoro algebra (with any central charge $c$). Nonetheless, if $r \not= s$ the decomposition (\ref{eq:notVirasoro}) can be refined to become Virasoro by introducing the traceless currents
\begin{equation}\label{eq:notrace}
    K^A - \frac{\mathrm{Tr}A}{n} E \quad \text{and} \quad
    J^X - \frac{\mathrm{STr} X}{r-s} E \,,
\end{equation}
which generate the current algebras $\widehat{\mathfrak{sl}}(n)_{s-r}$ and $\widehat{\mathfrak{sl}}(r|s)_n\,$, respectively. In this way one gets
\begin{equation}\label{eq:VirD3}
    T_{\rm free} = T_{\widehat{\mathfrak{sl}}(n)_{s-r}}
    + T_{\widehat{\mathfrak{gl}}(1)_{n(s-r)}} +
    T_{\widehat{\mathfrak{sl}}(r|s)_n} ,
\end{equation}
where the three summands are
\begin{align}
    T_{\widehat{\mathfrak{sl}}(n)_{s-r}} &= T_{KK} -
    \frac{1/2}{n+s-r} \, \frac{(EE)}{n} , \\
    T_{\widehat{\mathfrak{sl}}(r|s)_n} &= T_{JJ} -
    \frac{1/2}{n+s-r} \, \frac{(EE)}{s-r} , \\
    T_{\widehat{\mathfrak{gl}}(1)_{n(s-r)}} &= \frac{(EE)}{2n(s-r)} ,
\end{align}
and each of them individually obeys the OPE (\ref{eq:OPE-TT}). The central charges are
\begin{equation}
    \frac{(s-r)(n^2-1)}{n+s-r} , \quad \frac{n ((s-r)^2-1)}{n+s-r} , \quad 1,
\end{equation}
respectively. These add up to the central charge $n(s-r)$ of $T_{\rm free}$ as required.

Alas, our case of interest is $r = s$ and, there, the decomposition (\ref{eq:VirD3}) fails. The obstruction is that one cannot arrange for the currents $J^X$ at $r = s$ to have vanishing supertrace by subtracting a multiple of the center current $E$; see (\ref{eq:notrace}). What exists irrespective of whether $r \not= s$ or $r = s$ is a two-summand decomposition,
\begin{equation}\label{eq:VirD2}
    T_{\rm free} = T_{\widehat{\mathfrak{sl}}(n)_{s-r}} + T_{\widehat{\mathfrak{gl}}(r|s)_n} ,
\end{equation}
where the second term on the right-hand side is defined to be the sum of $T_{JJ}$ and the traceful part coming from $T_{KK}$:
\begin{equation}\label{eq:T-glrs}
    T_{\widehat{\mathfrak{gl}}(r|s)_n} = - \frac{(-1)^{|\alpha|}}{2n} (J^\alpha_{\;\; \beta} J^\beta_{\;\; \alpha}) + \frac{(EE)}{2n(n+s-r)}\, .
\end{equation}
A special case of relevance to our further development is
\begin{equation}\label{eq:T-glrr}
    T_{\widehat{\mathfrak{gl}}(r|r)_n} = - \frac{(-1)^{|\alpha|}}{2n} (J^\alpha_{\;\; \beta} J^\beta_{\;\; \alpha}) + \frac{(EE)}{2n^2}\, .
\end{equation}
Since both $T_{\rm free}$ and $T_{\widehat{\mathfrak{sl}}(n)_0}$ represent the Virasoro algebra (with $c = 0$ at super-dimension $r - s = 0$), so does the difference $T_{\widehat{\mathfrak{gl}}(r|r)_n} = T_{\rm free} - T_{\widehat{\mathfrak{sl}}(n)_0}$.

\subsection{Taking the coset by $\widehat{\mathfrak{sl}}(n)_0$} \label{sect:3.3}

We now inject an observation that could have been made earlier in our text: the terms coupling to the Gaussian random field $A$ in the Dirac Lagrangian (\ref{eq:DiracL}) can be expressed entirely in terms of the holomorphic $\mathfrak{gl}(n)$ currents $K_l^{\;m}$ and their anti-holomorphic analogs. Indeed, $A_{l,\,m}$ couples to $(K_L)_{l}^{ \;m}(z) = (c^{\ast L}_{l \alpha} c_L^{\alpha m})(z)$ and $\overline{A}_{ m, \,l}$ does to $(K_R)_{l}^{\;m}(\bar{z}) = (c^{\ast R}_{l \alpha} c_R^{ \alpha m}) (\bar{z})$.

As a preparation for Sect.\ \ref{sect:4}, we are now interested in the strong-disorder limit of a widely fluctuating random gauge field $A\, , \bar{A}$. In that limit, the Gaussian $A$-integral has the effect of a Dirac $\delta$-function, setting the scaling dimensions of the currents $K_{L,\,R}$ to zero and removing them from the low-energy effective theory. More precisely, by the assumption of traceless $A$, it is the currents $K$ in $\mathfrak{sl}(n) \subset \mathfrak{gl}(n)$ that are killed. One can work through this step of elimination in complete detail, using non-Abelian bosonization \cite{Witten} and a functional integral version of the Goddard-Kent-Olive coset construction \cite{Gawedzki}, but we will not dwell on that here. In fact, the present model has been analyzed before \cite{MCW,BernLeCl,Tsvelik}; we can therefore be brief.

In short, the theory after $A$-averaging is a coset conformal field theory with the energy-momentum tensor $T_{\widehat{ \mathfrak{gl}} (r|r)_n}$ given in Eq.\ (\ref{eq:T-glrr}); in the present setting of a Howe pair or level-rank duality, one also speaks of a ``conformal embedding'' \cite{diFrancesco}. Since the OPE between the $\mathfrak{gl} (r|r)$ currents $J^{\alpha}_{\; \beta}$ and the $\mathfrak{sl} (n)$ currents $K^A - ( \mathrm{Tr} A / n) E$ is trivial, the latter have vanishing scaling dimension in this CFT. All the currents $J^{\alpha}_{\;\; \beta}$ (including $E = (-1)^{|\alpha|} J^{\alpha}_{\;\; \alpha}$) still have the conformal weights of a holomorphic current,
\begin{equation}\label{eq:dim-J}
    (h(J),\bar{h}(J)) = (1,0) ,
\end{equation}
just as they do in the free theory.

We turn to the fundamental $\mathfrak{gl}(r|r)$-multiplet of fields defined by
\begin{equation}
    M^{\alpha}_{\;\; \beta} (z,\bar{z}) = c^{\alpha l}_L (z) c^{\ast R}_{l \beta}(\bar{z}) .
\end{equation}
These are spinless with conformal weights
\begin{equation}\label{eq:dim-Q}
    h(M) = \bar{h}(M) = \frac{1}{2n^2} \,,
\end{equation}
a result that can already be found in \cite{MCW}. We note that the non-zero value of $h(M)$ comes solely from the second term $(EE)/2n^2$ on the right-hand side of (\ref{eq:T-glrr}). Indeed, applying the first term $T_{JJ}$ of (\ref{eq:T-glrr}) to, say $c_L^{\alpha l} \equiv c^{\alpha l}$, one has
\begin{equation}
    T_{JJ}(z) c^{\alpha l}(w) = \frac{J^\alpha_{\;\; \beta}(z)}{2n(z-w)} \,  c^{\beta l}(w) + \ldots ,
\end{equation}
and by
\begin{equation}
    J^\alpha_{\;\; \beta}(z) c^{\beta l}(w) \sim
    \frac{(-1)^{|\beta|+1}}{z-w} \, c^{\alpha l} (w)
\end{equation}
one arrives at a free sum over the index $\beta$ which gives $\sum_\beta (-1)^{|\beta|} = r - s = 0$ by supersymmetry.

Let us inject a remark here to connect more closely with the literature. For the special case of $n = 2$ there exists an alternative description that has been publicized in \cite{BernLeCl,Tsvelik}. That alternative derives from the accidental isomorphism $\mathfrak{sl}(2) \cong \mathfrak{sp}(2)$ of complex Lie algebras, which leads to an accidental relation between current algebras. Indeed, $\mathfrak{sp}(n)$ forms a Howe pair together with $\widetilde{\mathfrak{osp}} (2r|2s) \equiv \mathfrak{spo}(2r|2s)$ inside $\mathfrak{osp}(2nr|2ns)$. (The tilde indicates orthogonal bosons and symplectic fermions, while in standard $\mathfrak{osp}$ the adjectives are interchanged.) The Howe pair property puts $\widehat{\mathfrak{sp}}(n)_{2s-2r}$ in level-rank duality with $\widehat{\mathfrak{spo}}(2r|2s)_{n}$. Now the current algebras $\widehat{\mathfrak{sp}}(n)_{2s-2r}$ and $\widehat{\mathfrak{sl}} (n)_{s-r}$ coincide for $n = 2$ and $r = s$. As a result, the corresponding level-rank duals have the same energy-momentum tensor:
\begin{equation*}
    T_{\widehat{\mathfrak{gl}}(r|r)_2} =
    T_{\rm free} - T_{\widehat{\mathfrak{sl}}(2)_0} =
    T_{\rm free} - T_{\widehat{\mathfrak{sp}}(2)_0} =
    T_{\widehat{\mathfrak{spo}}(2r|2r)_2} .
\end{equation*}
The latter theory is discussed in \cite{BernLeCl,Tsvelik} under the name of $\widehat{\mathfrak{osp}}(2r|2r)_{-2}\,$.

We now ask: what is the classical Lagrangian corresponding to $T_{\widehat{ \mathfrak{gl}}(r|r)_n}$ (and its analog in the anti-holomorphic sector)? To answer that, we observe that $T_{\widehat{ \mathfrak{gl}}(r|r)_n}$ was obtained from the free Dirac/ghost theory by gauging with respect to the $\mathfrak{sl}(n)$ degrees of freedom. Among some other symmetry group actions, that free theory carries a (partly anomalous) action by symmetries $G_L \times G_R$ where $G_L = G_R = \mathrm{GL}(r|r)$. Since $\mathfrak{gl}(r|r)$ and $\mathfrak{gl}(n) \supset \mathfrak{sl}(n)$ have the Howe pair property of centralizing each other in the free-particle algebra $\mathfrak{osp}(2nr|2nr)$, our CFT with energy-momentum tensor $T_{\widehat{\mathfrak{gl}}(r|r)_n}$ still carries the same action by $\mathrm{GL}(r|r)_L \times \mathrm{GL}(r|r)_R\,$. Taken together with the formulas (\ref{eq:dim-J}, \ref{eq:dim-Q}) for the scaling dimensions, this means that our putative coset CFT is actually the $\mathrm{GL}(r|r)$ Wess-Zumino-Witten (WZW) model at level $k = n$. Some relevant information about it is collected in the next subsection.

\subsection{$\mathrm{GL}(r|s)_n$ WZW model}
\label{sect:GL-WZW}

For the moment, we again relax the condition $r = s$ and consider the general case of any pair $r, s$ (intending to return to $r = s$ when appropriate).

First of all, in view of dissenting proposals in the literature (see \cite{Troost} for a recent reference), we must offer some basic clarification as to what is meant by a $\mathrm{GL}(r|s)$ WZW model. Notwithstanding its misleading name, such a field theory does \emph{not} have any Lie supergroup $\mathrm{GL}(r|s)$ or $\mathrm{U}(r|s) \subset \mathrm{GL}(r|s)$ for its real target space (before complexification). Rather, the target space is what we call a Riemannian symmetric superspace \cite{suprev}. Indeed, for $r = 0$ (fermionic replicas only) the target space is well known to be the compact Lie group $\mathrm{U}(s) \equiv X_1\,$. On other hand, for $s = 0$ (bosonic replicas only) it has to be the non-compact dual of $\mathrm{U}(r)$, namely the symmetric space $X_0 \equiv \mathrm{GL}(r, \mathbb{C}) / \mathrm{U}(r) \simeq \mathrm{Herm}^+(r)$ of positive Hermitian matrices $M_{00}\,$,
\begin{equation}
    M_{00} = g g^\dagger = g u u^\dagger g^\dagger , \quad g \in \mathrm{GL}(r,\mathbb{C}), \quad u \in \mathrm{U}(r) .
\end{equation}
There are many reasons why the correct identification is $X_0 = \mathrm{GL}(r,\mathbb{C}) / \mathrm{U}(r)$ and not $X_0 \stackrel{?}{=} \mathrm{U}(r)$ as would be the case for the Lie supergroup $\mathrm{U} (r|s)$. To give one reason, $T_{\widehat{\mathfrak{gl}} (r|s)_n}$ remembers from $T_{\rm free}$ (through the coset construction) the space of states for the bosonic ghosts in (\ref{eq:DiracL}). This space is very much larger than the corresponding space for the Dirac fermions in (\ref{eq:DiracL}), as every mode of excitation can be occupied not just once but an arbitrary number of times (by the absence of the Pauli principle). The choice $X_0 \stackrel{?}{=} \mathrm{U}(r)$ would fail to capture this drastic enlargement of the Hilbert space. To give a second reason, the metric tensors of the symmetric spaces $X_0 = \mathrm{GL} (r,\mathbb{C}) / \mathrm{U}(r)$ and $X_1 = \mathrm{U}(s)$ combine by the supertrace form to an invariant metric tensor on $X_0 \times X_1$ which is Riemannian. Here the emphasis is on the adjective ``Riemannian'', as this property is what is necessary to have a sign-definite action functional and thus a sensible functional integral. To give yet another reason, the sign in (\ref{eq:T-glrs}) of the $JJ$ terms for the boson-boson sector ($|\alpha| = |\beta| = 0$) is opposite to that for the fermion-fermion sector ($|\alpha| = |\beta| = 1$). Since the latter translates to a WZW model of compact type, one infers from the functional integral version of the coset construction \cite{Gawedzki} that the former (with a Sugawara energy-momentum tensor of the opposite sign) must be a WZW model of non-compact type.

Mathematically speaking, the full target space $X$ of the $\mathrm{GL} (r|s)$ WZW model is a cs-supermanifold \cite{Bernstein} which arises by taking exterior powers of a vector bundle over $X_0 \times X_1$ with standard fiber $V$,
\begin{equation}
    V = V_{01} \oplus V_{10}\, , \quad
    V_{01} = \mathrm{Hom}(\mathbb{C}^s , \mathbb{C}^r) , \quad
    V_{10} = \mathrm{Hom}(\mathbb{C}^r , \mathbb{C}^s) .
\end{equation}
The vector bundle is associated to the direct product of principal bundles $\mathrm{GL}(r,\mathbb{C}) \to X_0$ and $\mathrm{U}(s) \times \mathrm{U}(s) \to X_1$ by the natural action on $V$ of the direct product of structure groups $\mathrm{U}(r) \times \mathrm{U}(s)$. In physics language, the target space $X$ consists of supermatrices
\begin{equation}
    M = \begin{pmatrix} M_{00} &M_{01} \cr M_{10} &M_{11} \end{pmatrix}
\end{equation}
where the left upper block $M_{00}$ is a positive Hermitian matrix of size $r \times r$, the right lower block $M_{11}$ is a unitary matrix of size $s \times s$, and the off-diagonal blocks $M_{01}$ and $M_{10}$ are rectangular matrices of sizes $r \times s$ and $s \times r$ with complex Grassmann variables as matrix entries. The connection to the said vector-bundle picture is made by the re-parametrization
\begin{equation}
    M = \begin{pmatrix} g &0 \cr 0 &u_L \end{pmatrix}
    \begin{pmatrix} \mathbf{1}_r &N_{01} \cr N_{10} &\mathbf{1}_s \end{pmatrix}
    \begin{pmatrix} g^\dagger &0 \cr 0 &u_R^\dagger \end{pmatrix} ,
\end{equation}
which is only determined up to the action
\begin{align}
    &g \to g\, u_0^{-1} , \quad u_L \to u_L\, u_1^{-1} , \quad
    u_R \to u_R\, u_1^{-1} , \cr &N_{01} \to u_0 N_{01}\, u_1^{-1} , \quad N_{10} \to u_1 N_{10}\, u_0^{-1} ,
\end{align}
of the structure group $\mathrm{U}(r) \times \mathrm{U}(s) \ni (u_0\, , u_1)$.

The WZW action functional is the usual one. For level $k = n$ one has
\begin{align}
    S^{\rm WZW}_{k=n} [M] &= \frac{\mathrm{i}\, n}{4 \pi} \int_\Sigma \mathrm{STr}\, \big( M^{-1} \partial M \wedge M^{-1} \bar\partial M \big) + \mathrm{i}\, n\, \Gamma[M] \,, \label{eq:S-WZW1} \\ \Gamma[M] &= \frac{1}{12 \pi} \int_\Sigma \! B \,, \quad \mathrm{d}B = \mathrm{STr}\, ( M^{-1} \mathrm{d} M)^{\wedge 3} ,
    \label{eq:S-WZW2}
\end{align}
where $\mathrm{d} = \partial + \bar\partial$ is the exterior derivative, and the domain $\Sigma$ is a closed Riemann surface. A standard remark is that no space-time metric appears here, as no more than a complex structure for $\Sigma$ is needed to decompose $\mathrm{d} = \partial + \bar\partial = dz \, \partial_z + d\bar{z} \, \partial_{\bar z}\,$. Note that $\Gamma[M]$ is real-valued on $X_1$ but imaginary-valued on $X_0\,$. Let us also record the observation that the third cohomology group of the odd-odd sector $X_1 = \mathrm{U}(s)$ is non-trivial (for $s \geq 2$). This has the well-known consequence that the anomalous weight $\exp(- \mathrm{i}n \Gamma[M])$ is well-defined only for integer values of the level number $n$.

Compared with the Dirac picture, a major change has occurred in that the functional integral is no longer totally oscillatory. In fact, by the Riemannian geometry of $X_0 \times X_1$ the numerical part of the first term on the right-hand side of (\ref{eq:S-WZW1}) is real and positive. (Note that $(\mathrm{i}/2)\, dz \wedge d\bar{z} = d^2 r > 0$.) What remains unchanged from before is that the functional integral needs regularization for the non-compact degrees of freedom. According to the standard rules \cite{Witten} of non-Abelian bosonization, the total particle number of the infinitesimal chemical potential term $-\mu N$ takes the form
\begin{equation}\label{eq:S-reg}
    S_{\rm reg}[M] = \int d^2 r\; \mathrm{STr}\, (M + M^{-1}) .
\end{equation}
As a quick check, observe that for a positive number $m = \mathrm{e}^\varphi$ one has
\begin{equation*}
    m + m^{-1} = 2 \cosh\varphi \,,
\end{equation*}
so the regulator term $S_{\rm reg}$ does indeed do the required job of cutting off the infinity due to the non-compact zero modes in the WZW field $M$. We also note that the functional integral regularized by $S_{\rm reg}$ still has all the requisite target-space supersymmetries to make the partition function trivial:
\begin{equation}
    Z = \int \mathrm{e}^{- S + \mu S_{\rm reg}} = 1  \quad (\mu < 0).
\end{equation}
Please be warned that other definitions (cf.\ \cite{Troost}) of the $\mathrm{GL}(r|r)$ WZW model give $Z = 0\,$, which is unacceptable and unphysical for our purposes.

Let us finish this subsection with two more remarks on (\ref{eq:S-WZW1}). Firstly, no term corresponding to $(EE)/2n(n+s-r)$ in the definition (\ref{eq:T-glrs}) of $T_{\widehat{\mathfrak{gl}}(r|s)_n}$ appears in the WZW Lagrangian. That term becomes negligible in the semiclassical limit of large $n$ and is to be viewed as a quantum correction arising in the Sugawara construction of the energy-momentum tensor. Secondly, the WZW model (\ref{eq:S-WZW1}) for level $n = 1$ is a free theory for any $(r,s)$ and, in particular, for $r = s$. Indeed, since the current algebra $\widehat{\mathfrak{sl}} (n)_{s-r}$ for $n = 1$ is void, the relation (\ref{eq:VirD2}) gives
\begin{equation}\label{eq:3.62}
    T_{\rm free} = T_{\widehat{\mathfrak{gl}}(r|s)_1} \,.
\end{equation}
This means in particular that the level $n = 1$ theory for $r = s$, when properly understood, has free-field correlation functions. Different statements exist in the literature, cf.\ \cite{Troost, SchomSaleur}, where the questionable choice of a Lie supergroup is made for the target space.

\section{Inferring the IQHE critical theory}\label{sect:4}

Having discussed the toy model (\ref{eq:DiracL}) in some detail, we now return to our task proper: constructing the CFT for the critical point of the IQHE plateau transition. Two avenues look especially inviting. For one, we could pursue the approach suggested in \cite{LFSG}, by adding further random-field perturbations to the Lagrangian (\ref{eq:DiracL}) to drive the system to the universal fixed point of interest. In this approach we would apply non-Abelian bosonization to the perturbations and analyze them in the WZW model. A second approach is to revert to the SUSY vertex model and advance the analysis there.

In either investigation, it is necessary to pay due attention to the difference between the retarded and advanced degrees of freedom (originating from the retarded and advanced single-electron Green's functions) of the theory. We recall that $\Sigma_3 = +1$ in the retarded sector and $\Sigma_3 = -1$ in the advanced sector. To avoid the danger of being misled by low-dimensional accidents, we will continue to work with an arbitrary replica number, as long as it does not make our notation overly cumbersome. Thus we consider $r \geq 1$ replicas of retarded bosons, retarded fermions, advanced bosons, and advanced fermions, $r$ of each kind. For the WZW target space $X$ this means that we double $r \to 2r$ and work over the base space
\begin{equation}
    X_0 = \mathrm{GL}(2r,\mathbb{C}) / \mathrm{U}(2r) \simeq \mathrm{Herm}^+(2r) , \quad X_1 = \mathrm{U}(2r) .
\end{equation}

\subsection{A telling observation}\label{sect:tell}

We recall from (\ref{eq:dim-Q}) that the field $M$ of the $\mathrm{GL} (2r| 2r)_n$ WZW model has conformal weights $h = \bar{h} = (2n^2)^{-1}$. Now, from the phenomenology of the IQHE plateau transition (as an Anderson transition in symmetry class $A$) we know with certainty that the fundamental field of the RG fixed-point theory we are seeking must have vanishing scaling dimension \cite{EversMirlin}. Indeed, that dimension is the scaling exponent of the local density of states, which is neither vanishing nor divergent but constant at any Anderson-transition critical point in class $A$. Motivated by this fact, we review in the present section a one-parameter CFT deformation by which the dimension of our field $M$ can be tuned to zero, $h = \bar{h}= 0$. This deformation was already described by Mudry, Chamon and Wen \cite{MCW} a long time ago; while it does not lead us to anything useful \emph{per se}, it will turn out to be of significance when put into the proper context.

By the Chaudhuri-Schwartz criterion \cite{CS-crit}, a conformal field theory with current algebra can be deformed by adding to the Lagrangian a truly marginal perturbation $\int d^2 r\, E_L E_R\,$ made from Abelian left-moving and right-moving currents $E_L$ and $E_R\,$. In our case, such an Abelian deformation exists owing to the presence of the central generator $E$ of $\mathfrak{gl}(2r|2r)$. The pertinent formulas are as follows. Recalling (\ref{eq:OPE-gl(r|s)}), consider the modified current-current operator product expansion
\begin{equation}\label{eq:OPE-d}
    J^X(z) J^Y(w) \sim \frac{\langle X , Y \rangle_{n,\gamma}}{(z-w)^2} + \frac{J^{[X,\,Y]}(w)}{z-w} \,,
\end{equation}
where the invariant bilinear form has been deformed by a parameter $\gamma$:
\begin{equation}\label{eq:CK-d}
    \langle X , Y \rangle_{n,\gamma} = - n\, \mathrm{STr}(XY) + \gamma \, \mathrm{STr}(X) \, \mathrm{STr}(Y) .
\end{equation}
(Note that $n^2 \gamma$ is denoted by $g_A / \pi$ in \cite{MCW}.) The deformed energy-momentum tensor, say the holomorphic part $T(z)$, is determined by requiring the OPE
\begin{equation}\label{eq:OPE-TJ}
    T(z) J^X(w) \sim \frac{J^X(w)}{(z-w)^2} + \frac{\partial_w J^X(w)}{z-w} \,,
\end{equation}
expressing the CFT principle that a holomorphic current $J^X$ has to be a Virasoro primary field with conformal weights $(1,0)$. It then follows that
\begin{equation}\label{eq:T-gamma}
    T_{\widehat{\mathfrak{gl}} (2r|2r)_{n,\gamma}} = - \frac{(-1)^{ |\alpha|}}{2n} (J^\alpha_{\;\; \beta} J^\beta_{\;\; \alpha}) + \frac{1 - \gamma}{2n^2} (E_L E_L) ,
\end{equation}
which still satisfies the OPE for a Virasoro algebra with central charge $c = 0$. Now, as was observed earlier, the conformal weights of $M$ stem entirely from the summand $(E_L E_L)$ of $T$; hence they vanish if we set $\gamma = 1$.

To be sure, the vanishing of the scaling dimension of $M$ at $\gamma = 1$ ought to be discarded as coincidental unless we can offer a convincing physical interpretation of the additional term $\mathrm{STr}(X)\, \mathrm{STr}(Y)$ in the current-current OPE (\ref{eq:OPE-d}). Let us therefore anticipate that such an interpretation does in fact exist in the modified scenario developed below. There, the appearance of the second summand in (\ref{eq:CK-d}) will be explained by the existence of \underline{two} sets of log-correlated operators in the critical theory. In conventional parlance, they correspond to the two types of basic correlator that can be made from retarded ($+$) and advanced ($-$) Green's functions in the microscopic model:
\begin{equation*}
    \Sigma({\bf r},{\bf r}^\prime) = \big\langle G^+ ({\bf r}, {\bf r}^\prime) \, G^- ({\bf r}^\prime , {\bf r}) \big\rangle \quad \text{and} \quad \Upsilon({\bf r},{\bf r}^\prime) = \big\langle G^+ ({\bf r}, {\bf r}) \, G^- ({\bf r}^\prime , {\bf r}^\prime) \big\rangle .
\end{equation*}
In the case of our network model, these take the form of
\begin{align}
    &\Sigma_{\ell_1 , \,\ell_2} = \mathbb{E} \left( \big| \langle \ell_1 \vert (1 - T)^{-1} \vert \ell_2 \rangle \big|^2 \right) , \label{eq:def-Sigma} \\ &\Upsilon_{\ell_1 , \,\ell_2} = \mathbb{E} \left( \langle \ell_1 \vert T (1 - T)^{-1} \vert \ell_1 \rangle \, \langle \ell_2 \vert T^\dagger (1 - T^\dagger)^{-1} \vert \ell_2 \rangle \right) , \label{eq:def-Xi}
\end{align}
where $T = Q U$ stands for the time-evolution operator $U = U_{\rm r} U_{\rm s}$ made sub-unitary by inserting \cite{BWZ2} one or more point contacts, $Q = 1 - \sum \vert \ell_{\rm c} \rangle \langle \ell_{\rm c} \vert$, or by the presence of a homogeneous absorbing background $Q = \mathrm{e}^{\mu}$ ($\mu \to 0-$) as before, or similar. The symbol $\mathbb{E}(\ldots)$ still denotes the disorder average. At the critical point, both correlators (\ref{eq:def-Sigma}) and (\ref{eq:def-Xi}) become logarithms of the distance $\vert \ell_1 - \ell_2 \vert$, and the deformed current-current OPE (\ref{eq:OPE-d}) will ultimately predict their universal amplitude ratio.

\subsection{The current algebra conundrum}\label{sect:puzzle}

If the reader is intrigued and encouraged by the observation that an important scaling dimension can be tuned to the desired value, then we must hasten to caution that the $\widehat{ \mathfrak{gl}}(2r|2r)_{n, \gamma}$ current algebra (\ref{eq:OPE-d}) is beset, for our purposes, with a serious flaw (for any $\gamma$). Using the traditional language of Green's functions $G^\pm$, we can phrase the flaw as follows.

Recall that the microscopic foundation of our functional integral or statistical mechanics problem defines a signature $\Sigma_3$ that distinguishes between the retarded $(+)$ and advanced $(-)$ sectors of the theory. Now if a correlation function is of mono-type $\mathbb{E}( G^+ G^+ \cdots G^+ )$ or $\mathbb{E}( G^- G^- \cdots G^- )$, i.e., probes only one of the two sectors, then it is well known to be trivial in the infrared limit; cf.\ \cite{EversMirlin}. In the network model the triviality is immediate because any product of, say, only retarded Green's functions $\langle \ell \vert (1 - T)^{-1} \vert \ell^\prime \rangle$ collapses to $\langle \ell \vert \ell^\prime \rangle$ by $\mathbb{E}(T^p) = 0$ (for any integer $p > 0$) due to the random $\mathrm{U}(1)$ phase average. In our field-theoretical reformulation the collapse is brought about by the supersymmetries in the unprobed sector. Of course the collapse carries over from Green's functions to current-current correlators. Thus all correlation functions $\langle J^{++} J^{++} \cdots J^{++} \rangle_\mathcal{V}$ in the SUSY vertex model, and also in the infrared limit of the Dirac/ghost system (\ref{eq:DiracL}) with IQHE-generic random perturbations, are trivial, and so are the $\langle J^{--} J^{--} \cdots J^{--} \rangle_\mathcal{V}\,$, in stark contrast to the OPE (\ref{eq:OPE-d}).

One might now entertain the idea that one should kill these very currents by means of a variant of the Goddard-Kent-Olive coset construction (or by gauging the WZW model). Alas, any such attempt is doomed to fail. For one reason, one would immediately run into a conflict with the global $\mathrm{U}(r,r|2r)$ symmetry. For another, correlation functions of mixed type $\langle J^{++} J^{--} \rangle$ do not suffer from SUSY collapse but, as we shall see, are actually critical; it is only the mono-cultures $\langle J^{++} \cdots J^{++} \rangle$ and $\langle J^{--} \cdots J^{--} \rangle$ that are trivial.

In view of this puzzling state of affairs, it is not clear at all how one might proceed with the assumption of a current algebra for the vertex model, or for the Dirac/ghost Lagrangian (\ref{eq:DiracL}) with generic perturbations. We shall therefore revisit our microscopic model for guidance, consulting, in Sect.\ \ref{sect:Kubo}, a physical observable of theoretical and experimental interest: conductivity.

Our strategy from here onwards is to complement known results and physical intuition for the network model with Ward identities in the SUSY vertex model, and vice versa. As a model of unitary quantum mechanics with conserved probability, the network model has a conserved $\mathrm{U}(1)$ charge current. On general grounds one expects the conservation law of that current to be enhanced -- at the critical point where conformal invariance emerges in the infrared limit -- by a \emph{second} conservation law to yield a pair of holomorphic and anti-holomorphic $\mathrm{U}(1)$ currents. By transferring that Abelian current to the SUSY vertex model, we will identify a conserved current which is non-Abelian, albeit not $\mathfrak{gl}(2r|2r)$. As a rewarding return, the known constraints on non-Abelian current algebras then improve our understanding of conserved currents in the network model. In a related context, such a scenario of doubling of conservation laws (or symmetry doubling) due to emerging conformal invariance was pointed out by Affleck \cite{Affleck}. Let us now review and adapt that scenario for our purposes.

\subsection{Symmetry doubling reviewed}\label{sect:Affleck}

As a preparation, we recall some standard facts from differential calculus in the continuum. Let $I(C)$ denote the total charge current passing through a $(d-1)$--dimensional surface $C$ in $d$ space dimensions. To express the total current as an invariantly defined integral $I(C) = \int_C j\,$, one equips $C$ with an outer (or transverse) orientation while the current density $j$ is modeled as a twisted differential form \cite{Bott-Tu} of degree $d-1$. Assuming the (d.c.) situation of a stationary current flow, the continuity equation of charge conservation states that $j$ is closed: $\mathrm{d} j = 0$. It then follows by Stokes' theorem that the total d.c.\ current through any boundary $C = \partial \Sigma$ is zero.

In our two-dimensional setting, the surface $C$ is a curve with outer orientation and $j$ is a twisted $1$-form. Assuming the $2d$ space, $X$, to be isotropic, let $R$ be a clockwise or counterclockwise rotation of tangent vector fields by $\pi / 2$. On $1$-forms the rotation $R\,$, which is also known as a complex structure of $X$, determines a Hodge star operator, $\star\,$. Choosing standard Cartesian coordinates $x$ and $y$ for the Euclidean plane $X$, one has $\star\, \mathrm{d}x = \mathrm{d}y$ and $\star\, \mathrm{d}y = -\mathrm{d}x$ if $R$ goes counterclockwise. Given $R\,$, we orient $X$ by declaring that for any tangent vector $e \not= 0$ the ordered pair $(e,Re)$ is positively oriented (or constitutes a positive system). The choice of complex structure also serves to simplify the integration data $C$ and $j$: using $R\,$, we convert the outer orientation of $C$ into an inner orientation (by an arrow pointing along the curve $C$) and we correspondingly turn $j$ into an untwisted form. The latter is done by the convention that a line segment of $C$ makes a positive contribution to $\int_C j$ if the direction of growth of the form $j$ constitutes a positive system with the arrow of the line segment. In Cartesian coordinates $x, y$ and assuming $R$ to rotate counterclockwise, we write
\begin{equation*}
    j = j_x \star {\rm d}x + j_y \star {\rm d}y =
    j_x \, {\rm d}y - j_y \, {\rm d} x .
\end{equation*}
The current $\int_C j$ through a short line segment $C$ from $y$ to $y + \delta y$ then is $j_x \, \delta y$; and for a short line segment from $x$ to $x + \delta x$ it is $- j_y\, \delta x$.

Given a Hodge star operator $\star$ one associates with $j$ its Hodge dual \begin{equation*}
    \star^{-1} j = - \star j = j_x \, {\rm d} x + j_y \, {\rm d} y .
\end{equation*}
More generally, $j = j_\mu \star {\rm d}x^\mu$ and $\star^{-1} j = j_\mu {\rm d}x^\mu$. The integral $\int_C \star^{-1} j$ is still invariantly defined; its (un-)physical meaning is that of the current flowing \emph{along} $C$. (Using the traditional language of vector calculus, one would say that the current-density vector field is line-integrated along the curve $C$.)

While $\mathrm{d} j = 0$ (or $\partial^\mu j_\mu = 0$) always holds true for a conserved current, there exists no reason for the current to be curl-free ($\mathrm{d} \star j = 0$ or $\epsilon^{\mu\nu} \partial_\mu j_\nu = 0$) in general. Indeed, our network model away from the critical point has circulating currents; in the strong localization regime on one side of the phase transition, currents flow around the elementary plaquettes with one sense of circulation; on the other side they flow around those with the opposite circulation. Yet, at the critical point separating the two phases with opposite circulation, the two opposing tendencies should balance out, and we therefore expect the circulating currents to vanish (around contractible domains) on average over the disorder and after coarse graining to eliminate non-universal behavior on short scales. If so, the critical network-model current $j$ after disorder averaging and coarse graining satisfies two continuity equations: $\mathrm{d} j = 0$ and $\mathrm{d} \star j = 0$.

A quantitative argument for this heuristic scenario is the following \cite{Affleck}. In the Euclidean plane $X$ with translation invariance, consider the two-point correlation function of a conserved current,
\begin{equation}
    \big\langle j_\mu ({\bf r}) j_\nu ({\bf r^\prime}) \big\rangle \equiv C_{\mu\nu}({\bf r} - {\bf r}^\prime) ,
\end{equation}
for an (as yet) unspecified statistical mechanical system at criticality, where we use the generic notation $\langle \, \cdots \rangle$ for statistical averages. By rotational invariance in the scaling limit, such a $2d$ correlation function has to be a sum of two $\mathrm{O}(2)$-equivariant tensors:
\begin{equation}\label{eq:4.69}
    C_{\mu \nu}({\bf x}) = x_\mu x_\nu \, f(|{\bf x}|^2) +
    \delta_{\mu\nu} \, g(|{\bf x}|^2) .
\end{equation}
(Note that the skew-symmetric tensor $\epsilon_{\mu\nu}$ cannot appear here as the bulk of the critical system also has an emerging reflection or parity symmetry; for example, in Pruisken's non-linear sigma model the  parity transformation is implemented by $\theta \to -\theta$, which is a bulk symmetry for $\theta = \pm \pi$.) Scale invariance at the critical point constrains the current-current correlation function (\ref{eq:4.69}) to be homogeneous of degree $-2$:
\begin{equation}
    C_{\mu \nu}({\bf x}) = \frac{x_\mu x_\nu}{|{\bf x}|^4} \, f_0 +  \frac{\delta_{\mu\nu}}{|{\bf x}|^2} \, g_0 \,,
\end{equation}
where $f_0$ and $g_0$ are two constants. The continuity equation for the conserved current ($\partial^\mu j_\mu = 0$) then fixes their ratio $g_0 / f_0$ to be $-1/2$:
\begin{equation}\label{eq:CCFn}
    C_{\mu \nu}({\bf x}) = \frac{x_\mu x_\nu - \delta_{\mu\nu} |{\bf x}|^2 / 2}{ |{\bf x}|^4}\, f_0 \,.
\end{equation}

To appreciate the consequences, one introduces $z = x + \mathrm{i}y$, ${\bar z} = x - \mathrm{i} y$ and decomposes the current into its complex $10$- and $01$-components:
\begin{equation*}
    j_z = {\textstyle{\frac{1}{2}}} (j_x - \mathrm{i} j_y) , \quad
    j_{\bar z} = {\textstyle{\frac{1}{2}}} (j_x + \mathrm{i} j_y) .
\end{equation*}
It then follows directly from Eq.\ (\ref{eq:CCFn}) that the cross correlation vanishes:
\begin{equation}
    \big\langle j_z(z, {\bar z}) j_{\bar z}(0,0) \big\rangle = 0 \,,
\end{equation}
and that the diagonal correlations are holomorphic and anti-holomorphic:
\begin{equation}\label{eq:4.74}
    \big\langle j_z(z, {\bar z}) j_z(0,0) \big\rangle = \frac{n}{z^2} \,, \quad \big\langle j_{\bar z}(z, {\bar z}) j_{\bar z}(0,0) \big\rangle = \frac{n}{{\bar z}^2} \,,
\end{equation}
for some parameter $n$. Finally, in a unitary theory one infers from Eqs.\ (\ref{eq:4.74}) the stronger result that
\begin{equation}\label{eq:4.74mz}
    \partial_{\bar z} j_z = 0 = \partial_z j_{\bar z}
\end{equation}
holds under the statistical expectation value $\langle \,\cdots \rangle$ (and away from other operators inserted into the correlation function).

Note that for our purposes there exists a caveat, since the SUSY vertex model fails to be unitary. For that reason, we are going to apply (actually, adapt) the symmetry doubling argument to the $\mathrm{U}(1)$ conserved current of the network model instead. Later, we will use what we learned for the network model in order to manufacture a trustworthy argument for some of the conserved currents of the SUSY vertex model.

To wrap up the current section, we write the result (\ref{eq:4.74mz}) in a coordinate-free form:
\begin{equation}\label{eq:4.75}
    \bar\partial j^{10} = 0 = \partial j^{01} ,
\end{equation}
where $\partial + \bar\partial = \mathrm{d}\,$, $\partial - \bar\partial = - \mathrm{i}\, \mathrm{d}\, \star$ (on $1$-forms), and $j = j^{10} + j^{01}$ with
\begin{equation}\label{eq:4.76mz}
    j^{10} = {\textstyle{\frac{1}{2}}} (j + \mathrm{i} \star j) = j_z \star {\rm d}z \,, \quad j^{01} = {\textstyle{\frac{1}{2}}} (j - \mathrm{i} \star j) = j_{\bar z} \star {\rm d}\bar{z} \,.
\end{equation}
By subtracting the two equations in (\ref{eq:4.75}) one sees that current conservation $\mathrm{d} j = 0$ is augmented by a second conservation law $\mathrm{d} \star j = 0$ as claimed.

\subsection{Symmetry doubling argument adapted}
\label{sect:SD-adapted}

Let us now see whether we can apply the generic argument above to our specific case of the Chalker-Coddington network model. The first question is how to define a good notion of conserved network-model current $j$. We require two properties: (i) $j$ must conform to the principle of conformal invariance at the critical point; (ii) it must translate to a local field in the SUSY vertex model (the latter property is needed for our eventual goal of making the extension to a non-Abelian current algebra). To meet these requirements, let $\psi_{\rm c} = U \psi_{\rm c}$ be a stationary network-model state that satisfies incoming-wave boundary conditions at a point contact \cite{BWZ1, BWZ2}. The wave intensity $\ell \mapsto | \psi_{\rm c}(\ell) |^2$ satisfies Kirchhoff's nodal rule, and we can therefore derive a conserved current $j$ from it. The details of this lattice construction will be presented in the next subsection; here we take the short-cut of simply assuming that we are handed $j$ in its continuum limit.

Given $j$, consider the disorder average
\begin{equation*}
    \mathbb{E} \left( j_\mu({\bf r}) j_\nu({\bf r}^\prime) \right)
\end{equation*}
of a product of two local currents. Although this looks like a two-point function, it is actually a three-point function, since the current $j$ remembers its definition by the incoming-wave boundary conditions for $\psi_{\rm c}$ at the point contact. Hence the previous argument using rotational symmetry and scale invariance does not apply \emph{verbatim} but needs to be adapted.

Our goal is to argue that the conservation law ${\rm d} j = 0$ implies a second conservation law $\mathbb{E} ( \bullet \, {\rm d} \star j) = 0$, at criticality. To that end, we set up a calculation in local coordinates around any point $o$ of the Euclidean plane. Let $j$ be Taylor-expanded in Cartesian coordinates $x, y$ based at that point:
\begin{equation}
    j = (a_0 + a_1 x + a_2 y + \ldots) \star{\rm d}x + (b_0 + b_1 x + b_2 y + \ldots) \star{\rm d}y ,
\end{equation}
where the ellipses indicate terms which are at least quadratic in $x, y$ and hence negligible for our purpose of computing first-order derivatives at $o$. The constant part $j^{(0)} = a_0 \star {\rm d} x + b_0 \star {\rm d} y$ clearly satisfies both ${\rm d}j = 0$ and ${\rm d}\star j = 0$. We therefore drop it and focus on the terms with coefficients that are linear in $x , y$. Current conservation ($\partial^\mu j_\mu = 0$) implies that $a_1 + b_2 = 0$. At linear order, this leaves a combination
\begin{equation}\label{eq:4.76}
    j^{(1)} = a j^{(h_1)} + b j^{(h_2)} + c j^{(e)}
\end{equation}
of three linearly independent conserved currents:
\begin{equation}
    j^{(h_1)} = y\, {\rm d}x + x\, {\rm d}y , \quad j^{(h_2)} = - x\, {\rm d}x + y\, {\rm d}y , \quad j^{(e)} = x\, {\rm d}x + y\, {\rm d}y ,
\end{equation}
where we assumed that $\star\, {\rm d}x = {\rm d} y$ and $\star\, {\rm d}y = - {\rm d} x$. The first two correspond to the vector fields $v^{(h_1)} = x \partial_x - y \partial_y$ and $v^{(h_2)} = y \partial_x + x \partial_y\,$, which generate hyperbolic flows with fixed point $o$. The third one corresponds to $v^{(e)} = y \partial_x - x \partial_y$ generating an elliptic flow (namely, rotation) around $o$.

The non-vanishing circulation of such an elliptic flow breaks parity or invariance under reflection at any line through $o$. Since parity is a symmetry in the bulk of the critical network model after disorder averaging -- in case of doubt, please inspect Fig.\ \ref{fig:CC} -- we expect that the coefficient $c$ of the last term in Eq.\ (\ref{eq:4.76}) has zero disorder average, $\mathbb{E}(c) = 0$, at criticality. Note also that $\star j^{(h_1)} = j^{(h_2)}\,$. Hence current conservation ${\rm d}j^{(h_1)} = {\rm d} j^{(h_2)} = 0 $ goes along with ${\rm d} \star j^{(h_1)} = {\rm d} \star j^{(h_2)} = 0$. We thus arrive at the desired result
\begin{equation}
    \mathbb{E} \big( \bullet ({\rm d} \star j)_o \big) = 0 ,
\end{equation}
at criticality, and for $o$ in the bulk and away from any operator insertion $\bullet$.

In summary, we have gone to some length to make a convincing argument underpinning the following scenario. As a model of unitary quantum mechanics, the Chalker-Coddington network model has a $\mathrm{U}(1)$ charge current which is conserved or divergence-free; that property holds on and off the critical point (and already before disorder averaging). At the plateau-transition critical point, and after disorder averaging, the continuum limit of that $\mathrm{U}(1)$ current also becomes \emph{curl-free}. This fact is key to the rest of the paper; in fact, we will use it to demonstrate that some (but not all) divergence-free currents of the SUSY vertex model become curl-free at criticality. To that end, we now take a detour to introduce the current that enters the current-current correlation function for the Kubo conductivity.

\subsection{Bi-local conductivity tensor}\label{sect:Kubo}

According to the Kubo theory of linear response, the electrical conductance is a current-current correlation function. More precisely, the d.c.\ conductance, $G$, associates with a pair of homology cycles $C_1$ and $C_2$ a quantum statistical expectation value $G(C_1,C_2) \propto \langle I(C_1) I(C_2) \rangle$ where $I(C) = \int_C j$ is a certain operator for the electrical current crossing the hypersurface $C$. Expressed in suitable physical units, the number $G(C_1,C_2)$ is the linear response current flowing across $C_1$ when the system is driven by an electrical voltage along a cycle dual (by the intersection pairing) to $C_2\,$. For non-interacting electrons, and in particular for our network model, the current-current correlation function for the d.c.\ conductance can be reduced to an explicit and simple form, cf.\ \cite{barstone}, as explained in this section.

\begin{figure}
    \begin{center}
        \epsfig{file=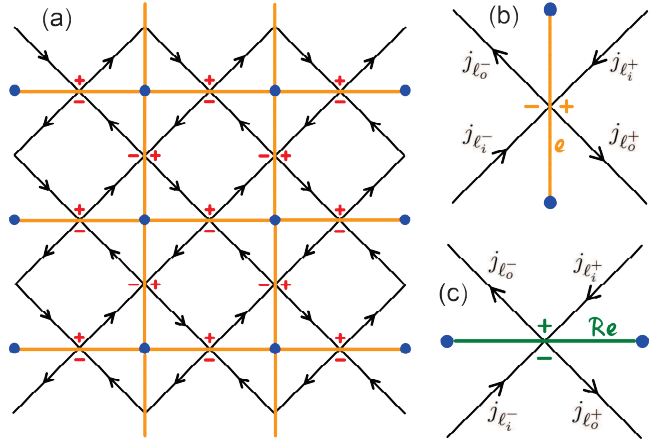,height=6cm}
    \end{center}
    \caption{(a) The auxiliary (or medial) lattice $\Gamma$. The $0$-cells (blue) and the $1$-cells (orange) of $\Gamma$ are in bijection with the plaquettes circulating clockwise and the nodes of the primary lattice, respectively. Each $1$-cell of $\Gamma$ is given an outer orientation by an assignment of plus/minus to its two sides. The $2$-cells of $\Gamma$ correspond to the plaquettes circulating counterclockwise. (b) Illustration of the $1$-cochain $e \mapsto j(e) = j_{\ell_i^-} - j_{\ell_o^-} = j_{\ell_o^+}-j_{\ell_i^+}$. (c) Illustration of the $1$-cochain $e \mapsto (\star^{-1} j)(e) \equiv j(R e) = j_{\ell_o^-} - j_{\ell_i^+} = j_{\ell_i^-} - j_{\ell_o^+}$.}
    \label{fig:Gamma}
\end{figure}

In a $2d$ continuum $C_1$ and $C_2$ would be curves. To formulate a discrete analog of $\int_C j$ we discretize $C$ as a $1$-chain on an auxiliary lattice $\Gamma$, the so-called medial lattice (of the square lattice of the network model), with $0$-cells and $1$-cells that are in bijection with a checkerboard of network-model plaquettes and connecting nodes, respectively; see Fig.\ \ref{fig:Gamma}a\,.
For each $1$-cell of $\Gamma$ we choose a plus side and a minus side to fix an outer orientation.

Let now the network be in a steady state with current distribution $\ell \mapsto j_\ell$ which is stationary, so that Kirchhoff's rule holds at every node. Having associated the $1$-cells of the medial lattice $\Gamma$ with nodes of the primary network, we may assign to any such $1$-cell, $e$, the current crossing it. This is done in the obvious way [Fig.\ \ref{fig:Gamma}b]: the network-model node for $e$ joins four links $\ell$ of the primary lattice; if these are indexed by $i/o$ for incoming/outgoing and by $+/-$ for the plus/minus side of $e$, then Kirchhoff's nodal rule states that $j_{\ell_i^+} + j_{\ell_i^-} = j_{\ell_o^+} + j_{\ell_o^-}$ and the stationary current flowing across $e$ is
\begin{equation}\label{eq:def-j(e)}
    j(e) \equiv j_{\ell_i^-}-j_{\ell_o^-} = j_{\ell_o^+}-j_{\ell_i^+}\,.
\end{equation}
In this way we re-interpret $j$ as a $1$-cochain on the medial lattice $\Gamma$. The integral $\int_C j$ is then given by the $1$-cochain $j$ paired with the $1$-chain $C$:
\begin{equation}
    \int_C j = \sum_{e \in C} j(e) .
\end{equation}
Note that the $1$-cochain $e \mapsto j(e)$ is closed ($\mathrm{d} j = 0$), as it was constructed by a scheme of coarse graining that respects the law of current conservation. Put differently, the total current $(\mathrm{d} j)_A$ flowing into any $2$-cell $A$ of $\Gamma$ is zero.

So far, we have left open the details of what current distribution $\ell \mapsto j_\ell$ we have in mind. We might follow \cite{BWZ2} and Sect.\ \ref{sect:SD-adapted} to set $j_\ell \equiv \vert \psi(\ell) \vert^2$ where $\psi = U \psi$ is a stationary state of quasi-energy zero for the network model with incoming-wave boundary conditions at a point contact. However, that is not the choice we want to make here. Instead, we are going to identify the conserved current $j$ with (either one of) the current operators in the current-current correlation function $\langle j j \rangle$ for the conductance $G(C_1,C_2)$.

For that purpose, consider the squared Green's function
\begin{equation}\label{eq:sqGF}
    \sigma_{\ell_1,\,\ell_2} = \big| \langle \ell_1 \vert (1 - T)^{-1} \vert \ell_2 \rangle \big|^2 \qquad (T = QU)
\end{equation}
between two distant links $\ell_1, \ell_2$ of the primary lattice. It has the property of being doubly closed; in other words, Kirchhoff's rule holds w.r.t.\ both of its arguments. To check that statement, say for the case of the right argument $\ell_2$ (and fixed left argument $\ell_1$), we sum $\sigma_{\ell_1 , \, \ell_2}$ over the two links $\ell_2 = \ell_{\rm in}$ that are incoming to any given node of the primary network and compare with the sum over the two links $\ell_2 = \ell_{\rm out}$ that are outgoing from the same node. The two sums agree -- that's Kirchhoff's rule for $\ell_2 \mapsto \sigma_{\ell_1 ,\, \ell_2}$ interpreted as a current distribution. Its proof simply utilizes the unitarity of $U$:
\begin{align*}
    &\sum\nolimits_{\ell_{\rm in}} \langle \ell_1 \vert (1-T)^{-1} \vert \ell_{\rm in} \rangle \langle \ell_{\rm in} \vert (1-T^\dagger)^{-1} \vert \ell_1 \rangle \cr
    &= \langle \ell_1 \vert (1-T)^{-1} U \left( \sum\nolimits_{\ell_{\rm in}} \vert \ell_{\rm in} \rangle \langle \ell_{\rm in} \vert \right) U^\dagger (1 - T^\dagger)^{-1} \vert \ell_1 \rangle \cr
    &= \sum\nolimits_{\ell_{\rm out}} \langle \ell_1 \vert (1-T)^{-1} \vert \ell_{\rm out} \rangle \langle \ell_{\rm out} \vert (1 - T^\dagger)^{-1} \vert \ell_1 \rangle .
\end{align*}
Here, assuming that $\ell_1$ and $\ell_{\rm in}$ are distant from each other and from any contact links, we used that $\langle \ell_1 \vert (1-T)^{-1} \vert \ell_{\rm in} \rangle = \langle \ell_1 \vert (1-T)^{-1} U \vert \ell_{\rm in} \rangle$.

In a paragraph above, we described how to convert a current distribution $\ell \mapsto j_\ell$ into a $1$-cochain $e \mapsto j(e)$. Following that blueprint, we now turn the squared Green's function $(\ell_1 , \ell_2) \mapsto \sigma_{\ell_1,\,\ell_2}$ into a double $1$-cochain $(e,e^\prime) \mapsto \sigma(e,e^\prime)$ on pairs $e, e^\prime$ of $1$-cells, again on the medial lattice $\Gamma$. That double $1$-cochain $\sigma$ on $\Gamma$ is the network-model analog for the Fermi-surface part of the non-local response function of conductivity; cf.\ Eq.\ (52) of \cite{barstone}. On the grounds of that correspondence, the dimensionless conductance associated with two cycles $C_1$ and $C_2$ on $\Gamma$ comes out to be the double sum
\begin{equation}\label{eq:Kubo-cond}
    G(C_1 , C_2) = \sum_{e \in C_1} \sum_{e^\prime \in C_2} \sigma(e , e^\prime) .
\end{equation}
By Kirchhoff's rule for the squared Green's function, the non-local response function $\sigma$ is doubly closed. Therefore, the conductance $G(C_1,C_2)$ depends on the cycles $C_1, C_2$ only through their homology classes, as required.

Let us add a couple of remarks here to preclude any misunderstandings.
Firstly, in continuous position space with continuous-time dynamics generated by a Hamiltonian, one obtains the non-local d.c.\ conductivity tensor by hitting each argument of the squared Green function with the first-order vector differential operator of momentum or velocity. In contrast, the conductivity tensor of the network model is directly given by the squared Green function (\ref{eq:sqGF}), without any derivatives entering. This is because the Hilbert space of the network model is spanned by states on links (not on nodes). In fact, since the links $\ell$ are directed, the squared wave function $\ell \mapsto \vert \psi(\ell) \vert^2$ has the simultaneous meaning of a probability \emph{current} density.

Secondly, the expression (\ref{eq:Kubo-cond}) is essentially the Kubo-Greenwood formula (not to be confused with the Landauer-B\"uttiker formula!) for the conductance as a current-current correlation function; cf.\ \cite{barstone}. The cycles $C_1$ and $C_2$ enclose two terminals ($1$ and $2$) that are employed for the purpose of measuring a conductance. Because the non-local conductivity response function $\sigma(e,e^\prime)$ is doubly closed (as a double $1$-cochain) or doubly divergence-free (as a bi-vector field), the cycles $C_1$ and $C_2$ can be pushed around at will in their respective homology classes (i.e., subject to the condition of staying away from the terminals and from each other), without changing the conductance $G(C_1,C_2)$. Of course, we are not saying that $G(C_1,C_2)$ is independent of the sample geometry.

\subsection{Critical conductivity}\label{sect:critical}

What are the implications at criticality? As it stands, the non-local response function $(e,e^\prime) \mapsto \sigma(e,e^\prime)$ of conductivity is a random variable due to its dependence on the $\mathrm{U} (1)$ random phase factors in $U_{\rm r}\,$. By the act of taking the disorder average, the response function becomes translation-invariant provided that a spatially homogeneous regularization scheme is adopted; in formulas we have that
\begin{equation}
    \mathbb{E} \big( \sigma(e,e^\prime) \big) =
    \mathbb{E} \big( \sigma(t_a \, e , t_a \, e^\prime) \big)
\end{equation}
holds for any lattice translation $t_a$ of $(e,e^\prime)$ to an equivalent pair $(t_a \, e  , t_a \, e^\prime)$.

Moreover, at the phase-transition critical point we expect conformal invariance to emerge as a symmetry in the infrared limit. To fathom the consequences thereof, we recall from differential calculus in the continuum (see Sect.\ \ref{sect:Affleck}) that (i) a complex structure in two dimensions is a rotation $R$ by $\pm \pi/2$, which determines (ii) a Hodge star operator $\star$ and (iii) a decomposition of the current density $j = j^{10} + j^{01}$ by Eqs.\ (\ref{eq:4.76mz}). The continuum line integral $\int_C j^{01}$ splits as
\begin{equation}\label{eq:int-J}
    \int_C j^{10} = \frac{1}{2} \int_C j - \frac{\mathrm{i}}{2} \int_C \star^{-1} j
\end{equation}
into the current $\int_C j$ flowing across $C$ and the current $\int_C \star^{-1} j = - \int_C \star j$ circulating along $C$. Recall that if $j$ is closed, then the transverse current $\int_C j$ depends on $C$ only through its homology class. On the other hand, if $j$ is co-closed ($\mathrm{d} \star j = 0$), then it is the longitudinal current that enjoys the property of invariance $\int_{C_1} \star j = \int_{C_2} \star j$ for homologous curves (i.e., for $C_1 - C_2 = \partial S$ a boundary).

Let us now discuss a lattice version of the Hodge-dual $\star^{-1} j$, first for the illustrative example of a $1$-cochain $e \mapsto j(e)$ and afterwards for the relevant case of our double $1$-cochain $(e,e^\prime) \mapsto \sigma(e, e^\prime)$ of conductivity. On the lattice as in the continuum, the complex structure $R$ is rotation by $\pm \pi/2$. We defined the values of the $1$-cochain $e \mapsto j(e)$ by the current $j(e)$ across $e$. In the same vein, we now introduce on $\Gamma$ another $1$-cochain $e \mapsto (\star^{-1} j)(e)$ by
\begin{equation}
    (\star^{-1} j)(e) \stackrel{\rm def}{=} j(R e) ,
\end{equation}
where $j(R e)$ still means the current across the rotated $1$-cell $R e$; by definition it is the same as the current \emph{along} the unrotated $1$-cell $e$. (Note that $e$ is a $1$-cell of $\Gamma$, whereas $Re$ is a $1$-cell of the lattice \emph{dual} to $\Gamma$.) Referring for notation to Fig.\ \ref{fig:Gamma}c and the text vicinity of Eq.\ (\ref{eq:def-j(e)}), that current is
\begin{displaymath}
    (\star^{-1} j)(e) = j_{\ell_o^-} - j_{\ell_i^+} = j_{\ell_i^-} - j_{\ell_o^+} \,,
\end{displaymath}
if $\ell_o^-$ and $\ell_i^+$ are the network-model links on the plus side of the rotated $1$-cell $R e$; otherwise it is the negative thereof.

\begin{figure}
    \begin{center}
        \epsfig{file=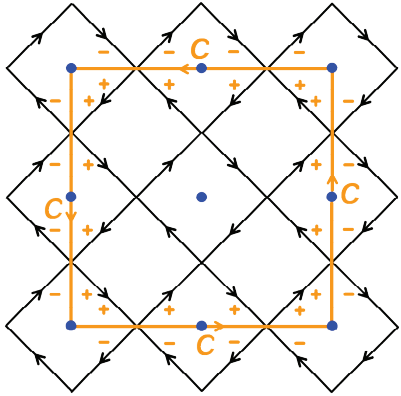,height=6cm}
    \end{center}
    \caption{Illustration of the lattice sum $\int_C \star^{-1} j$ for the current circulation. For ease of drawing, the outer orientation of $C$ is converted into an inner orientation, say by the counterclockwise sense of circulation. The lattice sum $\int_C \star^{-1} j$ is then seen to be a weighted sum of primary-lattice currents $j_\ell$ with weights $1+1$, $1$, $0$, $-1$ as indicated by the orange symbols $+/-$ at the links $\ell$.} \label{fig:circulate}
\end{figure}

For a path $C$ on the medial lattice $\Gamma$, consider now the sum
\begin{equation}
    \int_C \star^{-1} j\; \stackrel{\rm def}{=}\; \sum_{e \in C} j(Re).
\end{equation}
If $C$ is a cycle, that sum computes (a lattice approximation to) the current circulation around $C$, cf.\ Fig.\ \ref{fig:circulate};
in general, it is not zero. However, at the critical point with conformal invariance, we do expect the current circulation to vanish after disorder averaging and in the scaling limit of large and contractible cycles $C$; the argument for that was spelled out in detail in Sect.\ \ref{sect:SD-adapted}. Thus the continuity equation $\mathrm{d} j = 0$ should be augmented with a second conservation law $\mathrm{d} \star j = 0$, valid in expectation $\mathbb{E} (...)$ and after coarse graining to eliminate lattice effects present for short wavelengths. The lattice currents $j^{10} = (j + \mathrm{i} \star j) / 2$ and $j^{01} = (j - \mathrm{i}\star j) /2$ will then be the parents of holomorphic and anti-holomorphic currents in the continuum.

For future reference, we put on record that the lattice currents $j^{10}$ and $j^{01}$ are given by
\begin{equation}
    e \mapsto j^{10}(e) = \big(j(e) - \mathrm{i}\, j(R e) \big)/2
\end{equation}
and similar for $j^{01}$ (with $- \mathrm{i} \to + \mathrm{i}$). We also note that, using the notation of Figs.\ \ref{fig:Gamma}b and \ref{fig:Gamma}c, the expression for $j^{10}(e)$ can be rewritten as
\begin{displaymath}
    j^{10}(e) = \frac{1+\mathrm{i}}{4} j_{\ell_o^+}
    + \frac{-1+\mathrm{i}}{4} j_{\ell_i^+}
    + \frac{-1-\mathrm{i}}{4} j_{\ell_o^-}
    + \frac{1-\mathrm{i}}{4} j_{\ell_i^-} \;.
\end{displaymath}
An invariant formulation of the same quantity is
\begin{equation}
    j^{10}(e) = 2^{-3/2} \sum\nolimits_\ell \mathrm{e}^{- \mathrm{i} \theta_e(\ell)} j_\ell \,,
\end{equation}
where the sum is over the four links $\ell$ joined by the node of $e$ and $\theta_e (\ell)$ denotes the angle of positive rotation (as determined by the choice of complex structure $R$) from the $e$-perpendicular axis (pointing from minus to plus by the outer orientation of $e$) to the direction of $\ell$.

We finally apply the conformal invariance argument (or symmetry doubling scenario for a secondary conservation law to emerge) to the key object of our endeavor: the non-local conductivity response function $\sigma$ at criticality. To begin, we recall that $(e,e^\prime) \mapsto \sigma(e, e^\prime)$ is a double $1$-cochain on $\Gamma$ and denote its disorder average by $\Sigma(e , e^\prime) = \mathbb{E}(\sigma(e,e^\prime))$. Earlier, we deduced from Kirchhoff's nodal rule for the squared Green's function that $\Sigma$ is closed with respect to both of its arguments -- this property already held before taking the disorder average and it still does so afterwards. Now, by virtue of the emerging conformal invariance at the critical point and by the reasoning of Sect.\ \ref{sect:SD-adapted}, we expect $\Sigma$ also to become co-closed with respect to both of its arguments, in the continuum limit. Altogether, this then implies that the tensor component $\Sigma_{zz} \equiv \Sigma^{10,10}$ becomes holomorphic with respect to both arguments:
\begin{equation}\label{eq:77-mz}
    \bar\partial \Sigma_{zz} (\bullet,\bullet^\prime) = 0 =
    \bar\partial^\prime \Sigma_{zz} (\bullet,\bullet^\prime) .
\end{equation}
Here $\Sigma_{zz}(\bullet,\bullet^\prime)$ means the continuum limit
of the linear combination
\begin{equation}\label{eq:78-mz}
    \Sigma_{zz} (e,e^\prime) = \Sigma \big(e - \mathrm{i} R e \,, \, e^\prime - \mathrm{i} R e^\prime \big) / 4
\end{equation}
of lattice response functions and their $\star\,$-duals, and our concise notation exploits the fact that a double $1$-cochain is a $\mathbb{C}$-bilinear function of the two chains in its arguments. By the same reasoning, $\Sigma_{\bar{z} \bar{z}}$ (defined in the analogous way) becomes anti-holomorphic w.r.t.\ both arguments.

If we specialize to the continuum of the Euclidean plane with complex coordinate $w = x + \mathrm{i} y$ (or the Riemann sphere with complex stereographic coordinate $w$), then the (anti-)holomorphic part of the response function can be presented in explicit form. Indeed, the rotational invariance of $\Sigma_{zz} = \Sigma_{zz}(w,w^\prime) \, \mathrm{d}w \otimes \mathrm{d}w^\prime$ in the continuum limit implies
\begin{equation}
    \Sigma_{zz}(w,w^\prime) = \mathrm{e}^{2\mathrm{i}\theta} \Sigma_{zz} (\mathrm{e}^{\mathrm{i}\theta} w , \mathrm{e}^{\mathrm{i}\theta} w^\prime)
\end{equation}
for any rotation angle $\theta$. In combination with translational invariance, this determines the holomorphic tensor component to be of the form
\begin{equation}\label{eq:planar-cond}
    \Sigma_{zz} (w,w^\prime) = \frac{n}{(w - w^\prime)^2} \,,
\end{equation}
with a constant $n$ that at the present stage could be any positive real number. The singularity on the diagonal $w = w^\prime$ reflects a singularity of $\Sigma$ that is immanent to its microscopic definition by the squared Green's function (\ref{eq:sqGF}). The expression for $\Sigma_{\bar{z} \bar{z}}$ is similar, with $(w-w^\prime)$ replaced by $(\bar{w} - \bar{w}^\prime)$.

As a disclaimer, let us stress that the formula (\ref{eq:planar-cond}) assumes the translation-invariant regularization of the Green's function $\langle \ell_1 \vert (1-T)^{-1} \vert \ell_2 \rangle$, $T = QU$, by an infinitesimal absorbing background $Q = \mathrm{e}^{\mu}$ ($\mu \to 0-$), as stated at the outset of the present subsection. For the real-world purpose of defining and computing a conductance, one must replace the absorbing background by a number $n_{\rm c}$ of terminals ($n_{\rm c} \geq 2$). If these are taken to be point contacts, the non-local response function of conductivity becomes an ($n_{\rm c} + 2$)-point function. The latter has the singularity $\Sigma_{zz}(w,w^\prime) \sim (w - w^\prime)^{ -2} $ for $w \to w^\prime$ but also further singularities when $w$ or $w^\prime$ approaches a terminal point.

\subsection{Response function in the vertex model}
\label{sect:resp-VM}

{}From the preceding section we take away the key message that the non-local conductivity response function $\Sigma$ for the network model at criticality has a $(10,10)$ tensor component $\Sigma_{zz}$ which is doubly holomorphic and exhibits the short-distance singularity (\ref{eq:planar-cond}). Guided by this insight we now seek an expression for $\Sigma$ as a correlator of operators in the SUSY vertex model of Sect.\ \ref{sect:SUSY-VM}. Our motivation for doing so is that the resulting operators might be candidates for holomorphic currents of the CFT to be identified.

The first step is to produce operators $V(\ell)$ and $W(\ell)$ expressed in terms of the boson and fermion Fock operators at the link $\ell\,$, such that
\begin{equation}
    \Sigma_{\ell_1 , \, \ell_2} = \mathrm{STr}_\mathcal{V} \; V(\ell_1) W(\ell_2) = \langle V(\ell_1) W(\ell_2) \rangle_\mathcal{V} \,,
\end{equation}
where $\Sigma_{\ell_1 , \, \ell_2} = \mathbb{E} (\sigma_{\ell_1 , \, \ell_2})$ is the disorder-averaged squared Green's function introduced in Eq.\ (\ref{eq:def-Sigma}). There exist numerous choices of such operators, and they already exist in the minimal theory with only one replica. For simplicity of notation, let us specialize to that case ($r = 1$). The good objects to consider then are
\begin{align}
    &V^0_{\;\; 2} = - b_+ b_- \,, \quad
    V^0_{\;\; 3} = b_+ f_- \,, \quad
    V^1_{\;\; 2} = - f_+ b_- \,, \quad
    V^1_{\;\; 3} = f_+ f_- \,, \cr
    &W^{\! 2}_{\; 0} = b_-^\dagger b_+^\dagger \,, \quad
    W^{\! 2}_{\; 1} = b_-^\dagger f_+^\dagger \,, \quad
    W^{\! 3}_{\; 0} = f_-^\dagger b_+^\dagger \,, \quad
    W^{\! 3}_{\; 1} = f_-^\dagger f_+^\dagger \,,
    \label{eq:92}
\end{align}
where we employ the notation introduced in Sect.\ \ref{sect:Read}. All operators carry the same link argument $\ell\,$, which has been omitted. The desired correlator $\Sigma_{\ell_1 , \, \ell_2}$ for $\ell_1 \not= \ell_2$ is obtained by suitable pairings of these:
\begin{align}
    \Sigma_{\ell_1 , \, \ell_2} &= - \big\langle V^{0}_{\;\; 2}(\ell_1) W^{\! 2}_{\; 0}(\ell_2) \big\rangle_\mathcal{V} = + \big\langle V^{1}_{\;\; 3}(\ell_1) W^{\! 3}_{\; 1}(\ell_2) \big\rangle_\mathcal{V} \cr &= - \big\langle V^{1}_{\;\; 2}(\ell_1) W^{\! 2}_{\; 1}(\ell_2) \big\rangle_\mathcal{V} = + \big\langle V^{0}_{\;\; 3}(\ell_1) W^{\! 3}_{\; 0}(\ell_2) \big\rangle_\mathcal{V} \,. \label{eq:82-mz}
\end{align}
To verify that assertion, e.g.\ for the pair of $V^{0}_{\;\; 2}$ with $W^{\! 2}_{\; 0}\,$, one starts by using the Wick contraction rule in the free theory (before disorder averaging):
\begin{align*}
    - \langle V^{0}_{\;\; 2} (\ell_1) W^{\! 2}_{\; 0}(\ell_2)
    \rangle_\mathcal{F} &= \big\langle b_+^\dagger (\ell_2) b_+^{\vphantom{\dagger}} (\ell_1) \big\rangle_\mathcal{F} \, \big\langle b_-^{\vphantom{\dagger}}(\ell_1) b_-^\dagger (\ell_2)  \big\rangle_\mathcal{F} \,,
\end{align*}
together with the formulas (\ref{eq:basic-Wick}) for the basic Wick contractions. The claimed relation (\ref{eq:82-mz}) then follows immediately by taking the disorder average on both sides of the equation.

As a direct consequence \cite{BWZ2} of the global $\mathrm{U}(r,r|2r)$ symmetry of the SUSY vertex model, each of the operators $V^{\! \alpha}_{\;\; \beta}$ and $W^{\! \alpha}_{\; \beta}$ obeys Kirchhoff's nodal rule in the sense of Sect.\ \ref{sect:Kubo}. One might therefore think that, by following the blueprint of Sect.\ \ref{sect:critical} to construct from $V$ and $W$ operator-valued closed $1$-cochains which become co-closed at criticality, one could produce the desired holomorphic and anti-holomorphic currents. However, such a direct attempt does not deliver the optimal return. Indeed, the big advantage of the SUSY vertex model, as compared with the network model, is the existence of a boson-fermion \emph{multiplet} of operators, offering the possibility of a non-Abelian current algebra. Yet, the basic Lie algebra structure needed for a non-Abelian current algebra is absent from the present setup, as the $V$'s and $W$'s do not close under the Lie superbracket. Hence we are going to modify the ansatz (\ref{eq:92}).

\subsection{Current algebra from response function}\label{sect:4.6}

The idea for a modified ansatz that does deliver is very simple: change the basis of Fock operators by mixing the retarded and advanced sectors! Such a change of basis was crucial for the progress made in \cite{BWZ2}, and it turns out to be key here as well. Thus at every link $\ell$ of the primary network we now take linear combinations $B_\pm(\ell)$ and  $C_\pm (\ell)$ of the fundamental bosons $b_\pm(\ell)$ and $b_\pm^\dagger (\ell)$:
\begin{align}
    &B_+ = \big( b_+^\dagger - \mathrm{e}^{\mathrm{i}\vartheta_0}\, b_- \big) / \sqrt{2} \,, \quad B_- = \big( b_+ + \mathrm{e}^{-\mathrm{i} \vartheta_0} \, b_-^\dagger \big) / \sqrt{2} \,, \cr
    &C_+ = \big( b_+^\dagger + \mathrm{e}^{\mathrm{i}\vartheta_0}\, b_- \big) / \sqrt{2} \,, \quad C_- = \big( b_+ - \mathrm{e}^{-\mathrm{i} \vartheta_0} \, b_-^\dagger \big) / \sqrt{2} \,, \label{eq:new-bos}
\end{align}
and we make a similar transformation also for the fermions:
\begin{align}
    &F_+ = \big( f_+^\dagger + \mathrm{e}^{\mathrm{i} \vartheta_1} f_- \big) / \sqrt{2} \,, \quad F_- = \big( f_+ + \mathrm{e}^{-\mathrm{i} \vartheta_1} f_-^\dagger \big) / \sqrt{2} \,, \cr
    &G_+ = \big( f_+^\dagger - \mathrm{e}^{\mathrm{i}\vartheta_1} f_- \big) / \sqrt{2} \,, \quad G_- = \big( f_+ - \mathrm{e}^{-\mathrm{i} \vartheta_1} f_-^\dagger \big) / \sqrt{2} \,. \label{eq:new-ferm}
\end{align}
The unitary factors $\mathrm{e}^{\mathrm{i} \vartheta_0}$ and $\mathrm{e}^{ \mathrm{i} \vartheta_1}$ are arbitrary but fixed (independent of $\ell$). Operators denoted by the same letter constitute canonical pairs:
\begin{equation}\label{eq:canon-rel}
    [B_- , B_+] = 1\,, \quad [C_- , C_+] = 1\,, \quad [B_- , C_+] = 0\,, \quad [F_- , F_+] = 1\,,
\end{equation}
and so on (the bracket of two fermion operators is the anti-commutator). The relations under Hermitian conjugation in Fock-Hilbert space are diagonal for the fermions but off-diagonal for the bosons:
\begin{equation}\label{eq:Herm-conj}
    F_-^\dagger = F_+ \,, \quad G_-^\dagger = G_+\,, \quad
    B_-^\dagger = C_+ \,, \quad B_+^\dagger = C_- \,.
\end{equation}

By forming such products as $B_- B_+\,$, taking the left factor from the minus-set and the right factor from the plus-set, one gets $4^2 = 16$ quadratic operators. [In the case of $r$ replicas their number would be $(4r)^2 = 16 r^2$.] Arranged as a supermatrix (with ordering even-odd-even-odd), they are
\begin{equation}\label{eq:mat-gl22}
    \mathcal{O} \equiv \begin{pmatrix}
    B_- B_+ &B_- F_+ &B_- C_+ &B_- G_+ \cr
    F_- B_+ &F_- F_+ &F_- C_+ &F_- G_+ \cr
    C_- B_+ &C_- F_+ &C_- C_+ &C_- G_+ \cr
    G_- B_+ &G_- F_+ &G_- C_+ &G_- G_+
    \end{pmatrix} .
\end{equation}
By the canonical bracket relations (\ref{eq:canon-rel}), these quadratic expressions realize the (co-adjoint representation of the) Lie superalgebra $\mathfrak{gl}(2|2)$ at the given link $\ell$, with the matrix position of each operator encoding its behavior w.r.t.\ the Lie superbracket. More explicitly, if $X$ and $Y$ are parameter supermatrices arranged in the same way as $\mathcal{O}$, then
\begin{equation}\label{eq:Lalg-iso}
    \big[ \mathrm{STr}(X \mathcal{O}) , \mathrm{STr}(Y \mathcal{O}) \big] = \mathrm{STr}\big( [X,Y] \mathcal{O} \big).
\end{equation}

The true significance of the matrix arrangement (\ref{eq:mat-gl22}) is that it groups the quadratic operators into four blocks, each of which will acquire a distinct meaning. Here comes the working hypothesis that we intend to explain in the sequel: the right upper block bosonizes to a Wess-Zumino-Witten field, $M$; the left lower block bosonizes to $M^{-1}$. The left upper block gives rise to the holomorphic currents $J \leftrightarrow \partial M \cdot M^{-1}$, and the right lower block plays the same role on the anti-holomorphic side: $\bar{J} \leftrightarrow M^{-1} \bar\partial M$.

In the first step, we focus on the operators in the left upper block of the matrix (\ref{eq:mat-gl22}), which are made from $B_\pm$ and $F_\pm\,$. By construction, these generate a subalgebra $\mathfrak{gl}(1|1) \subset \mathfrak{gl}(2|2)$. [For a number $r \geq 1$ of replicas, this would be a subalgebra $\mathfrak{gl}(r|r) \subset \mathfrak{gl}(2r|2r)$.] For simplicity of notation, we introduce the symbol $O^{\alpha}_{\;\; \beta} $ ($\alpha, \beta = 0,1$) for our $\mathfrak{gl}(1|1)$ multiplet of operators:
\begin{align}
    &O^{0}_{\;\; 0} =\; : \! B_- B_+ \! : \,, \quad O^{0}_{\;\; 1} = B_- F_+ \,, \cr &O^{1}_{\;\; 0} = F_- B_+\,,\quad O^{1}_{\;\; 1} = \; :\! F_- F_+ \!: \,.
\end{align}
As usual, the double colon means normal-ordering; i.e., we subtract the expectation value in the Fock vacuum. This step ensures that all one-point functions vanish: $\langle O^{\alpha}_{\;\; \beta}(\ell) \rangle_\mathcal{V} = 0\,$.

Next we express the two-point correlation functions of these operators as disorder averages in the network model. As before, the calculational method is to use the Wick contraction rule for the free-field correlators $\langle \ldots \rangle_\mathcal{F}$ and then take the disorder average. Let us state the outcome of this straightforward calculation in a concise manner. For that we recall our index-free notation $O^X \equiv \mathrm{STr} (OX) = \mathrm{STr}(XO) \equiv (-1)^{|\alpha|} X^{\! \alpha}_{\; \beta}\, O^{\beta}_{\;\; \alpha}$ where $X^{\! \alpha}_{\; \beta}$ are the matrix elements of a parameter supermatrix $X$ over $\mathfrak{gl}(1|1)$. Assuming $\ell_1 \not= \ell_2$ we then find
\begin{align}
    \langle O^X (\ell_1) O^Y (\ell_2) \rangle_\mathcal{V} = &- \mathrm{STr}(XY)\, \big( \Sigma_{\ell_1 , \, \ell_2} + \Sigma_{\ell_2 , \, \ell_1} \big) / 4 \cr &- \mathrm{STr}(X)\, \mathrm{STr}(Y) \big( \Upsilon_{\ell_1 , \, \ell_2} + \Upsilon_{\ell_2 , \, \ell_1} \big) / 4 \,, \label{eq:87}
\end{align}
where $\Upsilon_{\ell_1,\,\ell_2}$ was defined in Eq.\ (\ref{eq:def-Xi}).
Our logic now proceeds as follows.

We first look at two odd operators $O^{0}_{\;\; 1}(\ell_1)$ and $O^{1}_{\;\; 0}(\ell_2)$, in which case the summand in the second line of (\ref{eq:87}) is absent and we simply get
\begin{equation}\label{eq:89-mz}
    \big\langle O^{0}_{\;\; 1} (\ell_1) O^{1}_{\;\; 0} (\ell_2) \big\rangle_\mathcal{V} = \big( \Sigma_{\ell_1 , \, \ell_2} + \Sigma_{\ell_2 , \, \ell_1} \big) / 4 \,.
\end{equation}
We then recall the procedure that takes the lattice quantity $\Sigma_{ \ell_1 ,\, \ell_2}$ to its holomorphic continuum limit $\Sigma_{zz}$ in Eqs.\ (\ref{eq:77-mz}, \ref{eq:78-mz}). (Notice that the symmetrization from $\Sigma_{\ell_1 , \, \ell_2}$ to $\Sigma_{\ell_1 , \, \ell_2} + \Sigma_{\ell_2 , \, \ell_1}$ makes no difference in that limit, thanks to rotational invariance.) In view of the equality (\ref{eq:89-mz}) we expect that the same procedure applied to the lattice quantities $O^{0}_{\;\; 1} (\ell_1)$ and $O^{1}_{\;\; 0} (\ell_2)$ gives rise to holomorphic currents $J^{0}_{\;\; 1}$ and $J^{1}_{\;\; 0}$ in the continuum. The expression (\ref{eq:planar-cond}) for the critical response function then implies that the operator product expansion for these currents has the leading singularity shown in (\ref{eq:OPE-d}, \ref{eq:CK-d}).

For more general choices of $O^{X,Y}$ the term in the second line of (\ref{eq:87}) will also be present. To isolate it, we may take $O^X = O^Y = O^{0}_{\;\; 0} + O^{1}_{\;\; 1} \equiv O^N$ to be the operator representing the superparity $N = E_{\! 0}^{\; 0} - E_{\! 1}^{\; 1} \in \mathfrak{gl}(1|1)$ with $\mathrm{STr} N = 2$ and $\mathrm{STr} (N^2) = 0$, so that Eq.\ (\ref{eq:87}) reduces to
\begin{equation}
    \langle O^N (\ell_1) O^N (\ell_2) \rangle_\mathcal{V} = - ( \Upsilon_{\ell_1 , \, \ell_2} + \Upsilon_{\ell_2 , \, \ell_1}).
\end{equation}
Now $N \in \mathfrak{gl}(1|1)$ lies in $\mathfrak{gl}(2|2)$, the complexification of our Lie superalgebra $\mathrm{Lie} \, \mathrm{U} (1,1|2)$ of global symmetries. The operator-valued distribution $\ell \mapsto O^N(\ell)$ then satisfies Kirchhoff's nodal rule under the statistical expectation $\langle O^N(\ell) \, \cdots \rangle_\mathcal{V} \,$. Indeed, that rule follows as a Ward identity from the fact that the statistical weight $\rho(U_{\rm s})$ of the SUSY vertex model factors on nodes of the primary network and for each such node separately commutes with the global action of the symmetry group $\mathrm{U}(1,1|2)$.

Given Kirchhoff's rule for $O^N(\ell)$, there is now a clear path to continue the analysis: (i) From the distribution $\ell \mapsto O^N(\ell)$ we construct the operator-valued $1$-cochain $e \mapsto j^N(e)$ on the medial lattice $\Gamma$ (see Sect.\ \ref{sect:Kubo}). (ii) Using the Hodge star operator, we isolate the $10$--component of $j^N$ (see Sect.\ \ref{sect:critical}). (iii) We take the continuum limit at the critical point and denote the limit of the $10$--component of $j^N$ by $J^N$. The resulting continuum current $J^N$ is holomorphic; therefore the singularity of its operator product expansion with itself must be
\begin{equation}
    J^N(z) J^N(w) \sim \frac{\gamma}{(z-w)^2} \,,
\end{equation}
with an undetermined constant $\gamma$. Altogether, we see that our reasoning has reproduced the leading singularity of the OPE (\ref{eq:OPE-d}, \ref{eq:CK-d}) (for $r = 1$, but the case $r \geq 1$ is no different). Now recall that we want the theory to contain a primary field $M$ with vanishing conformal weights. From the discussion in Sect.\ \ref{sect:tell} we know that this will happen if $\gamma = 1$; we are thus led to \emph{postulate} that value for $\gamma$. As a direct consequence we can make a prediction: the scaling limit of the network-model correlator $\Upsilon_{\ell_1 , \, \ell_2}$ projected to its $(10,10)$-component $\Upsilon_{zz}$ is proportional to $\Sigma_{zz}$ with a definite amplitude ratio:
\begin{equation}\label{eq:amp-ratio}
    \Sigma_{zz} / \Upsilon_{zz} = - n / \gamma = - n \,.
\end{equation}
We offer this as a prediction to be verified by numerical simulation.

In order to complete the picture, we ought to check that our lattice currents conform to the next-to-leading singularity in the OPE (\ref{eq:OPE-d}). In the continuum theory, that singularity translates to the statement
\begin{equation}\label{eq:OPE-next}
    (2\pi \mathrm{i})^{-1} \oint_C dz \, J^X(z) J^Y(w) = J^{[X,Y]} (w) ,
\end{equation}
where $C$ is any integration contour that encloses $w$ (but no points of further operator insertions if such are present). It does not seem easy to check (\ref{eq:OPE-next}) directly for the SUSY vertex model in its present form as a statistical sum (or ``path integral''). However, it is textbook knowledge \cite{Polchinski} that the path integral relation (\ref{eq:OPE-next}) simply reflects the non-Abelian algebra of conserved charges in the quantum theory. That correspondence suggests to switch to the transfer-matrix (or ``quantum'') formulation of the SUSY vertex model. The calculation then simplifies if one assumes deformability of the transfer matrix to its anisotropic limit as the Hamiltonian of a quantum (super-)spin chain with Hilbert space $\ldots \otimes \mathcal{V}_\ell^{\vphantom{\ast}} \otimes \mathcal{V}_\ell^\ast \otimes \mathcal{V}_\ell^{\vphantom{\ast}} \otimes \mathcal{V}_\ell^\ast \otimes \ldots\,$. In fact, in the spin-chain limit verification of the lattice precursor to the continuum relation (\ref{eq:OPE-next}) becomes a straightforward matter of checking the bracket relations of the Lie superalgebra of operators displayed in the matrix of (\ref{eq:mat-gl22}). [On the dual space $\mathcal{V}_\ell^\ast$, the operators (\ref{eq:new-bos}, \ref{eq:new-ferm}) act by the co-representation.] This algebra is closed for the left upper block (as well as the right lower block), which is why we made the change of basis (\ref{eq:new-bos}, \ref{eq:new-ferm}) in the first place.

We finish this subsection with a few remarks. (i) The construction of the anti-holomorphic currents $\bar{J}$ can be done in essentially the same way; for that we take the $\mathfrak{gl}(1|1)$ algebra generated by the left upper block in (\ref{eq:mat-gl22}) and replace it by the $\mathfrak{gl}(1|1)$ algebra of the right lower block. (ii) One easily checks that the holomorphic currents $J$ have a trivial OPE with the anti-holomorphic currents $\bar{J}$. (iii) By the relations (\ref{eq:Herm-conj}) under Hermitian conjugation, the modes of the holomorphic boson-boson current $J^{0}_{\;\; 0}$ are adjoint to those of the anti-holomorphic current $\bar{J}^{0}_{\;\; 0}\,$, whereas in the case of the fermion-fermion currents $J^{1}_{\;\; 1}$ and $\bar{J}^{1}_{\;\; 1}$ they are adjoint to themselves. (iv) While the number $n$ in Eq.\ (\ref{eq:planar-cond}) could have been any positive number, it is now seen to be the quantized level of a current algebra with compact sector $\mathfrak{u}(r)$, which is non-Abelian for $r > 1$; as such it must be a positive integer. (v) Although we had to postulate the value $\gamma = 1$ here, we will offer a constructive reason for it in Sect.\ \ref{sect:SBS}.

\subsection{Wess-Zumino-Witten field}\label{sect:bosonize}

Our attention now turns to the right upper block of the matrix array
(\ref{eq:mat-gl22}); we abbreviate its entries as
\begin{equation}\label{eq:M-WZW}
    \begin{pmatrix}  B_- C_+ &B_- G_+ \cr F_- C_+ &F_- G_+\end{pmatrix}
    = \begin{pmatrix} M^{0}_{\;\; 0} &M^{0}_{\;\; 1} \cr
    M^{1}_{\;\; 0} &M^{1}_{\;\; 1} \end{pmatrix} .
\end{equation}
In \cite{BWZ2} we made a detailed study of these operators and the observable quantities derived from them (with different ordering conventions and $B \leftrightarrow C$, but it makes no difference); here we summarize the main message.

A special role is played by the element $B_- C_+$ in the left upper corner. This operator is a highest-weight element for the adjoint representation of $\mathfrak{gl}(2|2)$ (with Cartan subalgebra generated by the diagonal operators $B_- B_+\,, \ldots, G_- G_+$), but it is also highest-weight for the fundamental action from the left and the anti-fundamental action from the right. Moreover, it has the distinctive property of being a positive operator: $B_- C_+ = (C_+)^\dagger C_+ > 0$. Building on these properties, we argued in \cite{BWZ2} that a coarse-grained form of $(B_- C_+)^q$ has the infrared behavior of the vertex operator $\mathrm{e}^{q \varphi}$ for a Gaussian free field $\varphi$ with background charge. This proposal had been tested numerically \cite{BWZ1} by recognizing $\langle (B_- C_+)^q \rangle_\mathcal{V}$ as the moment $\mathbb{E}(\vert \psi \vert^{2q})$ of a critical (hence multifractal) stationary wavefunction $\psi = U \psi$.

We now upgrade that argument to the stronger proposal that $B_- C_+$ corresponds (after coarse graining and non-Abelian bosonization) to the boson-boson matrix element of a Wess-Zumino-Witten field $M$. Our first remark here is to recall that, for present purposes, an acceptable WZW target space must be a cs-supermanifold [inside the complex Lie supergroup $\mathrm{GL}(r|r)$] based on a Riemannian symmetric space $\mathrm{Herm}^+ (r) \times \mathrm{U}(r)$. This requirement is satisfied by the proposal (\ref{eq:M-WZW}) as $B_- C_+$ is positive Hermitian and the fermion-fermion pair $F_- G_+\,$, made from one left-mover ($F_-$) and one right-mover ($G_+$), bosonizes to a unitary by standard lore. For a second remark, note that our proposal means that $M$ will be a primary field not just for the Virasoro algebra but even for the underlying Kac-Moody algebra of currents; that will make for a rapid derivation of the multifractality spectrum in Sect.\ \ref{sect:multif}. Thirdly, note the dependence of $M$ on the unitary constants $\mathrm{e}^{ \mathrm{i} \vartheta_0}$ and $\mathrm{e}^{ \mathrm{i} \vartheta_1}$; in Sect.\ \ref{sect:SBS} these constants will be recognized as order parameters for a spontaneously broken symmetry.

The operator product expansion between the holomorphic current $J$ and the WZW fundamental field $M$ is
\begin{equation}\label{eq:OPE-JM}
    J^{\alpha}_{\;\; \beta}(z) \, M^{\gamma}_{\;\; \delta}(w,\bar{w}) \sim (-1)^{|\beta|+1} \frac{\delta_{\gamma}^{\beta}}{z-w} \, M^{\alpha}_{\;\; \delta}(w,\bar{w}) \,,
\end{equation}
and an analogous formula holds for the OPE of $\bar{J}$ with $M$:
\begin{equation}\label{eq:OPE-bJM}
    \bar{J}^{\alpha}_{\;\; \beta}(\bar{z}) \, M^{\gamma}_{\;\; \delta} (w,\bar{w}) \sim M^{\gamma}_{\;\; \beta} (w,\bar{w}) \frac{
    \delta_{\alpha}^{\delta}}{\bar{z} - \bar{w}} \, (-1)^{(|\alpha| + |\beta|) |\gamma| + |\alpha||\beta|} .
\end{equation}
As before, this is verified on the lattice by first computing the OPE in the free theory with the Wick contraction rules (\ref{eq:basic-Wick}) and then taking the disorder average -- with the \emph{proviso} that insertions of additional operators may be (and in fact must be) present. Let us motivate why the holomorphic current acts on the left while the anti-holomorphic current acts on the right. The basic Wick contractions in the transformed basis (\ref{eq:new-bos}, \ref{eq:new-ferm}) are
\begin{align}
    &\big\langle B_- (\ell_1) B_+ (\ell_2) \big\rangle_\mathcal{F} =
    \big\langle F_- (\ell_1) F_+ (\ell_2) \big\rangle_\mathcal{F}
    = \big\langle C_- (\ell_1) C_+ (\ell_2) \big\rangle_\mathcal{F} \cr
    =\; &\big\langle G_- (\ell_1) G_+ (\ell_2) \big\rangle_\mathcal{F}
    = \frac{1}{2} \big\langle \ell_1 \mid (1-T)^{-1} - T^\dagger (1-T^\dagger)^{-1} \mid \ell_2 \big\rangle \,,
\end{align}
reflecting the $\mathrm{U}(1,1|2)$ invariance of $\langle \, \cdots \rangle_\mathcal{F}\,$. All others go to zero by restoration of the global $\mathrm{U}(1,1|2)$ symmetry in the limit of vanishing regularization. For example, using $T = \mathrm{e}^\mu U$ and $T^\dagger = U^{-1} \mathrm{e}^\mu$ one has
\begin{align*}
    &\big\langle B_+ (\ell_1) C_- (\ell_2) \big\rangle_\mathcal{F}
    = \frac{1}{2} \big\langle \ell_2 \mid T (1-T)^{-1} + (1-T^\dagger)^{-1} \mid \ell_1 \big\rangle \cr
    &= \frac{1}{2} \big\langle \ell_2 \mid U (\mathrm{e}^{-\mu} - U)^{-1} - U (\mathrm{e}^\mu - U)^{-1} \mid \ell_1 \big\rangle \stackrel{\mu\to 0}{\longrightarrow} 0 \,.
\end{align*}
In order for the last inference (sending $\mu \to 0$) to be a rigorous statement, the role of regulator must be taken over by an operator insertion (say, that of a point contact) in the correlation function $\langle \cdots \rangle_\mathcal{F}$ to cut off the geometric sum $(1 - U)^{-1} = \sum_n U^n$. The left-right structure of the OPEs (\ref{eq:OPE-JM}) and (\ref{eq:OPE-bJM}) then follows because the holomorphic current $J$ is constructed from operators $B_\pm \,, F_\pm \,$, the anti-holomorphic current $\bar{J}$ from $C_\pm \,, G_\pm \,$, while $M$ is made of one from $B_- \,, F_-$ (left factor) and one from $C_+ \,, G_+$ (right factor).

\subsection{Deformation of WZW model}\label{sect:deform}

We now ask what is the Lagrangian of the conformal field theory defined by the operator product expansions (\ref{eq:OPE-d}, \ref{eq:CK-d}, \ref{eq:OPE-JM}, \ref{eq:OPE-bJM})? The answer to that question begins with the reminder that the currents of the standard $\mathrm{GL}(r|r)_n$ WZW model (without deformation) satisfy (\ref{eq:OPE-d}) with $\gamma = 0$. To introduce the deformation parameter $\gamma$, we add to the WZW action functional (\ref{eq:S-WZW1}) a current-current interaction for the central generator $E$:
\begin{equation}\label{eq:def-WZW}
    S_{n,\gamma}^{\rm WZW}[M] = S_{n}^{\rm WZW}[M] - \frac{\mathrm{i}\, \gamma}{4\pi} \int_\Sigma \mathrm{STr}(M^{-1} \partial M ) \wedge \mathrm{STr}(M^{-1} \bar\partial M) .
\end{equation}
The equations of motion derived from this action functional are $\bar\partial J = \partial \bar{J} = 0$ where $J$ is the $\gamma$-deformed holomorphic current
\begin{equation}\label{eq:def-JX}
    J^Y = n \, \mathrm{STr}(Y \partial M\! \cdot \! M^{-1}) - \gamma \, \mathrm{STr}(Y)\, \mathrm{STr}(\partial M\! \cdot \! M^{-1}) ,
\end{equation}
and the same formula (with $\partial M\! \cdot \! M^{-1}$ replaced by $M^{-1} \bar\partial M$) gives $\bar{J}^Y$.

On general grounds \cite{KZ} and as a direct consequence of the OPE (\ref{eq:OPE-JM}), the integrated current $(2\pi\mathrm{i})^{-1} \oint J^X$ for a supermatrix-valued holomorphic function $X \equiv X(z)$ is the generator of left translations:
\begin{equation}
    \delta_X M^{\alpha}_{\;\; \beta} (z,\bar{z}) = - X^{\alpha}_{\;\; \nu}(z) M^{\nu}_{\;\; \beta} (z,\bar{z}) .
\end{equation}
From the expression (\ref{eq:def-JX}) one sees immediately that the deformed current responds to infinitesimal left translations as
\begin{equation}\label{eq:inf-leftT}
    \delta_X J^Y = - n \, \mathrm{STr}(Y \partial X) + \gamma \, \, \mathrm{STr}(Y) \, \mathrm{STr}(\partial X) + J^{[X,Y]} .
\end{equation}
It then follows that the holomorphic current $J$ obeys the operator product expansion (\ref{eq:OPE-d}, \ref{eq:CK-d}) with deformation parameter $\gamma$. Thus the action functional (\ref{eq:def-WZW}) does the required job of giving the path-integral representation of our deformed conformal field theory.

We turn to the energy-momentum tensor, $T(z)$. We recall that $T(z)$ is determined by the requirement (\ref{eq:OPE-TJ}) that the holomorphic current $J$ be a Virasoro-primary field of conformal weights $(1,0)$. By SUSY cancelation due to the equal number of bosonic and fermionic fields, the number of replicas $r$ versus $2r$ does not change the result for $T(z)$. Thus the expression (\ref{eq:T-gamma}) remains valid in the present case of a $\widehat{\mathfrak{gl}}(r|r)_{n,\gamma}$ current algebra.

We already know from Sect.\ \ref{sect:tell} that the deformation parameter $\gamma$ must be set to unity in order for $M$ to have conformal weights $(0,0)$ as required by the phenomenology of Anderson transitions in class $A$. This leaves the current algebra level $n$ as the only open parameter. In the next section we will demonstrate that the quantized level $n \in \mathbb{N}$ must be $n = 4$ to match the multifractality spectrum known from numerical simulations.

\subsection{Multifractal scaling exponents}\label{sect:multif}

The expression (\ref{eq:T-gamma}) for the energy-momentum tensor $T(z)$ simplifies to
\begin{equation}\label{eq:T(z)}
    T(z) = - \frac{(-1)^{ |\alpha|}}{2n} \, \big( J^{\alpha}_{\;\; \beta}(z) J^{\beta}_{\;\; \alpha}(z) \big)
\end{equation}
at deformation parameter $\gamma = 1$. Above, we argued that $M$ is a primary field for the Kac-Moody (or affine) Lie superalgebra $\widehat{\mathfrak{gl}}(r|r)_{n, \gamma}\,$, and is so, in particular, for $\gamma = 1$. We expect this property to carry over to a continuum of infinite-dimensional $\mathfrak{gl}(r|r)$-irreducible representations. Exactly what those are in our non-compact current algebra setting is a non-trivial question, at least from the mathematics perspective of serious analysis. Here we will content ourselves with a simple physics-style argument. We set $r = 1$ and look at any power $q \in \mathbb{N}$ of the boson-boson field component $M^{0}_{\;\; 0}\,$. (Standard heuristics indicates that the ``good'' powers are $q = 1/2 + \mathrm{i} \lambda$ with $\lambda \in \mathbb{R}$.) Building on the assumption that $(M^{0}_{\;\; 0})^q$ is an affine primary field, we obtain \cite{KZ} the OPE
\begin{equation}
    T(z) (M^{0}_{\;\; 0})^q (w,\bar{w}) \sim \frac{h_q}{(z-w)^2}
    \, (M^{0}_{\;\; 0})^q (w,\bar{w}) + ... \,, \quad h_q = \frac{\mathrm{Cas}_2(q)}{2n} \,,
\end{equation}
where $\mathrm{Cas}_2(q)$ is the quadratic $\mathfrak{gl}(1|1)$ Casimir element evaluated in the representation of $(M^{0}_{\;\; 0})^q$. To compute the latter, we use the formula
\begin{align}
    \mathrm{Cas}_2 &= - (-1)^{|\beta|} E_{\alpha}^{\;\; \beta} E_{ \beta}^{\;\; \alpha} = - (E_{0}^{\;\; 0})^2 + (E_{1}^{\;\; 1})^2
    - E_{1}^{\;\; 0} E_{0}^{\;\; 1} + E_{0}^{\;\; 1} E_{1}^{\;\; 0} \cr &= - (E_{0}^{\;\; 0})^2 + (E_{1}^{\;\; 1})^2 - 2 E_{1}^{\;\; 0} E_{ 0}^{\;\; 1} + E_{0}^{\;\; 0} + E_{1}^{\;\; 1} ,
\end{align}
with $E_{\alpha}^{\;\; \beta} = e_{\alpha} \otimes e^{\beta}$ the standard generators of $\mathfrak{gl}(1|1)$. Our primary field $(M^{0}_{\;\; 0})^q$ transforms as the $q^{\rm th}$ symmetric power of the fundamental vector $e_0\,$, which obeys the relations
\begin{equation}
    E_{1}^{\;\; 1} e_0^q = E_{0}^{\;\; 1} e_0^q = 0 \,, \quad
    E_{0}^{\;\; 0} e_0^q = q \, e_0^q \,, \quad
    (E_{0}^{\;\; 0})^2 e_0^q = q^2 e_0^q \,.
\end{equation}
Using these we obtain the Casimir eigenvalue $\mathrm{Cas}_2(q) = - q^2 + q$ and hence \begin{equation}
    h_q = - \frac{q(q-1)}{2n} \,.
\end{equation}
The total scaling dimension of $(M^{0}_{\;\; 0})^q$ then is
\begin{equation}\label{eq:Delta-q}
    \Delta_q = h_q + \bar{h}_q = \frac{q(1-q)}{n} \,,
\end{equation}
and we expect this to continue analytically to values of $q$ beyond the discrete set $\mathbb{N}$. In our previous work \cite{BWZ2} we identified $(M^{0}_{\;\; 0})^q$ as the SUSY vertex model operator for the disorder average $\mathbb{E} \big( \vert \psi(\ell) \vert^{2q} \big)$ of a multifractal wavefunction $\ell \mapsto \psi(\ell)$ of the network model at criticality. Thus $\Delta_q$ is what is known as the spectrum of multifractal scaling exponents. (Here it should be noted that, owing to our definition of critical wavefunctions $\psi_{\rm c}$ using point contacts \cite{BWZ1, BWZ2}, the multifractality spectrum for such properly defined $\psi_{\rm c}$ does not suffer from the phenomenon of termination or freezing.)

The conclusion (cf.\ \cite{BWZ2}) from recent numerics \cite{BWZ1} was that a good fit of the finite-size numerical data can be had with $\Delta_q = X q(1-q)$ for $X \equiv 1/n$ close to $0.26$. We now predict with confidence that careful finite-size scaling for large systems will converge to the result (\ref{eq:Delta-q}) with $n = 4$.

\subsection{Gaussian free fields}\label{sect:GFF}

In the following, the operator-theoretic result of the previous section will be reproduced by doing a functional integral computation. Our goal here is to establish a direct connection with earlier work \cite{BWZ2}, where we argued that the spectrum of multifractal scaling exponents can be understood from the CFT of a Gaussian free field with background charge.

Continuing to focus on the simple case of replica number $r = 1$, we make a Gauss decomposition:
\begin{equation}\label{eq:gauss}
    M = \begin{pmatrix} M^{0}_{\;\; 0} &M^{0}_{\;\; 1} \cr M^{1}_{\;\; 0} &M^{1}_{\;\; 1} \end{pmatrix} = \begin{pmatrix} 1 &0 \cr \eta &1 \end{pmatrix} \begin{pmatrix} \mathrm{e}^\varphi &0 \cr 0 &\mathrm{e}^{\mathrm{i}\theta} \end{pmatrix}
    \begin{pmatrix} 1 &\xi \cr 0 &1 \end{pmatrix} .
\end{equation}
After this change of variables the action functional (\ref{eq:def-WZW}) reads
\begin{align}
    S_{n,\gamma}^{\rm WZW} &= \frac{\mathrm{i}\,n}{4\pi} \int (\partial\varphi \wedge \bar\partial\varphi + \partial\theta \wedge \bar\partial\theta + 2\, \mathrm{e}^{\varphi - \mathrm{i}\theta} \partial\xi \wedge \bar\partial \eta) \cr &- \frac{\mathrm{i}\,\gamma}{4\pi} \int (\partial\varphi - \mathrm{i}\,\partial\theta) \wedge (\bar\partial \varphi - \mathrm{i}\,\bar\partial \theta) . \label{eq:4.123}
\end{align}
Since the action in the new variables is quadratic in the Grassmann fields $\xi$ and $\eta$, the latter can be eliminated by performing a Gaussian integral. In view of the expression (\ref{eq:4.123}) the result of that Gaussian integration might seem to be the determinant of the differential operator $- \partial \,  \mathrm{e}^{\varphi - \mathrm{i} \theta} \bar\partial\,$. However, the change of variables (\ref{eq:gauss}) also comes with a Jacobian, which heuristically is $\prod_{\bf r} \mathrm{e}^{-\varphi + \mathrm{i}\theta} ({\bf r})$. With this Jacobian taken into account, the Gaussian integral over $\xi, \eta$ results in the functional determinant
\begin{equation}
    Z_{\xi\eta}[\difang] = \mathrm{Det} \big( - \mathrm{e}^{-\difang /2} \, \partial_z \, \mathrm{e}^{\difang} \, \partial_{\bar z} \, \mathrm{e}^{ -\difang /2} \big) , \quad \difang \equiv \varphi - \mathrm{i}\,\theta .
\end{equation}
Its calculation is a standard field-theory problem. For a Riemann surface $\Sigma$ with scalar curvature $2$-form $\mathcal{R}$ the result (known as the Polyakov formula \cite{Polyakov, Alvarez}) is
\begin{equation}\label{eq:4.125}
    - \ln \frac{Z_{\xi\eta}[\difang]} {Z_{\xi\eta}[0]} = \frac{\mathrm{i}} {4\pi} \int_\Sigma \partial\difang \wedge \bar\partial\difang - \frac{Q}{8\pi} \int_\Sigma \difang \mathcal{R} \,,
\end{equation}
where the parameter $Q = - 1$ is called the background charge. Note that by the Gauss-Bonnet theorem one has $\int_\Sigma \mathcal{R} = 4 \pi \chi(\Sigma)$ with $\chi$ the Euler characteristic. In what follows we take $\Sigma$ to be the Euclidean plane compactified to a sphere, so that $\chi(\Sigma) = 2$ and $\int_\Sigma \mathcal{R} = 8\pi$.

For a check on our expressions, let us specialize to the case of $n = 1$ and $\gamma = 0$. According to the relation (\ref{eq:3.62}), we then expect to get the correlation functions of the supersymmetric theory (\ref{eq:T-free}) of a free Dirac field with bosonic ghost system describing the continuum limit of the Chalker-Coddington network model without disorder ($U \equiv U_{\rm s}$). In particular, the field $(M^{0}_{\;\; 0})^q = \mathrm{e}^{q \varphi}$ should have scaling dimension $q$. This can be verified as follows. The effective action without the background charge term is
\begin{equation*}
    S_{n=1,\gamma=0}^{\rm eff} = \frac{\mathrm{i}}{4\pi} \int \big( 2\, \partial\varphi \wedge \bar\partial\varphi - \mathrm{i}\, \partial\varphi \wedge \bar\partial\theta - \mathrm{i}\, \partial\theta \wedge \bar\partial\varphi \big) .
\end{equation*}
By inverting the matrix of the quadratic form for $S^{\rm eff}$,
\begin{equation*}
    \begin{pmatrix} 2 &-\mathrm{i} \\ -\mathrm{i} &0 \end{pmatrix}^{-1} = \begin{pmatrix} 0 &\mathrm{i} \\ \mathrm{i} &2 \end{pmatrix} ,
\end{equation*}
we get the two-point functions $\langle \varphi({\bf r}) \varphi({\bf r}^\prime) \rangle = 0$, $\langle \theta({\bf r}) \theta({\bf r}^\prime) \rangle = - 4 \ln |{\bf r} -{\bf r}^\prime |$, and
\begin{equation*}
    \langle \varphi({\bf r}) \theta({\bf r}^\prime) \rangle = - 2\mathrm{i}\, \ln |{\bf r} - {\bf r}^\prime | .
\end{equation*}

To make sense of the present calculation, we must regularize the field integral. Recall from Sect.\ \ref{sect:SD-adapted} that our preferred scheme of regularization in the network model is to create a stationary state $\psi_{\rm c} = U_{\rm s} \psi_{\rm c}$ satisfying incoming-wave boundary conditions at a point contact ${\rm c}\,$, say at position ${\bf o}$. In our field-theory representation, this regularization corresponds to inserting a point-contact operator $\pi_{\rm c}({\bf o})$ into the field integral:
\begin{equation*}
    \mathbb{E} \left( \vert \psi_{\rm c}({\bf r}) \vert^{2q} \right) = \big\langle \mathrm{e}^{q \varphi({\bf r})} \pi_{\rm c}({\bf o}) \big\rangle ,
\end{equation*}
where $\mathbb{E}(...)$ just means coarse graining for the continuum limit. By the modification of charge neutrality due to the anomalous term in Eq.\ (\ref{eq:4.125}), the point-contact operator must contribute the charges $(- Q - q\,, +\mathrm{i}\,Q)$ for the pair of fields $(\varphi, \theta)$ in order to neutralize the vertex operator $\mathrm{e}^{q \varphi}$. With the value $Q = - 1$ for the background charge, we thus have $\pi_c \sim \mathrm{e}^{(1-q)\varphi - \mathrm{i}\theta} = \mathrm{e}^{-q\varphi} \mathrm{e}^{\difang}$. Hence,
\begin{equation*}
    \mathbb{E} \left( \vert \psi_{\rm c}({\bf r}) \vert^{2q} \right) \sim \big\langle \mathrm{e}^{q \varphi({\bf r})} \mathrm{e}^{- q \varphi + \difang} ({\bf o}) \big\rangle = \mathrm{e}^{- \mathrm{i} q \langle \varphi({\bf r}) \theta ({\bf o}) \rangle} = \vert {\bf r} - {\bf o} \vert^{-2q} .
\end{equation*}
By comparing with $\mathbb{E} \left( \vert \psi_{\rm c}({\bf r}) \vert^{2q} \right) \sim \vert {\bf r} - {\bf o} \vert^{-2 \Delta_q^{\rm free}}$ we infer the desired result $\Delta_q^{\rm free} = q\,$, thus validating (for $r = 1$) our claim that the free Dirac/ghost system (\ref{eq:T-free}) bosonizes to the $\mathrm{GL}(r|r)_{n=1}$ WZW model.

We return to our subject proper, namely the multifractality spectrum of the Chalker-Coddington network model with $\mathrm{U}(1)$-invariant random phase disorder ($U = U_{\rm r} U_{\rm s}$). For that case we have learned to set $\gamma = 1$, which has the consequence that the first term on the right-hand side of (\ref{eq:4.125}) cancels the second line of (\ref{eq:4.123}). Thus the effective action simplifies to
\begin{equation}\label{eq:GFF}
    S_{n,\gamma=1}^{\rm eff} = \frac{\mathrm{i}\,n}{4\pi} \int_\Sigma \big( \partial\varphi \wedge \bar\partial\varphi + \partial\theta \wedge \bar\partial\theta \big) - \frac{Q}{8\pi} \int_\Sigma (\varphi - \mathrm{i}\,\theta) \mathcal{R} \,.
\end{equation}
What we see here is a decoupled system of two Gaussian free fields, one of compact and another one of non-compact type, each with background charge $Q = -1$. The non-compact sector ($\varphi$) coincides exactly (but for a change of sign convention for the background charge, and with the identification $1/n \equiv X$) with Eq.\ (223) of \cite{BWZ2}. By that coincidence we may use the results of \cite{BWZ2} to reproduce the parabolic spectrum (\ref{eq:Delta-q}).

Let us finish here with the remark that it is informative also to consider the field powers offered by the compact sector. From Sect.\ 2.3 of \cite{BWZ2} and the discussion of Sect.\ \ref{sect:bosonize} transcribed from $M^{0}_{\;\; 0} \sim B_- C_+$ to $M^{1}_{\;\; 1} \sim F_- G_+\,$, we expect that $(M^{1}_{\;\; 1})^p$ (for $p \in \mathbb{N}$) represents the totally skew Schur polynomial $e_p(K)$ for a certain matrix $K$ (see \cite{BWZ2}) constructed from at least $p$ critical wavefunctions of the network model. That field  $(M^{1}_{\;\; 1})^p$ is predicted to have scaling dimension $+p(p-1)/n$, which is sign-opposite to the scaling dimension $-q(q-1)/n$ of $(M^{0}_{\;\; 0})^q$. This prediction can be verified by a variant of the calculation done in Sect.\ \ref{sect:multif} or, more directly, by adapting the functional integral computation of the present section. For the latter purpose, it is convenient to reorder the Gauss decomposition (\ref{eq:gauss}) as
\begin{equation*}
    \begin{pmatrix} M^{0}_{\;\; 0} &M^{0}_{\;\; 1} \cr M^{1}_{\;\; 0} &M^{1}_{\;\; 1} \end{pmatrix} = \begin{pmatrix} 1 &\xi \cr 0 &1 \end{pmatrix} \begin{pmatrix} \mathrm{e}^\varphi &0 \cr 0 &\mathrm{e}^{\mathrm{i}\theta} \end{pmatrix}
    \begin{pmatrix} 1 &0 \cr \eta &1 \end{pmatrix} .
\end{equation*}
so that $M^{1}_{\;\; 1} = \mathrm{e}^{\mathrm{i}\theta}$ (without any additional terms involving the Grassmann fields $\xi$ and $\eta$). The calculation then is the same as before, except that the exponential weight factor $\mathrm{e}^{\varphi - \mathrm{i}\theta} \equiv \mathrm{e}^{\difang}$ in the WZW-action (\ref{eq:4.123}) gets replaced by its reciprocal, $\mathrm{e}^{-\difang}$. The net effect is to change the sign of the background charge ($Q \to -Q$) in Eq.\ (\ref{eq:4.125}). Thus the point-contact operator must now contribute $\pi_{\rm c} \sim \mathrm{e}^{- \mathrm{i} p \theta} \mathrm{e}^{-\difang}$ to neutralize the vertex operator $(M^{1}_{\;\; 1})^p = \mathrm{e}^{\mathrm{i}p\theta}$ in the presence of the background charge. Hence
\begin{equation*}
    \big\langle \mathrm{e}^{\mathrm{i} p \theta({\bf r})} \pi_{\rm c}({\bf o}) \big\rangle \sim \big\langle \mathrm{e}^{\mathrm{i} p \theta({\bf r})} \mathrm{e}^{-\mathrm{i} p \theta - \difang} ({\bf o}) \big\rangle = \mathrm{e}^{p(p-1) \langle \theta({\bf r}) \theta ({\bf o}) \rangle} = \vert {\bf r} - {\bf o} \vert^{-2 \Delta_p^\prime} ,
\end{equation*}
with $\Delta_p^\prime = p(p-1)/n\,$. This scaling exponent is another CFT prediction that should be verifiable by numerical simulations.

\subsection{Spontaneously broken symmetry}\label{sect:SBS}

At the present stage, two nagging questions remain open: (i) We are proposing the conformal field theory (\ref{eq:def-WZW}) as the continuum limit of the SUSY vertex model at criticality. Now the latter has the global Lie supergroup symmetry $U = \mathrm{U} (r,r|2r)$, but $U$ does not seem to act on our CFT. So, what happened to it? (ii) To reproduce the known phenomenology, we had to set the deformation parameter to $\gamma = 1$, which might appear to require fine tuning of the parameters of the microscopic theory. Hence we ask: is there a constructive (as opposed to phenomenological) reason for $\gamma = 1$?

As will be explained in this section, the two questions have the same answer: spontaneous symmetry breaking! To be specific, we will demonstrate that the deformation in Eq.\ (\ref{eq:def-WZW}) (with $\gamma = 1$) is the result of integrating out the stiff Goldstone modes due to a spontaneously broken symmetry.

To begin, let us note that the famous Mermin-Wagner-Coleman theorem holds under the assumption that the symmetry group is compact; it does not forbid the spontaneous breaking of \emph{non-compact} symmetries in dimension two or below. In fact, it is known \cite{Seiler} that bosonic non-linear sigma models with a non-compact target space inevitably suffer from spontaneous symmetry breaking in all space dimensions greater than zero. From the perspective of this fact, what needs explanation in the field theory of Anderson localization is not the occurrence of spontaneous symmetry breaking but rather the absence thereof. (Of course, the feature making the latter possible is the presence of fermionic degrees of freedom.) Anyway, there exists a consensus among the mesoscopic physics community that the symmetry $U$ of the non-linear sigma model of Anderson (de-)localization is spontaneously broken in the metallic phase of delocalized states in three space dimensions. The symmetry breaking happens because the stiffness (physically speaking, the scale-dependent conductivity) of the non-linear field flows to infinity under renormalization. Here we will argue that for the IQHE plateau transition, at its critical point with diverging localization length, a similar scenario takes place: the symmetry $U$ is partially broken because the field stiffness flows to infinity for some of the field degrees of freedom (the ``angular'' ones); the stiffness of the remaining field degrees of freedom (the ``radial'' ones) flows to a finite value -- as expected for a non-trivial RG-fixed point.

\subsubsection{Simple heuristic}\label{sect:4.13.1}

Given that the SUSY vertex model has the symmetry group $U = \mathrm{U} (r, r|2r)$, one expects the target space of the effective field theory to be a $U$-orbit (or, perhaps, a family of such orbits). In the non-linear sigma model (\ref{eq:NLsM}), the target is the $U$-orbit of the invertible element $\Sigma_3$ [or $\mathrm{i} \Sigma_3$ for an adjoint orbit in $\mathrm{Lie}(U)$] given in Eq.\ (\ref{eq:NLsM-Q}). For weak coupling, the non-linear sigma model is known to be renormalizable in perturbation theory; what happens under renormalization is that the ``radius'' $t(a)$ of the $U$-orbit shrinks: $\Sigma_3 \to t(a) \Sigma_3$ and $t(a)$ decreases with increasing short-distance cutoff $a$.

At the critical point of the plateau transition, we are facing the uncharted territory of a strongly coupled field theory where $t(a)$ is so small that perturbation theory ceases to apply. We will now break with the traditional wisdom to suggest that the renormalization group flow at strong coupling drives the target space toward a $U$-orbit consisting of \emph{nilpotent} (hence non-invertible) elements. In order to motivate our treatment in Sect. \ref{sect:4.13.2} of the full theory for an arbitrary number $r$ of bosonic and fermionic replicas, we first convey the main idea of the proposal at the example of the boson-boson sector for $r = 1$, where the symmetry group simplifies to the non-compact group $\mathrm{U} (1,1)$.

Consider then for any $z \in \mathbb{C} \setminus \{ 0 \}$ the $2 \times 2$ matrix
\begin{equation}\label{eq:nil-Q}
    Q = \frac{\mathrm{i}}{2} \begin{pmatrix} |z| &- \bar{z} \cr z &- |z| \end{pmatrix} = - \sigma_3\, Q^\dagger \sigma_3 \in \mathrm{Lie}\, \mathrm{U}(1,1) .
\end{equation}
Such matrices $Q$ are nilpotent: $Q^2 = 0$, and they constitute the $U$-orbit of the nilpotent element $Q_0 = (\mathrm{i} \sigma_3 + \sigma_2)/2$ by $Q = u Q_0 u^{-1}$ for $u \in \mathrm{U}(1,1) \equiv U$. One may view that $U$-orbit as a homogeneous space $U / N$ where the nilpotent one-parameter group $N = \mathrm{exp}( \mathbb{R} Q_0) \subset U$ is the isotropy group of $Q_0 \in \mathrm{Lie} \,U$. The $U$-invariant metric tensor on $U/N$ is expressed by
\begin{equation}\label{eq:deg-met}
    2 \mathrm{Tr}\, d Q^2 = |z|^2 (d\, \mathrm{arg} z)^2 , \quad \mathrm{arg} z = (2\mathrm{i})^{-1} \ln (z / \bar{z}) .
\end{equation}
We observe that this metric is degenerate: it vanishes in the direction of the radial variable $|z|$.

In what follows below, $|z|$ (or rather a supermatrix extension thereof) will be identified with the WZW field $M$. Of course, the metric tensor for that field does not vanish but flows to a finite limit at the RG-fixed point. Therefore, to make a sensible identification one needs to multiply the metric (\ref{eq:deg-met}) by a diverging factor to render the vanishing stiffness for $|z|$ finite. By that rescaling, the finite stiffness of the angular variable $\mathrm{arg} z$ becomes infinite, which provides the mechanism for spontaneous symmetry breaking.

To add some quantitative detail to this scenario, let $Q$ be modified as
\begin{equation}\label{eq:Q-eps}
    \widetilde{Q} = \frac{\mathrm{i}}{2} \begin{pmatrix} \sqrt{\varepsilon + |z|^2} &- \bar{z}\cr z &- \sqrt{\varepsilon + |z|^2} \end{pmatrix}
\end{equation}
for $\varepsilon > 0$ and scale the $U$-invariant metric tensor by $\varepsilon^{-1}$:
\begin{equation}
    2 \varepsilon^{-1} \mathrm{Tr}\, d \widetilde{Q}^2 = \varepsilon^{-1} |z|^2 (d\, \mathrm{arg} z)^2 + \frac{(d\, |z|)^2}{\varepsilon + |z|^2} \,.
\end{equation}
We see that for $\varepsilon \to 0$ the radial part $(\varepsilon + |z|^2)^{-1} (d\, |z|)^2$ of the metric converges to a finite limit, whereas the angular part $\varepsilon^{-1} |z|^2 (d\, \mathrm{arg} z)^2$ diverges. Hence our more detailed idea is to approach the nilpotent orbit $U/N$ by sending $\varepsilon \to 0$ in (\ref{eq:Q-eps}) while taking the stiffness parameter to infinity as $\varepsilon^{-1}$. Note the crucial feature that $U$-invariance is maintained in the process.

The divergent stiffness tends to lock the angular field $\mathrm{arg} z ({\bf r})$ into a constant value, say $\vartheta_0\,$, independent of the position ${\bf r}$. Thus $\vartheta_0 \equiv \mathrm{arg} z = {\rm const}$ plays the role of an order parameter for the spontaneously broken symmetry. What determines its numerical value? Now while the order parameter in the standard scenarios of spontaneous symmetry breaking takes a value that is determined by an (infinitesimal) external field or a boundary condition, the present situation is more intricate and needs explanation, as follows.

As we have seen, all our quantities of physical interest can be expressed in terms of the WZW field $M$ and the currents $\partial M \cdot M^{-1}$ and $M^{-1} \bar\partial M$. Recall also that the construction of $M \sim B_- C_+$ in Sect.\ \ref{sect:4.6} required a choice of unitary $\mathrm{e}^{\mathrm{i}\vartheta_0}$ for the definition (\ref{eq:new-bos}) of the transformed bosons $B_\pm$ and $C_\pm\,$. No physical meaning was attached to $\mathrm{e}^{\mathrm{i}\vartheta_0}$ at that stage, and there existed no preferred value for it. Yet, some choice had to be made.

{}From the vantage point of the current section, we can make the revealing observation that the factor $\mathrm{e}^{\mathrm{i}\vartheta_0}$ in (\ref{eq:new-bos}) is the same as our phase-locked factor $z / |z| = \mathrm{e}^{\mathrm{i} \vartheta_0}$. Indeed, by the Lie algebra isomorphism (\ref{eq:Lalg-iso}) from $\mathrm{Lie}\, \mathrm{U} (1,1)$ to its realization in terms of the fundamental bosons $b_\pm$ and $b_\pm^\dagger\,$, we have
\begin{equation*}
    \begin{pmatrix} 1 &-\mathrm{e}^{ -\mathrm{i}\vartheta_0}\cr \mathrm{e}^{ \mathrm{i} \vartheta_0} &-1 \end{pmatrix} \mapsto
    \mathrm{Tr} \begin{pmatrix} 1 &-\mathrm{e}^{ -\mathrm{i} \vartheta_0}\cr \mathrm{e}^{ \mathrm{i} \vartheta_0} &-1 \end{pmatrix} \begin{pmatrix} b_+ b_+^\dagger &b_+ b_-\cr - b_-^\dagger b_+^\dagger &- b_-^\dagger b_- \end{pmatrix} = 2 B_- C_+\, ,
\end{equation*}
and the nilpotent matrix on the left-hand side matches the nilpotent matrix in (\ref{eq:nil-Q}) for $z = |z|\, \mathrm{e}^{\mathrm{i} \vartheta_0}$. Thus the free parameter $\mathrm{e}^{\mathrm{i} \vartheta_0}$, whose existence had no good explanation in Sect.\ \ref{sect:4.6}, acquires a palpable meaning here: in order to express the physical observables (conductivities, wave-function moments and correlators, etc.) of the network model in the SUSY vertex model and ultimately in the effective continuum field theory, we must choose some arbitrary but fixed value of $\mathrm{e}^{\mathrm{i} \vartheta_0}$ in (\ref{eq:new-bos}). That choice determines the operator insertions in the field integral computing the physical observables, and it thus selects a phase ($\mathrm{arg} z = \vartheta_0$) for the field (\ref{eq:nil-Q}) on the nilpotent $U$-orbit. By the diverging stiffness of the phase field, the chosen value of $\mathrm{e}^{ \mathrm{i} \vartheta_0}$ gets promoted to the symmetry-breaking order parameter.

\subsubsection{Full implementation}\label{sect:4.13.2}

Next, we implement the heuristic ideas of Sect.\ \ref{sect:4.13.1} in the full theory with symmetry $U = \mathrm{U}(r,r|2r)$. For this, we observe that the Lie supergroup $U$ is equipped with a so-called Cartan involution $\theta : \; U \to U$,
\begin{equation}
    \theta(u) = \alpha u \alpha^{-1} , \quad \alpha^2 = - \mathbf{1} ,
\end{equation}
where $\alpha = \mathrm{i} \Sigma_3\,$, using the notation of Sect.\ \ref{sect:1.1}. The subgroup of $\theta$-fixed points in $U$ is the isotropy group $\mathrm{Fix}_U(\theta) \equiv K \subset U$ of the target space $U/K$ of the non-linear sigma model. Now, the order parameter of our symmetry-breaking scenario determines a second involutory automorphism:
\begin{equation}
    \eta(u) = \beta u \beta^{-1} , \quad \beta^2 = + \mathbf{1} , \quad \alpha\beta + \beta\alpha = 0 .
\end{equation}
In the simplified setting of Sect.\ \ref{sect:4.13.1} the expressions are $\alpha = \mathrm{i}\sigma_3$ and $\beta = \sigma_2 \cos \vartheta_0 - \sigma_1 \sin\vartheta_0\,$. Because $\alpha$ and $\beta$ anti-commute, the two involutions $\theta$ and $\eta$ commute. We remark that the automorphism $\theta$ is inner ($\alpha \in U$), whereas $\eta$ is outer ($\beta \notin U$). Note also that $(\alpha + \beta)^2 = 0$.

Before regularization (by $\varepsilon > 0$ --- see below) our supermatrix field $Q$ lives on the $U$-orbit of the nilpotent element $Q_0 = (\alpha + \beta)/2 :$
\begin{equation}
    Q = u Q_0 u^{-1} , \quad Q^2 = 0 .
\end{equation}
As before, the $U$-invariant metric tensor $\mathrm{STr}\, dQ^2$ on the nilpotent $U$-orbit is degenerate. To describe its null directions, we notice that the linear transformation $\alpha\beta$ has square $(\alpha \beta)^2 = +1$ and therefore two eigenvalues, $+1$ and $-1$. We denote the two eigenspaces as
\begin{equation}
    E_L \equiv E_{+1}(\alpha\beta) , \quad E_R \equiv E_{-1}(\alpha\beta) .
\end{equation}
{}From the commutation relations
\begin{equation*}
    [ \alpha\beta , \alpha \pm \beta ] = \pm 2 (\alpha \pm \beta)
\end{equation*}
we see that $\alpha + \beta$ (resp.\ $\alpha - \beta$) vanishes on $E_L$ (resp.\ $E_R$) and that the two maps
\begin{equation}
    {\textstyle{\frac{1}{2}}} (\alpha + \beta) :\;  E_R \to E_L \,, \quad
    {\textstyle{\frac{1}{2}}} (\beta - \alpha) :\;  E_L \to E_R \,,
\end{equation}
are isomorphisms that invert each other. Hence we have the following matrix representations with respect to the decomposition $\mathbb{C}^{2r|2r} = E_L \oplus E_R :$
\begin{equation}
    \alpha\beta = \begin{pmatrix} \mathbf{1} &0 \cr 0 &-\mathbf{1} \end{pmatrix} , \quad {\textstyle{\frac{1}{2}}} (\alpha + \beta)
    = \begin{pmatrix} 0 &\mathbf{1} \cr 0 &0 \end{pmatrix} , \quad
    {\textstyle{\frac{1}{2}}} (\beta -\alpha)
    = \begin{pmatrix} 0 &0 \cr \mathbf{1} &0 \end{pmatrix} .
\end{equation}
If $G = \mathrm{Fix}_U(\theta\eta)$ denotes the subgroup of elements in $U$ that commute with $\alpha\beta$, then
\begin{equation}
    G \ni g = \begin{pmatrix} g_L &0 \cr 0 &g_R \end{pmatrix} , \quad g Q_0 g^{-1} = \begin{pmatrix} 0 &M \cr 0 &0 \end{pmatrix} , \quad M = g_L^{\vphantom{-1}} g_R^{-1} ,
\end{equation}
and we conclude that the $U$-invariant metric tensor $\mathrm{STr}\, dQ^2$ vanishes on the ``radial'' subspace generated by $G$ acting on $Q_0\,$.
Our notation is to indicate that $M$ is the WZW field, and $G$ acting on $M$ by $M \mapsto g_L M g_R^{-1}$ is the WZW chiral symmetry group.

In order to generate the full orbit of $U$ on the nilpotent element $Q_0\,$, we may take
\begin{equation}\label{eq:par-Q}
    Q = h g\, Q_0\, (h g)^{-1} , \quad h = \exp \begin{pmatrix} 0 &Y \cr Y &0 \end{pmatrix} .
\end{equation}
Notice that $\theta(h) = h^{-1}$. Therefore, $h \in \exp(\mathfrak{p})$ where $\mathfrak{p}$ is the orthogonal complement of $\mathrm{Lie}\, K$ in $\mathrm{Lie}\, U$. By the definition of $U = \mathrm{U}(r,r|2r)$, the boson-boson part $Y_{00}$ of $Y$ is a Hermitian matrix while the fermion-fermion part $Y_{11}$ is anti-Hermitian. The $U$-invariant metric tensor expands as
\begin{equation}
    \mathrm{STr}\,d Q^2 = 2\,\mathrm{STr}\, (M dY)^2 + \mathcal{O}(Y^4).
\end{equation}
Thus $Y$ gives the non-null (hence stiff) directions. Higher-order terms in the supermatrix field $Y$ play a negligible role, as fluctuations in $Y$ will be suppressed by the diverging stiffness of $Y$.

We now move $Q$ off the nilpotent $U$-orbit by a small amount ($\varepsilon > 0$):
\begin{equation}
    \widetilde{Q} = hg\, Q_\varepsilon (hg)^{-1} , \quad
    Q_\varepsilon = Q_0 + \varepsilon (\alpha - \beta) / 2 \,.
\end{equation}
Note $\widetilde{Q}^2 = - \varepsilon$. The expansion of $\widetilde{Q}$ in powers of $Y$ is
\begin{equation}
    \widetilde{Q} = \begin{pmatrix} - M Y &M \cr - \varepsilon M^{-1} - Y M Y &Y M \end{pmatrix} + \ldots
\end{equation}
Then, scaling the $U$-invariant metric tensor by a diverging factor $\varepsilon^{-1}$ we get
\begin{equation}\label{eq:4.126}
    (2 \varepsilon)^{-1} \mathrm{STr}\, d \widetilde{Q}^2 = \mathrm{STr} \, (M^{-1} dM)^2 + \varepsilon^{-1} \mathrm{STr}\, (M dY)^2 + \mathcal{O} (Y^4).
\end{equation}
The first term on the right-hand side is recognized as the metric tensor of the WZW target space; its coupling will be finite at the RG-fixed point. The second term is a singular correction whose role is to restore the invariance under $U$ at the infinitesimal level (relative to the symmetry-broken state).

\subsubsection{Constructive explanation of $\gamma = 1$}
\label{sect:4.13.3}

We are now ready to explain the origin of the deformation parameter $\gamma = 1$. In a nutshell, our argument is this: (i) the global symmetry $U$ of the SUSY vertex model must be reflected in the CFT description of the RG-fixed point; (ii) the symmetry $U$ is spontaneously broken to the chiral symmetry $G$ of the WZW model (\ref{eq:def-WZW}); (iii) the variables $Y$ appearing in (\ref{eq:4.126}) are the Goldstone modes of the spontaneously broken symmetry; (iv) integration over the Goldstone modes $Y$ with diverging stiffness $\varepsilon^{-1}$ yields exactly the deformation in (\ref{eq:def-WZW}) with parameter $\gamma = 1$. The details are as follows.

First, we re-iterate the logic. As a theory of the IQHE plateau transition, the WZW model (\ref{eq:def-WZW}) {\it per se} is not satisfactory, as it carries no action of the symmetry group $U = \mathrm{U}(r,r|2r)$ of the SUSY vertex model. To re-instate the $U$-symmetry, we must incorporate the field degrees of freedom from $U$ that are not generated by the WZW chiral symmetry $G$. These are given by the supermatrices $Y$. Assuming the scenario of renormalization group flow to a nilpotent orbit $U Q_0 U^{-1}$ ($Q_0^2 = 0$), we have demonstrated that the $U$-invariant metric tensor $\mathrm{STr}\, dQ^2$ consists of the WZW metric tensor $\mathrm{STr} \, (M^{-1} dM)^2$ and the singular term $\varepsilon^{-1} \mathrm{STr} \, (M dY)^2$ (plus higher-order terms that become negligible as $Y \sim \sqrt{\varepsilon}\,$). Therefore, we shall now consider the action functional $S_n^{\rm WZW}[M]$ augmented by
\begin{equation}\label{eq:restore-U}
    \varepsilon^{-1} \int d^2 r \, \mathrm{STr}\, (M \partial_\mu Y)^2 .
\end{equation}

One may ask whether we are missing an additional $2$-form term due to the anomaly in $S_n^{\rm WZW}[M]$. The answer is no. By conformal invariance at the RG-fixed point, such a term would have to be proportional to
\begin{equation}\label{eq:zero-147}
    \pm \mathrm{i} \, \epsilon^{\mu\nu} \varepsilon^{-1} \int d^2 r \, \mathrm{STr}\, (M \partial_\mu Y) (M \partial_\nu Y) = 0 ,
\end{equation}
which vanishes identically by $\mathrm{STr}\, M \partial_\mu Y M \partial_\nu Y = \mathrm{STr}\, M \partial_\nu Y M \partial_\mu Y$.

Changing variables from $Q = u Q_\varepsilon u^{-1}$ to the pair $(M;Y)$ gives rise to a non-trivial Jacobian (or Berezinian). To accommodate it, we recall that $M = g_L^{\vphantom{-1}} g_R^{-1}$ for some (gauge) choice of $g_L$ and $g_R\,$, and we substitute $Y = g_R \widetilde{Y} g_L^{-1}$. Writing $A_\mu = g_L^{-1} \partial_\mu g_L^{\vphantom{-1}}$ and $B_\mu = g_R^{-1} \partial_\mu g_R^{\vphantom{-1}}$ we then obtain
\begin{equation}
    \mathrm{STr} \, (M \partial_\mu Y)^2 = \mathrm{STr} \, (\partial_\mu \widetilde{Y} + A_\mu \widetilde{Y} - \widetilde{Y} B_\mu)^2 .
\end{equation}
{}From this we see that the gauge-dependent variables $\widetilde{Y}$ are the good variables to use in that they come with a Jacobian of unity (independent of $M$). Another way to introduce $\widetilde{Y}$ is to say that we replace the parametrization (\ref{eq:par-Q}) by $\widetilde{Q} = g \widetilde{h}\, Q_{\varepsilon} (g \widetilde{h})^{-1}$ while substituting $\widetilde{h}(\widetilde{Y})$ for $h(Y)$.

We now add the vanishing term (\ref{eq:zero-147}) to the expression (\ref{eq:restore-U}) and compute the Gaussian functional integral
\begin{equation}\label{eq:GFIY}
    \delta S = - \ln \int \mathrm{e}^{- \frac{4}{\varepsilon} \int d^2 r \, \mathrm{STr}\, (M \partial_{\bar z} Y) (M \partial_z Y) }
\end{equation}
with respect to the fields $\widetilde{Y}$. (For later reference, please note the invariance under $z \leftrightarrow \bar{z}$.) The result is (the reciprocal of) a superdeterminant, taken over the $\mathbb{Z}_2$-graded vector space of the supermatrices $\widetilde{Y}$. To write that superdeterminant in good fashion, we introduce the symbols $\mathcal{L}$ and $\mathcal{R}$ for the operators of left and right translation:
\begin{equation}
    Y = g_R \widetilde{Y} g_L^{-1} \equiv
    \mathcal{L}_{g_R} \mathcal{R}_{g_L^{-1}} \widetilde{Y} .
\end{equation}
The result of doing the Gaussian integral is then expressed as
\begin{equation}\label{eq:151-mz}
    \delta S = - \ln \mathrm{SDet}^{-1/2} \left( -
    \mathcal{L}_{g_L^{-1}} \mathcal{R}_{g_R} \partial_{\bar z} \circ
    \mathcal{L}_{M} \mathcal{R}_{M} \partial_z \circ
    \mathcal{L}_{g_R} \mathcal{R}_{g_L^{-1}} \right) .
\end{equation}
We simplify by choosing the gauge $g_R^{-1} = g_L \equiv \widetilde{g}$ and abbreviating
\begin{equation*}
    \mathcal{L}_{\widetilde{g}} \mathcal{R}_{\widetilde{g}} \equiv \widehat{g} \,.
\end{equation*}
The expression (\ref{eq:151-mz}) then becomes
\begin{equation}
    \delta S = \frac{1}{2} \ln \mathrm{SDet} \left( - \widehat{g}^{\, -1} \partial_{\bar z} \circ \widehat{g}^{\, 2} \partial_z \circ \widehat{g}^{\, -1} \right) .
\end{equation}
Discarding zero modes (which, anyway, cannot be present in our situation  as the $U$-symmetry zero modes of the SUSY vertex model are removed by the insertion of point contacts, absorbing boundaries, or similar), this is the same as the superdeterminant of an operator of Dirac type:
\begin{equation}\label{eq:Dirac-det}
    \delta S = \frac{1}{2} \ln \mathrm{SDet} \begin{pmatrix} 0
    &\widehat{g}^{\, -1} \partial_{\bar z} \circ \widehat{g}
    \cr \widehat{g}\, \partial_z \circ \widehat{g}^{\, -1}
    &0 \end{pmatrix} .
\end{equation}
The gradient expansion of such a log-determinant is known \cite{Alvarez-NPB84} to produce a WZW action for $M$, taken in the representation given by $M \mapsto \mathcal{L}_M \mathcal{R}_M\,$.
In the next subsection we are going to show that the result is
\begin{equation}\label{eq:grad-exp}
    \delta S = - \frac{1}{8\pi} \int d^2 r \, \big( \mathrm{STr}\, M^{-1}\partial_\mu M \big)^2 .
\end{equation}
Comparing with the action functional $S_{n,\gamma}^{\rm WZW}[M]$ in (\ref{eq:def-WZW}), we recognize that what we have obtained is the truly marginal deformation term, with a coupling of the correct sign and magnitude for $\gamma = 1$.

Thus we have provided a quantitative detail supporting the hypothesis put forth at the beginning of this subsection: by integrating over the stiff Goldstone modes $Y$ due to the spontaneous symmetry breaking $G \hookrightarrow U = \mathrm{U}(r,r|2r)$, we have produced \emph{precisely} the deformation term in $S_{n,\gamma=1}^{\rm WZW}[M]$. In hindsight, this was to be expected. Indeed, in Sects.\ \ref{sect:deform} and \ref{sect:multif} we showed that the deformation term for $\gamma = 1$ corrects the energy-momentum tensor $T(z)$ in such a way as to yield a spectrum $\Delta_q = \Delta_{1-q}$ of multifractal scaling exponents; at the same time, we know \cite{GLMZ} that the invariance of $\Delta_q$ under Weyl reflections $q \mapsto 1-q$ is a characteristic feature of systems with the class-$A$ symmetry $U = \mathrm{U}(r,r|2r)$.

\subsubsection{Computation of the functional determinant}
\label{sect:4.13.4}

Here we supply a few details of the derivation of the effective action (\ref{eq:grad-exp}) from its definition in Eq.\ (\ref{eq:Dirac-det}). According to Eq.\ (3.25) of \cite{Alvarez-NPB84} (transcribed to our supersymmetric setting), the regular part of $\delta S$ defined in (\ref{eq:Dirac-det}) is
\begin{equation}\label{eq:154-mz}
    \delta S_{\rm reg} = - \frac{1}{8\pi} \int d^2 r \, \mathrm{STr}\, \mathcal{A}_\mu \mathcal{A}_\mu \,,
\end{equation}
where $\mathcal{A}_\mu$ is determined by decomposing the current
\begin{equation*}
    \begin{pmatrix} \widehat{g}^{\, -1} \partial_\mu \widehat{g}
    &0 \cr 0 &\widehat{g}\, \partial_\mu \widehat{g}^{\, -1}
    \end{pmatrix} = \mathcal{V}_\mu + \mathcal{A}_\mu
\end{equation*}
into its vector ($\mathcal{V}$) and axial ($\mathcal{A}$) parts:
\begin{equation}
    \mathcal{A}_\mu = \frac{1}{2} \begin{pmatrix} \widehat{g}^{\,-1} (\partial_\mu \, \widehat{g}^{\,2}) \widehat{g}^{\,-1} &0 \cr 0 &- \widehat{g}^{\,-1} (\partial_\mu\, \widehat{g}^{\,2}) \widehat{g}^{\,-1} \end{pmatrix} .
\end{equation}
Insertion into (\ref{eq:154-mz}) yields
\begin{equation}
    \delta S_{\rm reg} = - \frac{1}{16\pi} \int d^2 r \, \mathrm{STr} \, \big( \widehat{M}^{\,-1} \partial_\mu \widehat{M} \, \big)^2 , \quad \widehat{M} = \widehat{g}^{\,2} .
\end{equation}
It remains to compute the trace in the representation $M \mapsto \widehat{M} = \mathcal{L}_M \mathcal{R}_M :$
\begin{equation}\label{eq:158-mz}
    \mathrm{STr} \, \big( \widehat{M}^{\,-1} \partial_\mu \widehat{M}\, \big)^2 = \mathrm{STr} \, \big( \mathcal{L}_M^{-1} \partial_\mu \mathcal{L}_M + \mathcal{R}_M^{-1} \partial_\mu \mathcal{R}_M \big)^2 .
\end{equation}
The diagonal terms are zero by SUSY cancelation:
\begin{equation*}
    \mathrm{STr} \, \big( \mathcal{L}_M^{-1} \partial_\mu \mathcal{L}_M \big)^2 = 0 = \mathrm{STr} \, \big( \mathcal{R}_M^{-1} \partial_\mu \mathcal{R}_M \big)^2 .
\end{equation*}
Indeed, working in the standard basis $E_{\alpha}^{\; \beta} = e_\alpha \otimes e^{\beta}$ of $\mathrm{End}(\mathbb{C}^{\,r|r})$ we have
\begin{align*}
    \mathrm{STr}_{\mathrm{End}(\mathbb{C}^{r|r})} \big( \mathcal{L}_M^{-1} \partial_\mu \mathcal{L}_M \big)^2 &= \sum_{\alpha \beta} (-1)^{|\beta|} \mathrm{STr}_{\mathbb{C}^{r|r}} E_{\alpha}^{\; \beta} (M^{-1} \partial_\mu M)^2 E_{\beta}^{\; \alpha} \cr
    &= \mathrm{STr} (M^{-1} \partial_\mu M)^2 \times
    \mathrm{STr}\, \mathbf{1} = 0
\end{align*}
due to the equal number of bosonic and fermionic replicas ($\mathrm{STr}\, \mathbf{1} = r - r$).

The argument for the contribution from $\mathcal{R}_M$ is the same. What is left then is the cross term of (\ref{eq:158-mz}):
\begin{equation*}
    \mathrm{STr} \, \big( \widehat{M}^{\,-1} \partial_\mu \widehat{M}\, \big)^2 = 2\, \mathrm{STr} \, \big( \mathcal{L}_M^{-1} \partial_\mu \mathcal{L}_M ) (\mathcal{R}_M^{-1} \partial_\mu \mathcal{R}_M \big) = 2\, \big(\mathrm{STr}\, M^{-1} \partial_\mu M \big)^2 .
\end{equation*}
This already completes the computation of the regular term:
\begin{equation}
    \delta S_{\rm reg} = - \frac{1}{8\pi} \int d^2 r \, \big( \mathrm{STr}\, M^{-1} \partial_\mu M \big)^2 .
\end{equation}
The analogous calculation of the anomalous contribution (a.k.a.\ the WZW term) to the effective action $\delta S$ yields zero. In fact, a non-zero result would depend on the choice of complex structure for the Riemann surface $\Sigma$, in contradiction with the fact that our starting expression (\ref{eq:GFIY}) is invariant under $z \leftrightarrow {\bar z}$. Thus we arrive at the result for $\delta S$ given in Eq.\ (\ref{eq:grad-exp}).

\subsection{Mean conductance}\label{sect:mean-G}

In our reasoning of Sects.\ \ref{sect:Kubo}-\ref{sect:critical} the conductance played a prominent theoretical role, but we still have to make a tangible prediction for it. In the following, by ``conductance'' we mean the longitudinal conductance (as opposed to the Hall conductance). For the disordered network model (or, for that matter, other models of the IQHE plateau transition) the conductance is known to be a fluctuating quantity with a scale-invariant distribution at the critical point. Numerical results are available \cite{ChoFisher, KOK} for square geometries; in the discussion below, we single out the simple case of a square geometry with cylinder topology, i.e., periodic boundary conditions for one pair of opposite edges and open boundary conditions for the other pair.

We will make no attempt to compute the conductance distribution here but specialize to the mean conductance, $\langle G \rangle$. Early numerics for small systems \cite{ChoFisher} gave a value for $\langle G \rangle$ close to $1/2$. A more recent study \cite{KOK} resulted in a larger value, $\langle G \rangle \simeq 0.57$, where the fit of the finite-size numerical data was done with a scaling dimension $y \simeq - 0.56$ for the leading RG-irrelevant perturbation. Now, assuming the framework of Pruisken's non-linear sigma model, the mean conductance $\langle G \rangle$ for a square geometry coincides with the coupling $\sigma_{xx}$ in Eq.\ (\ref{eq:NLsM}), which is known to have the physical interpretation of longitudinal conductivity. Its classical or unrenormalized value in the network model is $\sigma_{xx} = 1/2$ from \cite{MRZ-network}. In our CFT description, the role of $\sigma_{xx}$ is taken over by the stiffness constant of the Gaussian free field with action functional (\ref{eq:GFF}). We thus expect
\begin{equation}\label{eq:pred-G}
    \langle G \rangle_{\rm CFT} = \frac{n}{2\pi} = 0.6366... \quad (\text{for} \; n=4).
\end{equation}

What should we make of the large discrepancy between this new analytical prediction and past numerical findings? For one thing, it is clear from the numerical data that $\langle G \rangle$ goes up with increasing system size; this trend is in line with the scenario of an RG flow from $\langle G \rangle_{{\rm NL}\sigma{\rm M}} = 1/2$ to $\langle G \rangle_{\rm CFT} = 2/\pi$. For another, the value $y \simeq - 0.56$ quoted in \cite{KOK} for the leading irrelevant exponent is unreasonably far from zero; more recently, finite-size scaling fits for related quantities have yielded values for $y$ much closer to zero; cf.\ \cite{SlevinOhtsuki}.

Concerning the last point, we can say the following. A natural candidate for an RG-irrelevant perturbation of the CFT (\ref{eq:def-WZW}) is given by the first term in the WZW action functional (\ref{eq:S-WZW1}). Employing the Kac-Moody currents $\bar{J} \propto M^{-1} \bar\partial M$ and $J \propto \partial M \cdot M^{-1}$ we can write it as
\begin{equation}\label{eq:S-pert}
    S_{\rm pert} \propto \int \mathrm{STr}\, \mathrm{Ad}(M) J \wedge \bar J ,
\end{equation}
where $\mathrm{Ad}(M): \; J \mapsto M J M^{-1}$ stands for the adjoint group action by the field $M$. Since the 1-forms $J = J(z) {\rm d} z$ and $\bar{J} = \bar{J} (\bar{z}) {\rm d}\bar{z}$ are scale-invariant as conserved currents, the perturbation $S_{\rm pert}$ owes its scaling dimension entirely to that of $\mathrm{Ad}(M)$; cf.\ \cite{KZ}. Now the adjoint representation of the Lie supergroup $\mathrm{GL}(r|r)$ is \emph{atypical}, by being degenerate with the trivial representation. In fact, from the expressions (\ref{eq:OPE-JM}) and (\ref{eq:T(z)}) one sees that the operator product expansion of the energy-momentum tensor with $\mathrm{Ad}(M)$ is
\begin{equation}
    T(z) \, M^{\alpha}_{\;\; \beta} (w,\bar{w}) \, (M^{-1})^{\gamma}_{\;\; \delta} (w,\bar{w}) \sim \frac{\delta_\delta^\alpha \, \delta_\beta^\gamma (-1)^{|\beta|}}{n (z-w)^2} + ...
\end{equation}
This entails that the perturbation (\ref{eq:S-pert}) is marginal, while feeding under renormalization into the Abelian deformation $\int \mathrm{STr} J \wedge \mathrm{STr} \bar{J}$ of Eq.\ (\ref{eq:def-WZW}).

There exist two further marginal perturbations of the CFT (\ref{eq:def-WZW}) that one might envisage; these are $\int \mathrm{STr}\, (J \wedge \bar{J})$ and $\int \mathrm{STr}\, \bar\partial J = - \int \mathrm{STr}\, \partial \bar{J}$. Now we appear to be running into a difficulty: we have as many as four marginal perturbations, whereas the Pruisken-Khmelnitskii scaling picture calls for only two perturbations (one relevant, and one irrelevant). Most likely, the resolution is that the (spontaneously broken) $\mathrm{U}(r,r|2r)$ symmetry confines the RG flow to a two-dimensional invariant subspace inside the four-dimensional space of RG-marginal couplings. For example, the perturbation $\int \mathrm{STr}\, (J \wedge \bar{J})$ can be ruled out immediately on symmetry grounds; indeed, $J$ operates in one space (namely, $BF$ in the notation of Sect.\ \ref{sect:4.6}), $\bar{J}$ operates in another (namely, $CG$), and therefore an expression such as $\mathrm{STr}\, (J \wedge \bar{J})$ is not invariantly defined. Put differently, there exists no way for $\mathrm{STr}\, (J \wedge \bar{J})$ to emerge by symmetry breaking from any $\mathrm{U}(r,r|2r)$-invariant expression. (Incidentally, the Chamon-Mudry-Wen perturbations $\mathrm{STr}\, M^q$ are forbidden for the same reason.) In contrast, the perturbation $\mathrm{STr}\, \mathrm{Ad} (M) J \wedge \bar{J}$ does qualify, as the said spaces for $J$ and $\bar{J}$ are exchanged by $M$ and $M^{-1}$.

To shed further light on the situation, recall that parity or reflection invariance is an emerging symmetry at the critical point. Now in the WZW model, parity acts by inversion of the complex structure ($z \leftrightarrow {\bar z}$) in conjunction with inversion of the WZW field ($M \leftrightarrow M^{-1}$); taken together, these two inversions exchange $J \leftrightarrow - \bar{J}$ and thus leave the action functional (\ref{eq:def-WZW}) invariant. (Notice that parity also reverses the orientation of the Riemann surface $\Sigma$.) Clearly, this operation of parity transformation changes the sign of $\int \mathrm{STr}\, \bar\partial J$, but preserves those of the other marginal perturbations.

We are finally ready to deliver the punch line. Starting from the SUSY vertex model near criticality, and taking the rough picture of Pruisken-Khmelnitskii scaling for granted, we expect the RG flow to explore a two-dimensional manifold containing the CFT (\ref{eq:def-WZW}) as a fixed point; our parity-breaking perturbation $\int \mathrm{STr}\, \bar\partial J$ presumably generates the direction of unstable flow. Now regardless of the exact details, we are confronted with the observation that all our perturbations near the RG-fixed point are \emph{marginal}! The situation gets complicated, of course, by the circumstance that the critical RG flow must first approach the nilpotent $U$-orbit in order to achieve the spontaneous symmetry breaking $G \hookrightarrow U = \mathrm{U}(r,r|2r)$. From our rough considerations in Sect.\ \ref{sect:SBS} we cannot infer the length scale for the symmetry breaking to be fully established. However, once that scale has been reached, the CFT (\ref{eq:def-WZW}) takes over and predicts the RG flow to vanish to linear order in the couplings around the fixed point! The RG then starts flowing only in quadratic order. This means that the putative scaling dimensions $1/\nu > 0$ and $y < 0$ do not have well-defined values; rather, they keep getting smaller and smaller as one goes closer and closer to the RG-fixed point. In this scenario, the ultimate approach to the CFT-value (\ref{eq:pred-G}) is expected to be very slow; larger system sizes and an improved finite-size scaling ansatz will be needed to verify it.

\section*{Acknowledgment}

I thank R.\ Bondesan and A.\ Saberi for collaboration related to this paper. Acknowledgments also go to J.T.\ Chalker, N.\ Nekrasov, Z.\ Komargodski, and M.\ Janssen for helpful discussions. I am grateful for the hospitality of the Simons Center for Geometry and Physics (State University of New York at Stony Brook, USA), where a substantial part of the paper took shape. This multi-year project was supported by the DFG grant ZI 513/2-1.


\begin{thebibliography}{99}
\bibitem{footnote1} I intend to demonstrate that fact in a follow-up paper using a novel functional integral representation of the network model.
\end{thebibliography}
\end{document}